\documentclass[]{aa}
\usepackage{txfonts}
\usepackage{natbib,twoopt}
\usepackage[breaklinks=true]{hyperref} %% to avoid \citeads line fills
\bibpunct{(}{)}{;}{a}{}{,} %% natbib format for A&A and ApJ
\makeatletter
\newcommandtwoopt{\citeads}[3][][]{\href{http://adsabs.harvard.edu/abs/#3}%
{\def\hyper@linkstart##1##2{}%
\let\hyper@linkend\@empty\citealp[#1][#2]{#3}}}
\newcommandtwoopt{\citepads}[3][][]{\href{http://adsabs.harvard.edu/abs/#3}%
{\def\hyper@linkstart##1##2{}%
\let\hyper@linkend\@empty\citep[#1][#2]{#3}}}
\newcommandtwoopt{\citetads}[3][][]{\href{http://adsabs.harvard.edu/abs/#3}%
{\def\hyper@linkstart##1##2{}%
\let\hyper@linkend\@empty\citet[#1][#2]{#3}}}
\newcommandtwoopt{\citeyearads}[3][][]%
{\href{http://adsabs.harvard.edu/abs/#3}
{\def\hyper@linkstart##1##2{}%
\let\hyper@linkend\@empty\citeyear[#1][#2]{#3}}}
\makeatother

\begin{document}

\title{An edge-on translucent dust disk around the nearest AGB star, L$_2$~Puppis\thanks{Based on observations made with ESO telescopes at Paranal Observatory, under ESO programs 090.D-0144(A), 074.D-0198(C) and an unreferenced VLTI/VINCI program.}}
\subtitle{VLT/NACO spectro-imaging from 1.04 to 4.05\,$\mu$m and VLTI interferometry}
\titlerunning{An edge-on translucent dust disk around the nearest AGB star, L$_2$~Puppis}
\authorrunning{P. Kervella et al.}
\author{
P.~Kervella\inst{1}
\and
M.~Montarg\`es\inst{1}
\and
S.~T.~Ridgway\inst{2}
\and
G.~Perrin\inst{1}
\and
O.~Chesneau\inst{3}
\and
S.~Lacour\inst{1}
\and
A.~Chiavassa\inst{3}
\and
X.~Haubois\inst{1}
\and
A.~Gallenne\inst{4}
}
\offprints{P. Kervella}
\mail{pierre.kervella@obspm.fr}
\institute{
LESIA, Observatoire de Paris, CNRS UMR 8109, UPMC, Universit\'e Paris Diderot,
5 place Jules Janssen, F-92195 Meudon, France
\and
National Optical Astronomy Observatories, 950 North Cherry Avenue, Tucson, AZ 85719, USA
\and
Laboratoire Lagrange, UMR 7293, Universit\'e de Nice-Sophia Antipolis, CNRS, Observatoire de la C\^ote d'Azur, Bd de l'Observatoire, B.P. 4229, F-06304 Nice cedex 4, France
\and
Universidad de Concepci{\'o}n, Departamento de Astronom\'{\i}a, Casilla 160-C, Concepci{\'o}n, Chile
}
\date{Received 18 December 2013; Accepted 4 March 2014}
\abstract
{As the nearest known AGB star ($d = 64$\,pc) and one of the brightest ($m_K \approx -2$), L$_2$ Pup is a particularly interesting benchmark object to monitor the final stages of stellar evolution. We report new serendipitous imaging observations of this star with the VLT/NACO adaptive optics system in twelve narrow-band filters covering the $1.0-4.0\,\mu$m wavelength range. These diffraction-limited images reveal an extended circumstellar dust lane in front of the star that exhibits a high opacity in the $J$ band and becomes translucent in the $H$ and $K$ bands. In the $L$ band, extended thermal emission from the dust is detected. We reproduced these observations using Monte Carlo radiative transfer modeling of a dust disk with the RADMC-3D code. We also present new interferometric observations with the VLTI/VINCI and MIDI instruments.
We measured in the $K$ band an upper limit to the limb-darkened angular diameter of $\theta_\mathrm{LD} = 17.9 \pm 1.6$\,mas, converting to a maximum linear radius of $R = 123 \pm 14\,R_\odot$. Considering the geometry of the extended $K$ band emission in the NACO images, this upper limit is probably close to the actual angular diameter of the star.
The position of L$_2$\,Pup in the Hertzsprung-Russell diagram indicates that this star has a mass of about $2\,M_\odot$ and is probably experiencing an early stage of the asymptotic giant branch. We did not detect any stellar companion of L$_2$\,Pup in our adaptive optics and interferometric observations, and we attribute its apparent astrometric wobble in the \emph{Hipparcos} data to variable lighting effects on its circumstellar material.
However, we do not exclude the presence of a binary companion,
because the large loop structure extending to more than 10\,AU to the northeast of the disk in our $L$ -band images may be the result of interaction between the stellar wind of L$_2$\,Pup and a hidden secondary object.
The geometric configuration that we propose, with a large dust disk seen almost edge-on, appears particularly favorable to test and develop our understanding of the formation of bipolar nebulae.
}

\maketitle

%__________________________________Introduction
\section{Introduction}

Evolved stars are important contributors to the enrichment of heavy elements in the interstellar medium, and more generally to the chemical evolution of the Universe. L$_2$\,Puppis (\object{HD 56096}, \object{HIP 34922}, \object{HR 2748}) is an asymptotic giant branch (AGB) semiregular variable. Its variability was discovered by Gould as early as 1872 \citepads{1907AnHar..55....1C}. Its M5III spectral type corresponds to an approximate effective temperature of $T_\mathrm{eff} = 3500$\,K, which is what we considered here. Its proximity \citepads[$\pi = 15.61 \pm 0.99$\,mas,][]{2007A&A...474..653V} makes it the closest AGB star and one of the brightest stars in the infrared sky.
\citetads{2007ApJS..173..137G} identified a periodic shift in the \emph{Hipparcos} astrometric position of L$_2$\,Pup with a 141-day period and a semimajor axis of 9.5\,mas. They attributed this displacement to the orbital reflex motion of the AGB star due to an unresolved companion. The corresponding orbital period is almost identical to the photometric variation period (140.6\,days) as listed in the General Catalogue of Variable Stars \citepads{2009yCat....102025S}.
\citetads{2002MNRAS.337...79B} explained the long-term (over decades) variability of the brightness of L$_2$\,Pup as the consequence of the obscuration of the star by circumstellar dust. These authors also pointed out that the period of L$_2$\,Pup has been remarkably stable over 75\,years of photometric observations, making it a semiregular variable of the SRa type (i.e. with a well-defined period), closely related to Miras.
\citetads{2005A&A...431..623L} obtained six radial velocity measurements spread around the maximum and minimum light phases, and estimated a radial velocity amplitude of 12 km.s$^{-1}$. The binarity hypothesis was discussed (and dismissed) by \citetads{2009A&A...498..489J}, based in particular on geometrical arguments on the linear size of the giant star. 
\citetads{2013ApJ...774...21M} recently discovered a 139-day periodic velocity centroid variation from SiO maser emission. They concluded that this variability points at the presence of differential illumination, or an asymmetric distribution of the circumstellar material around L$_2$\,Pup.
Alternatively, \citetads{2009MNRAS.394...51G} predicted that the centroid of SiO emission would shift in velocity according to the stage in the stellar cycle, as a consequence of the shock behavior and because the SiO maser region is located in a compact region, within a few stellar radii of the central object.
\citetads{2014A&A...561A..47O} recently observed L$_2$\,Pup in the thermal infrared domain ($N$ band) using the high spectral resolution mode of VISIR, and concluded that its spectrum cannot be reproduced satisfactorily using MARCS atmosphere models. However, this author considered a temperature of 2800\,K for the central star, which may be underestimated (see Sect.~\ref{photom}).

We present in Sect.~\ref{observations} our new NACO, VINCI and MIDI observations of L$_2$\,Pup, and in Sect.~\ref{radmc} the RADMC-3D radiative transfer model we propose to reproduce these observations. This model consists of a central star surrounded by an edge-on circumstellar disk. Section~\ref{discussion} is dedicated to a discussion of the evolutionary status, mass loss geometry, and possible binarity of L$_2$\,Pup.

%__________________________________Observations
\section{Observations and data processing\label{observations}}

\subsection{VLT/NACO imaging\label{nacoobs}}

\subsubsection{Instrumental setup\label{nacosetup}}

We observed L$_2$\,Pup on the night of 21 March 2013 using the Nasmyth Adaptive Optics System \citepads{2003SPIE.4839..140R} of the Very Large Telescope (VLT), coupled to the CONICA infrared camera \citepads{1998SPIE.3354..606L}, which is abbreviated as NACO. As its name indicates, NACO is installed at the Nasmyth focus of the Unit Telescope~4 (Yepun), located on the eastern side of the VLT platform. NAOS is equipped with a tip-tilt mirror and a deformable mirror controlled by 185 actuators, as well as two wavefront sensors: one for visible light, and one for the infrared domain. For our observations, we used the visible light wavefront sensor, as L$_2$\,Pup is bright in the R band \citepads[$m_R = 2.45$;][]{2002yCat.2237....0D}. 

%_________Table of NACO observations
\begin{table}
        \caption{Log of the NACO observations of L$_2$ Pup and its associated PSF calibrators, $\beta$\,Col and $\alpha$\,Lyn.}
        \centering          
        \label{naco_log}
        \begin{tabular}{lllll}
        \hline \hline
        \noalign{\smallskip}
        MJD & Star & Filter & DIT [ms] & AM \\
         & & [$\mu$m] & $\times$ NDIT & \\
         \hline         
        \noalign{\smallskip}
        56364.1751 & $\alpha$\,Lyn & 3.74 & 8.4 $\times$ 3900 & 2.189 \\
        56364.1757 & $\alpha$\,Lyn & 4.05 & 8.4 $\times$ 3900 & 2.194 \\
        56372.0463 & $\beta$\,Col & 1.75 & 7.2 $\times$ 4200 & 1.202 \\
        56372.0468 & $\beta$\,Col & 1.64 & 7.2 $\times$ 4200 & 1.204 \\
        56372.0476 & $\beta$\,Col & 1.28 & 7.2 $\times$ 4200 & 1.207 \\
        56372.0482 & $\beta$\,Col & 1.26 & 7.2 $\times$ 4200 & 1.210 \\
        56372.0489 & $\beta$\,Col & 1.24 & 7.2 $\times$ 4200 & 1.212 \\
        56372.0495 & $\beta$\,Col & 1.09 & 7.2 $\times$ 4200 & 1.215 \\
        56372.0501 & $\beta$\,Col & 1.08 & 7.2 $\times$ 4200 & 1.218 \\
        56372.0508 & $\beta$\,Col & 1.04 & 7.2 $\times$ 4200 & 1.220 \\
        56372.0515 & $\beta$\,Col & 2.12 & 7.2 $\times$ 4200 & 1.224 \\
        56372.0521 & $\beta$\,Col & 2.17 & 7.2 $\times$ 4200 & 1.226 \\
        56372.0534$^*$ & $\beta$\,Col & 3.74 & 8.4 $\times$ 3900 & 1.232 \\
        56372.0540$^*$ & $\beta$\,Col & 4.05 & 8.4 $\times$ 3900 & 1.235 \\
        \hline         
        \noalign{\smallskip}
        56372.0609 & L$_2$\,Pup & 1.75 & 7.2 $\times$ 4200 & 1.119 \\
        56372.0613 & L$_2$\,Pup & 1.75 & 7.2 $\times$ 4200 & 1.120 \\
        56372.0626 & L$_2$\,Pup & 1.64 & 7.2 $\times$ 4200 & 1.122 \\
        56372.0630 & L$_2$\,Pup & 1.64 & 7.2 $\times$ 4200 & 1.123 \\
        56372.0637 & L$_2$\,Pup & 1.28 & 7.2 $\times$ 4200 & 1.124 \\
        56372.0641 & L$_2$\,Pup & 1.28 & 7.2 $\times$ 4200 & 1.125 \\
        56372.0648 & L$_2$\,Pup & 1.26 & 7.2 $\times$ 4200 & 1.127 \\
        56372.0652 & L$_2$\,Pup & 1.26 & 7.2 $\times$ 4200 & 1.128 \\
        56372.0659 & L$_2$\,Pup & 1.24 & 7.2 $\times$ 4200 & 1.129 \\
        56372.0663 & L$_2$\,Pup & 1.24 & 7.2 $\times$ 4200 & 1.130 \\
        56372.0670 & L$_2$\,Pup & 1.09 & 7.2 $\times$ 4200 & 1.131 \\
        56372.0674 & L$_2$\,Pup & 1.09 & 7.2 $\times$ 4200 & 1.132 \\
        56372.0678 & L$_2$\,Pup & 1.09 & 7.2 $\times$ 4200 & 1.133 \\
        56372.0685 & L$_2$\,Pup & 1.08 & 7.2 $\times$ 4200 & 1.134 \\
        56372.0689 & L$_2$\,Pup & 1.08 & 7.2 $\times$ 4200 & 1.135 \\
        56372.0693 & L$_2$\,Pup & 1.08 & 7.2 $\times$ 4200 & 1.136 \\
        56372.0700 & L$_2$\,Pup & 1.04 & 7.2 $\times$ 4200 & 1.138 \\
        56372.0704 & L$_2$\,Pup & 1.04 & 7.2 $\times$ 4200 & 1.139 \\
        56372.0708 & L$_2$\,Pup & 1.04 & 7.2 $\times$ 4200 & 1.140 \\
        56372.0716 & L$_2$\,Pup & 2.12 & 7.2 $\times$ 4200 & 1.141 \\
        56372.0720 & L$_2$\,Pup & 2.12 & 7.2 $\times$ 4200 & 1.142 \\
        56372.0727 & L$_2$\,Pup & 2.17 & 7.2 $\times$ 4200 & 1.144 \\
        56372.0731 & L$_2$\,Pup & 2.17 & 7.2 $\times$ 4200 & 1.145 \\
        56372.0770 & L$_2$\,Pup & 3.74 & 8.4 $\times$ 3900 & 1.154 \\
        56372.0775 & L$_2$\,Pup & 3.74 & 8.4 $\times$ 3900 & 1.155 \\
        56372.0781 & L$_2$\,Pup & 4.05 & 8.4 $\times$ 3900 & 1.157 \\
        56372.0786 & L$_2$\,Pup & 4.05 & 8.4 $\times$ 3900 & 1.158 \\
        \hline
        \end{tabular}
        \tablefoot{MJD is the average modified julian date of the exposures. DIT is the individual exposure time of the NACO frames, and NDIT is the total number of frames recorded. AM is the airmass.
        $^*$ The observations of $\beta$\,Col in the L band (3.74 and 4.05\,$\mu$m) were affected by a centering problem and were therefore replaced by another PSF standard ($\alpha$\,Lyn).}
\end{table}

We employed 12 narrow-band filters\footnote{\url{http://www.eso.org/sci/facilities/paranal/instruments/naco/}} spread over the infrared $JHKL$ bands, with central wavelengths of 1.04, 1.08, 1.09, 1.24, 1.26, 1.28, 1.64, 1.75, 2.12, 2.17, 3.74, and 4.05\,$\mu$m. For the $JHK$ band filters (1.04 to 2.17\,$\mu$m), we selected the S13 camera that provides the smallest available pixel scale of $13.26 \pm 0.03$\,mas/pixel \citepads{2003A&A...411..157M, 2008A&A...484..281N}, giving a field of view of 13.6$\arcsec$. In the $L$ band (3.74 and 4.05\,$\mu$m), we used the L27 camera, which has a pixel scale of 27.05\,mas/pixel \citepads{2010A&A...511A..18S} and a field of view of $27.7\arcsec$. Because of the brightness of L$_2$\,Pup, we also used neutral-density filters, labeled ``{\tt ND2\_short}" for the wavelengths between 1.04 and 2.17\,$\mu$m and ``{\tt ND2\_long}" for the 3.74 and 4.05\,$\mu$m filters. These two filters have respective transmissions of about 1.3\% and 2\%. We selected the small 64 pixel ($JHK$) or 120 pixel ($L$) windows of the CONICA detector to obtain the highest frame frequency. The detector integration time (DIT) was set to the minimum possible: ${\rm DIT}=7.2$\,ms in the $JHK$ bands, and ${\rm DIT}=8.4$\,ms for the two L band filters, to freeze the residual atmospheric distortions of the images. 

%______________ Figure
\begin{figure}[]
        \centering
        \includegraphics[width=\hsize]{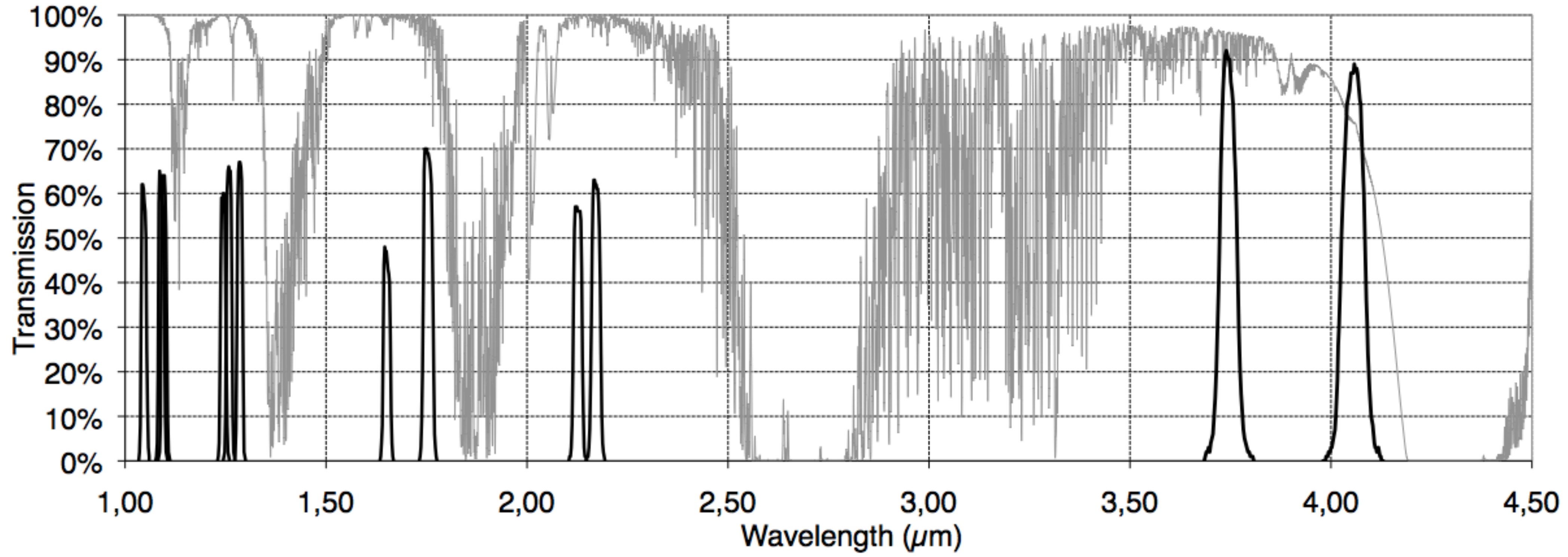} 
        \caption{Transmissions of the 12 CONICA narrow-band filters used for the observations of L$_2$\,Pup (black profiles) and of the atmosphere \citep[thin gray curve,][]{1992lord}.\label{nb_filters}}
\end{figure}

\subsubsection{Raw data reduction\label{rawreduction}}

Following \citetads{2009A&A...504..115K}, we processed the raw image cubes using a serendipitous imaging technique. The individual raw images were pre-processed (bias subtraction, flat-fielding, bad-pixel masking) using the Yorick\footnote{\url{http://yorick.sourceforge.net/}} and IRAF\footnote{\url{http://iraf.noao.edu}} software packages in a standard way. The quality of adaptive optics images is affected by a residual jitter of the star on the detector (tip-tilt), and by high-order residual wavefront distorsions. Taking advantage of the \emph{cube} mode of CONICA, our extremely short exposures allowed us to freeze the image jitter. After a precentering at the integer pixel level, the images were sorted based on their maximum intensity, used as a proxy of the Strehl ratio. The 50\% best images of each cube were then resampled up by a factor 4 (in the $JHK$ bands) or 2 (in the L band) using a cubic spline interpolation, and the star image was precisely centered using Gaussian fitting. The selected and centered image cubes were eventually averaged to obtain the master images of each star used in the following analysis. A more stringent selection (e.g. considering only 10\% of the frames) does not result in a significantly better image quality, as most of the residual image degradation is caused by the tip-tilt errors and not by higher order pertubations.

\subsubsection{Photometric calibration}

\begin{figure*}[]
        \centering
        \includegraphics[width=4.4cm]{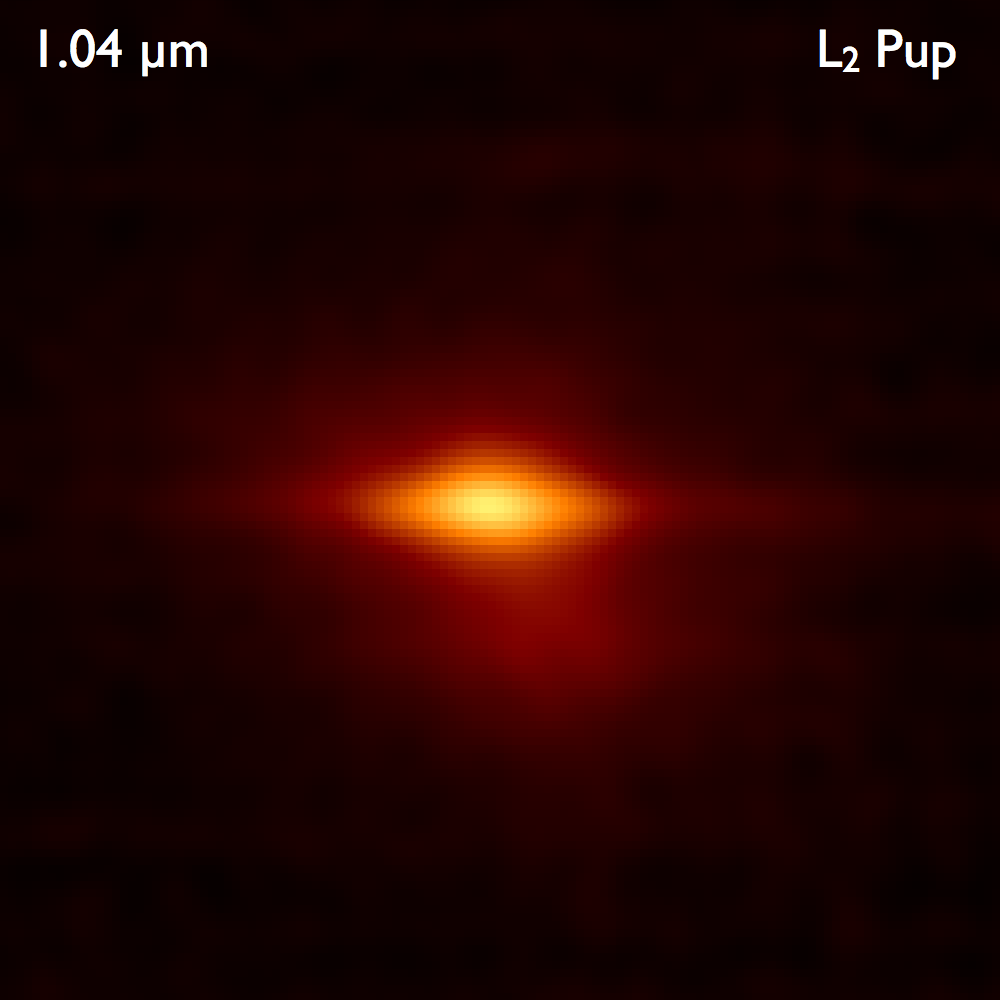} \includegraphics[width=4.4cm]{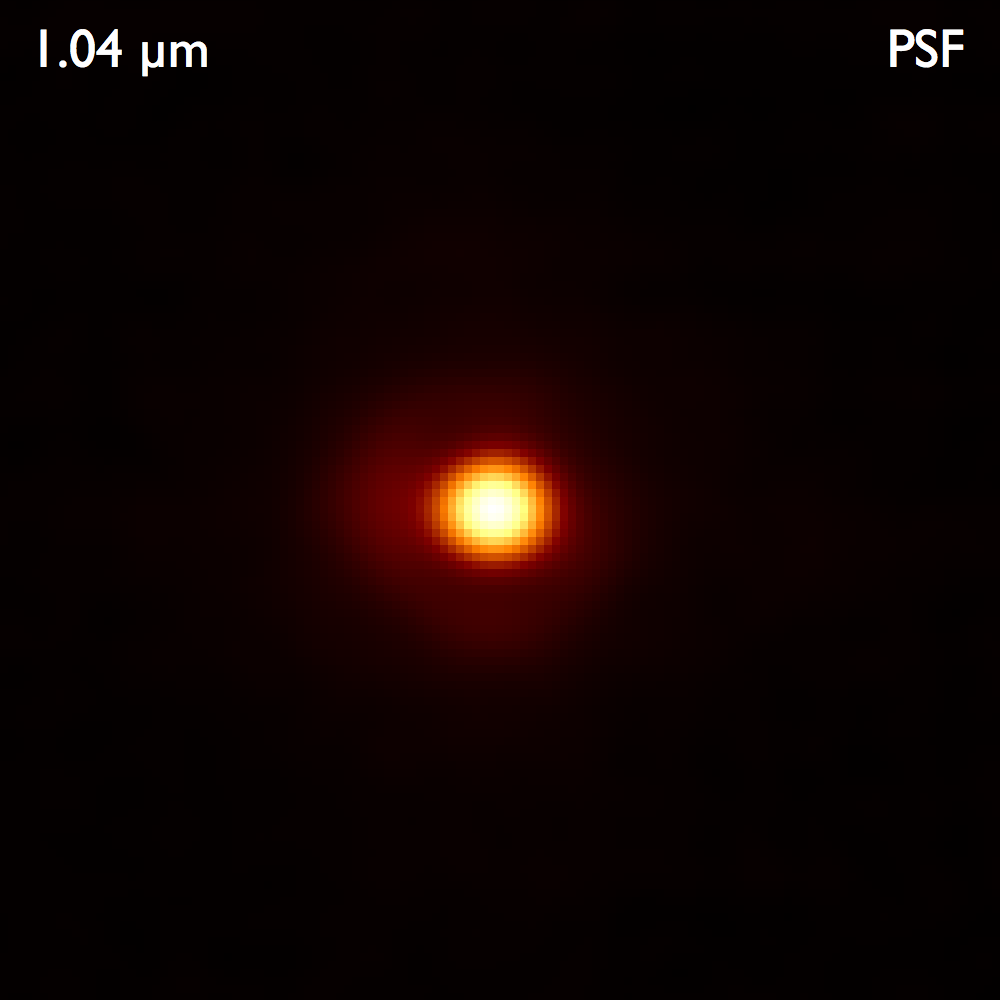} \vspace{1mm}
        \hspace{1mm}
        \includegraphics[width=4.4cm]{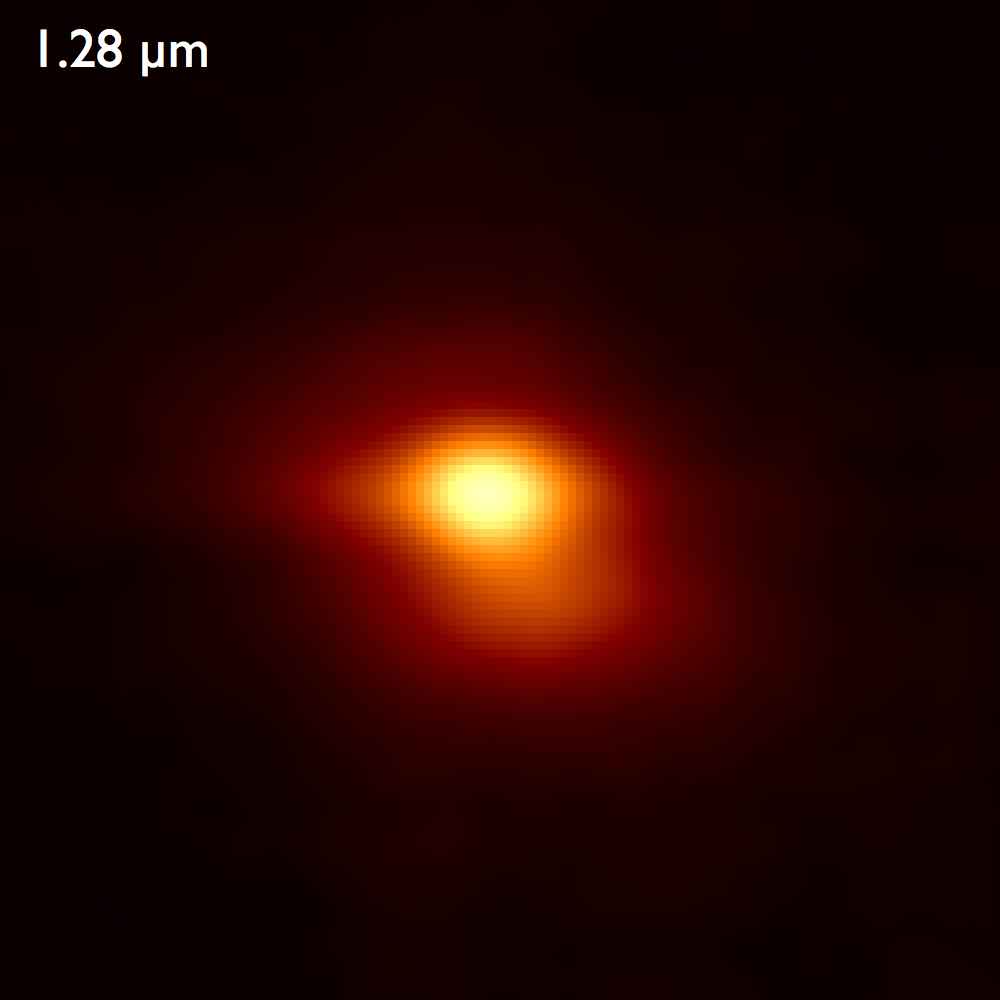} \includegraphics[width=4.4cm]{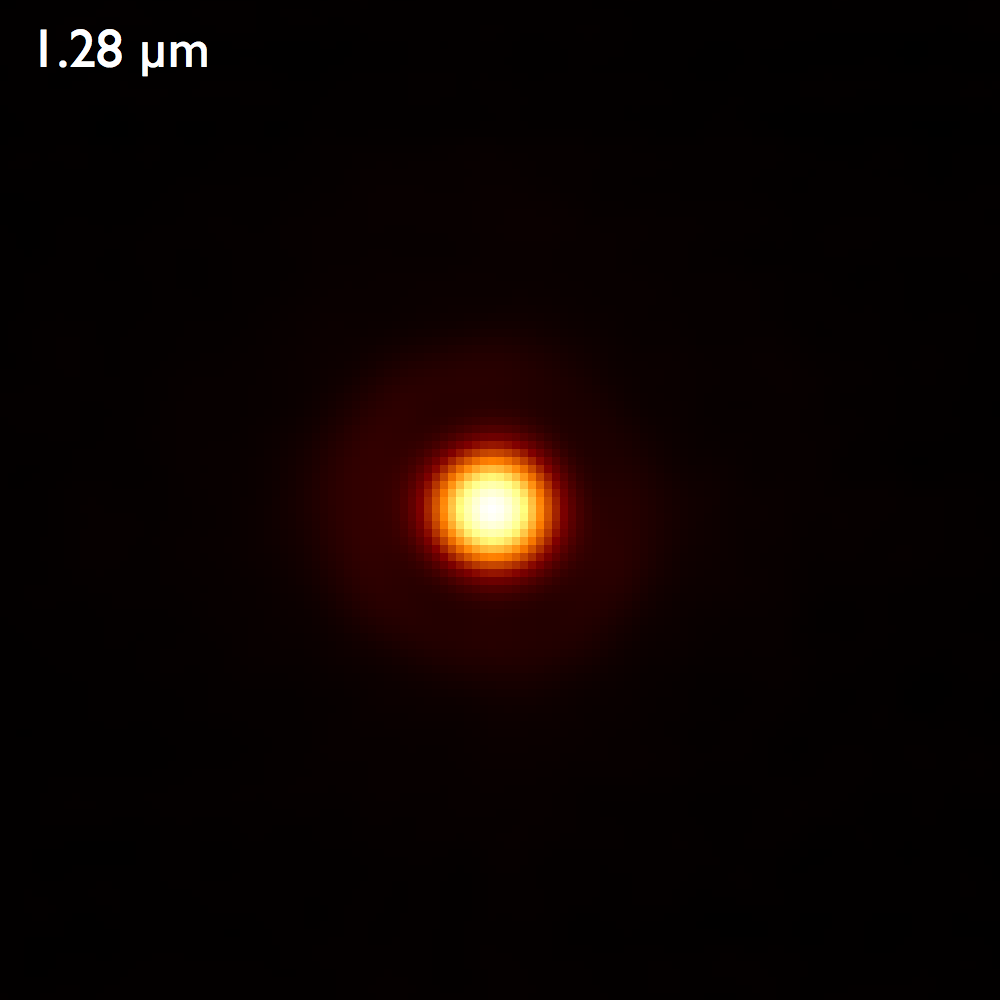}
        \vspace{1mm}
        \includegraphics[width=4.4cm]{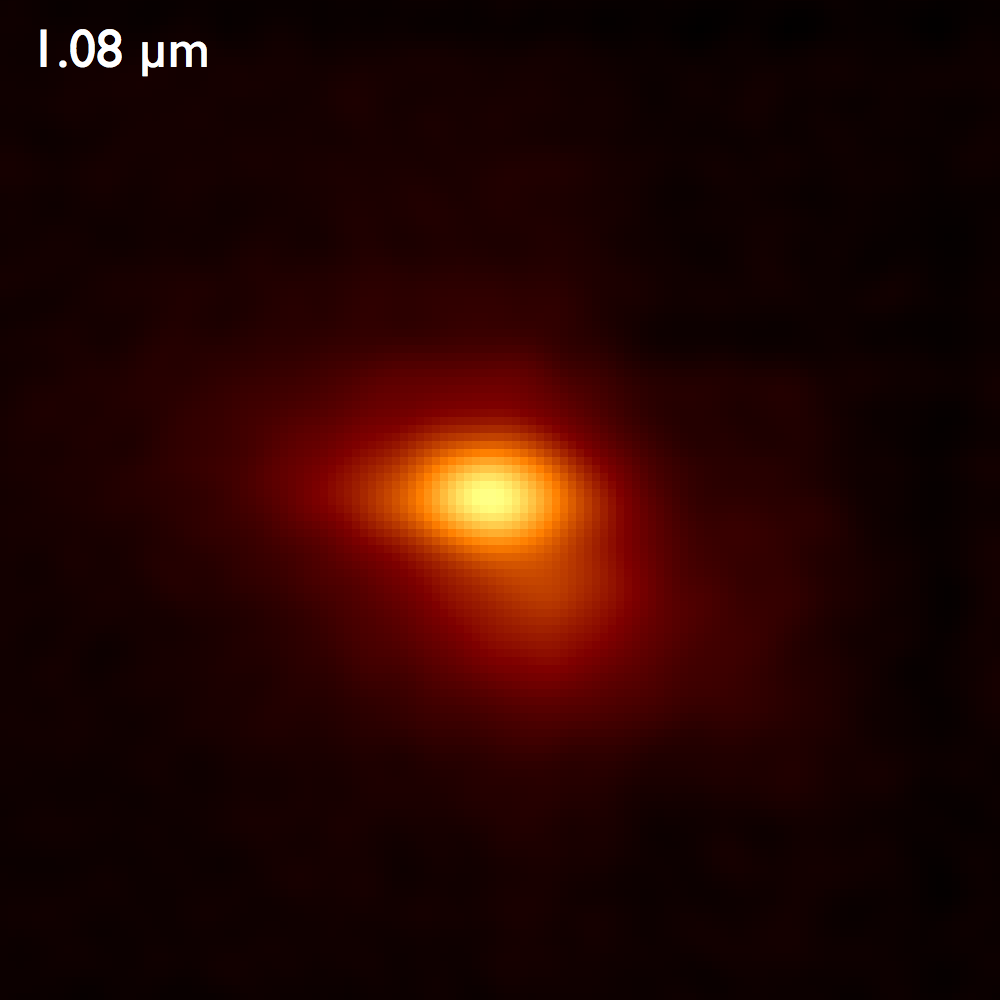} \includegraphics[width=4.4cm]{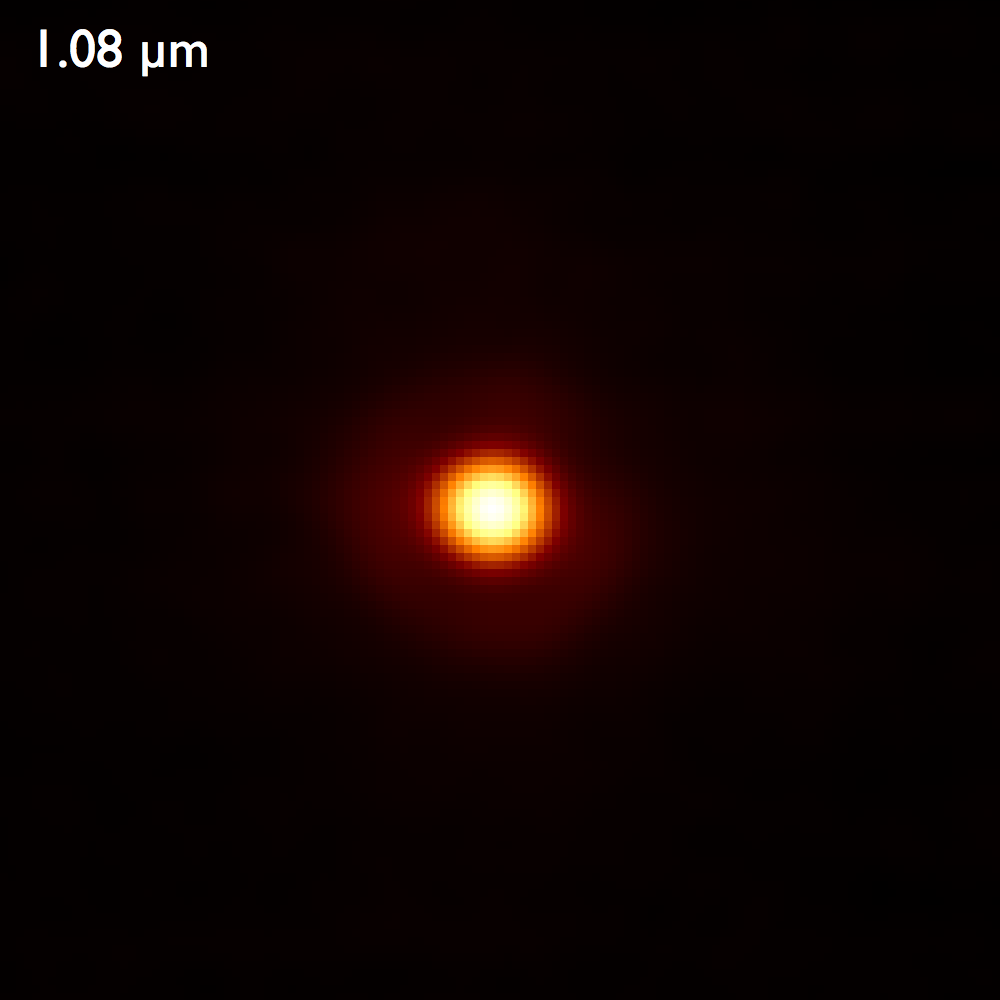} \hspace{1mm}
        \includegraphics[width=4.4cm]{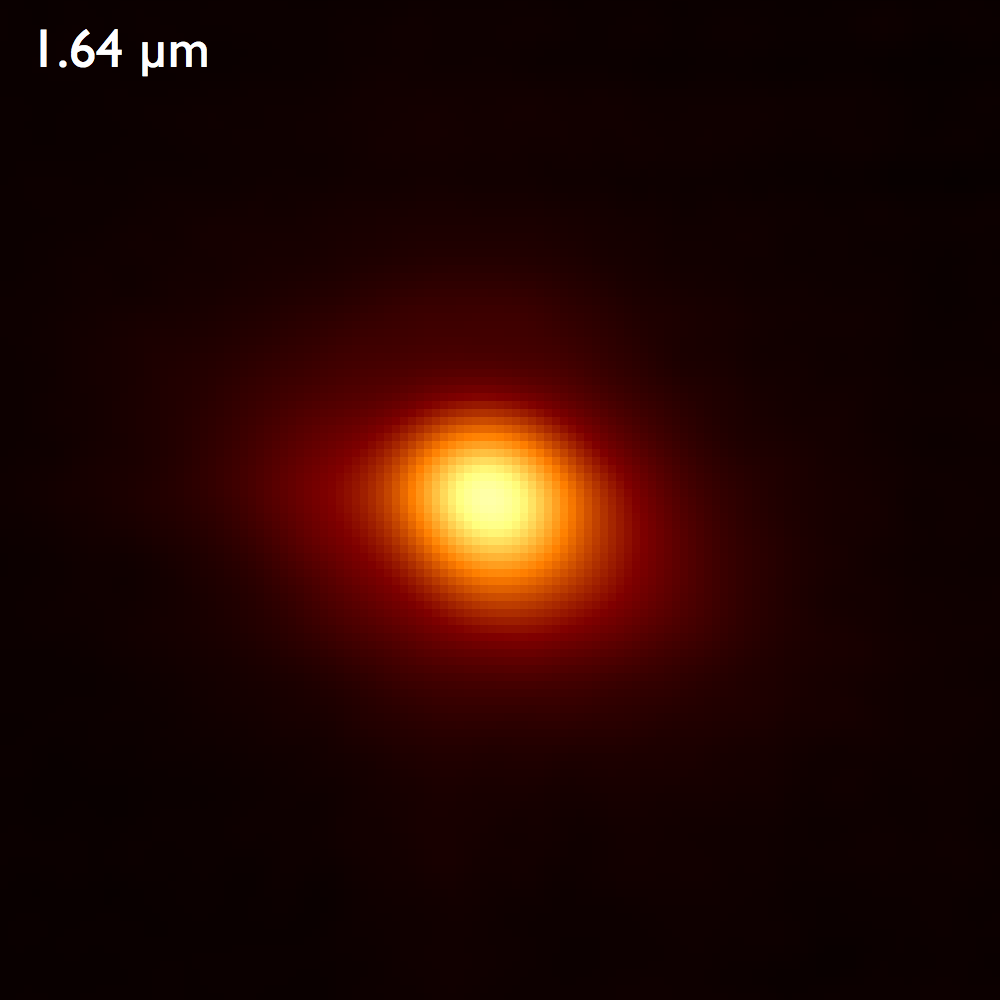} \includegraphics[width=4.4cm]{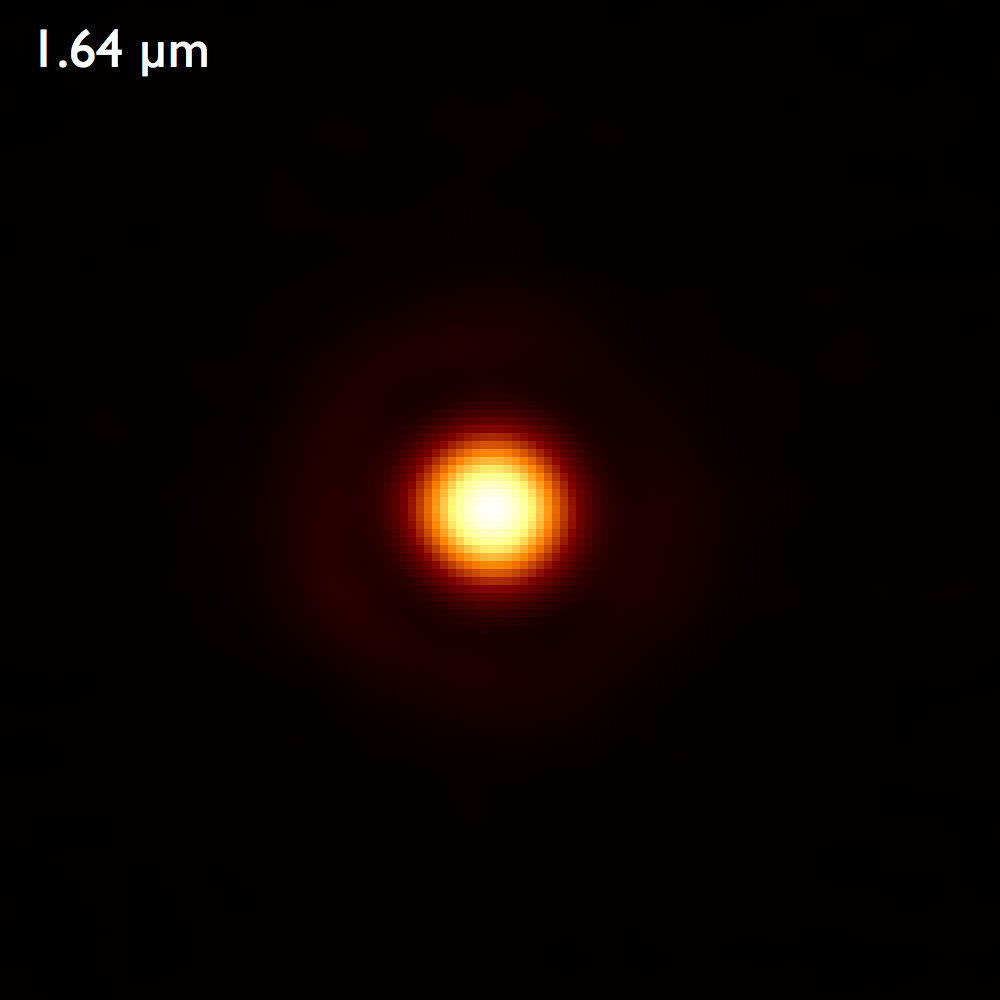}
        \includegraphics[width=4.4cm]{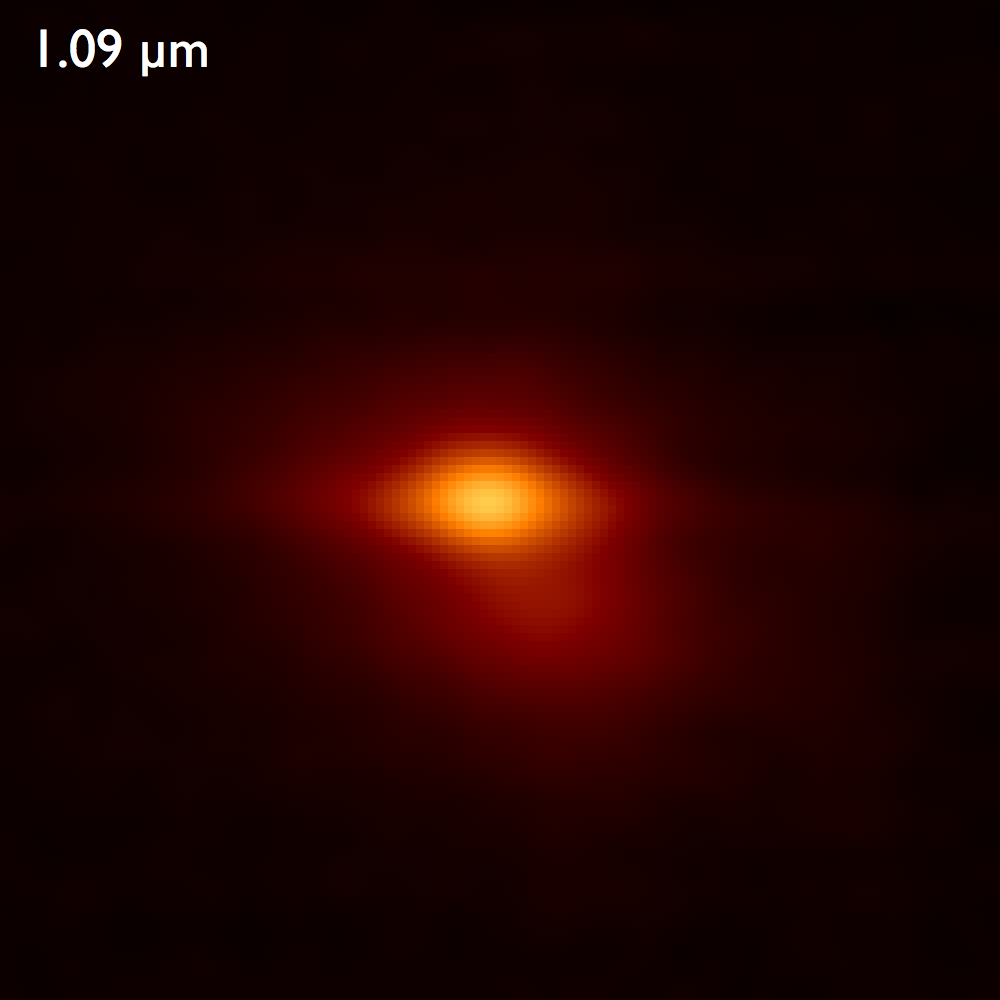} \includegraphics[width=4.4cm]{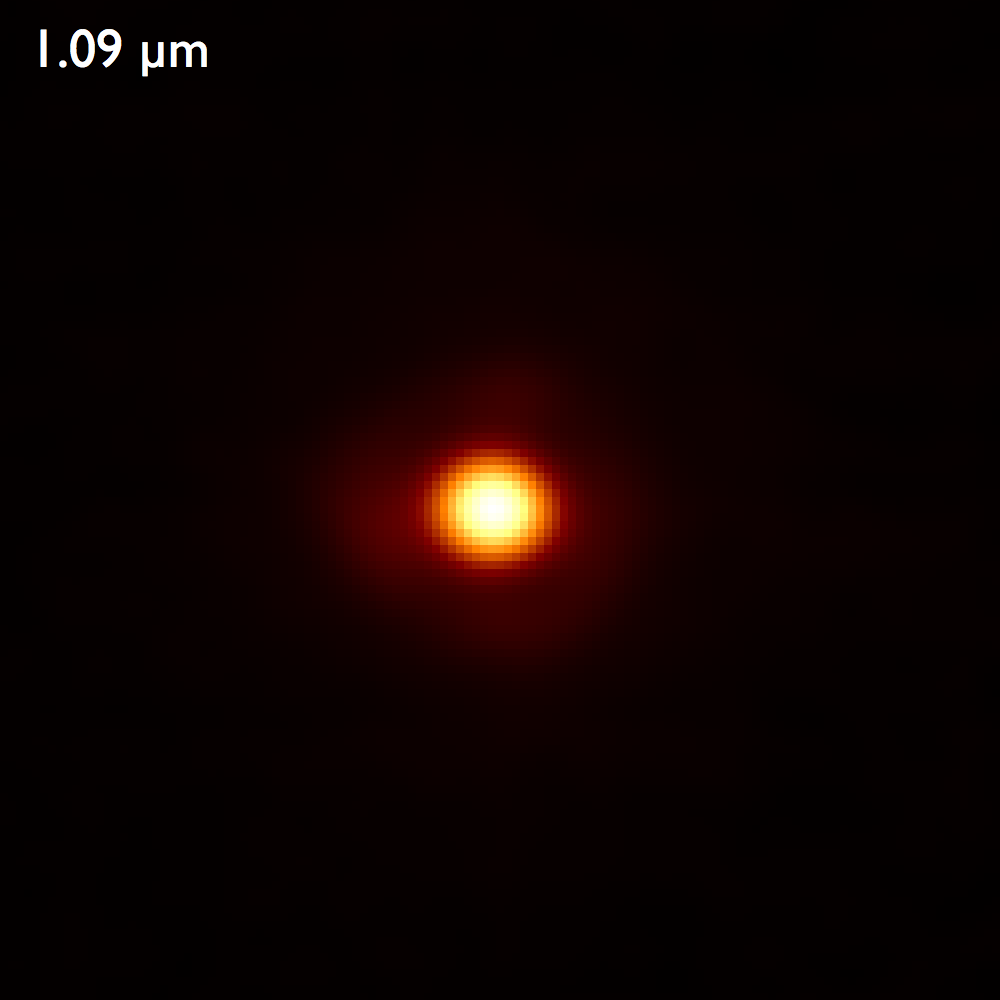} \vspace{1mm}
        \hspace{1mm}
        \includegraphics[width=4.4cm]{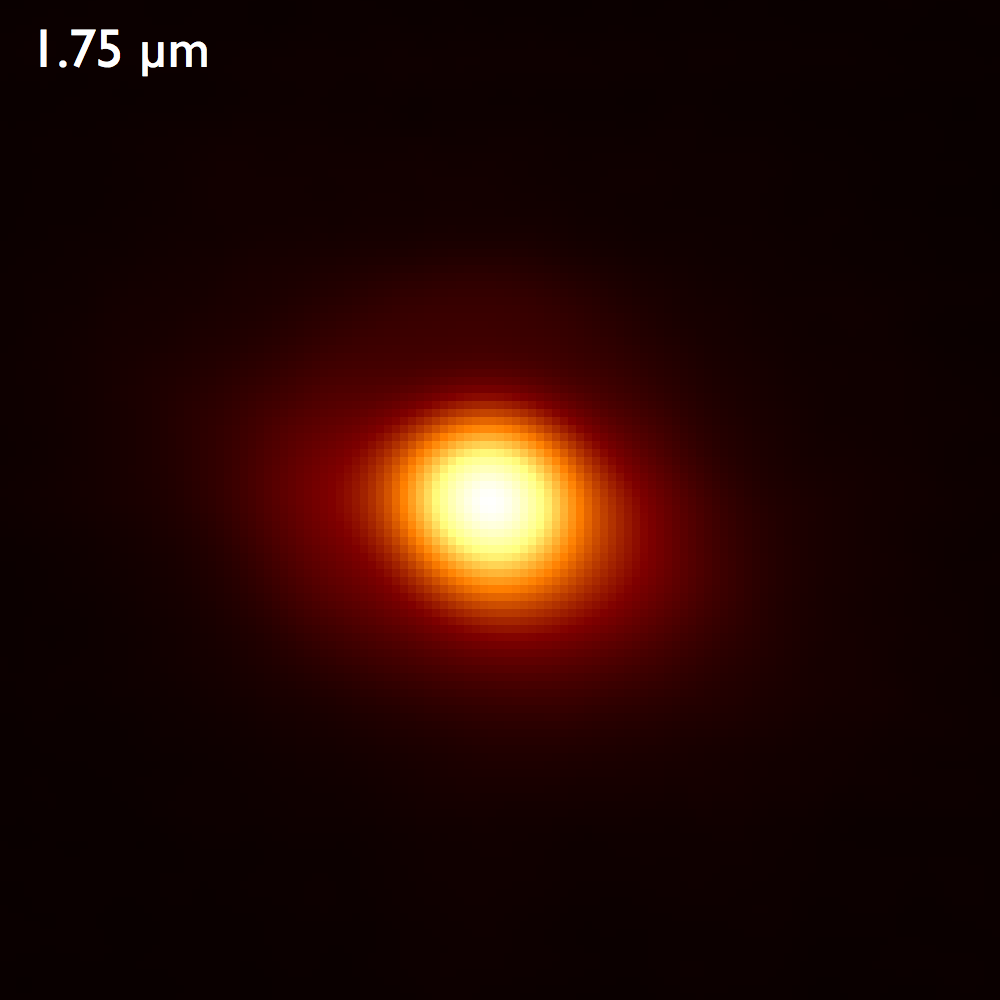} \includegraphics[width=4.4cm]{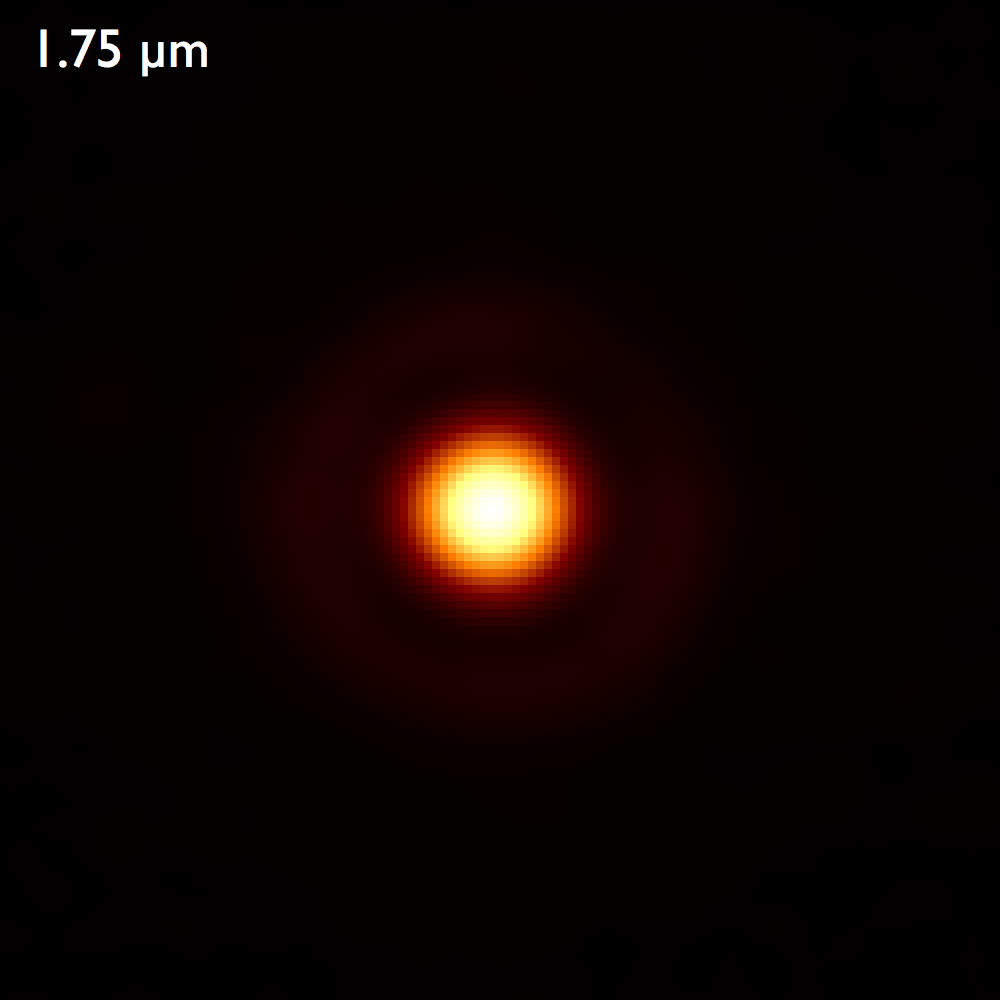}
        \includegraphics[width=4.4cm]{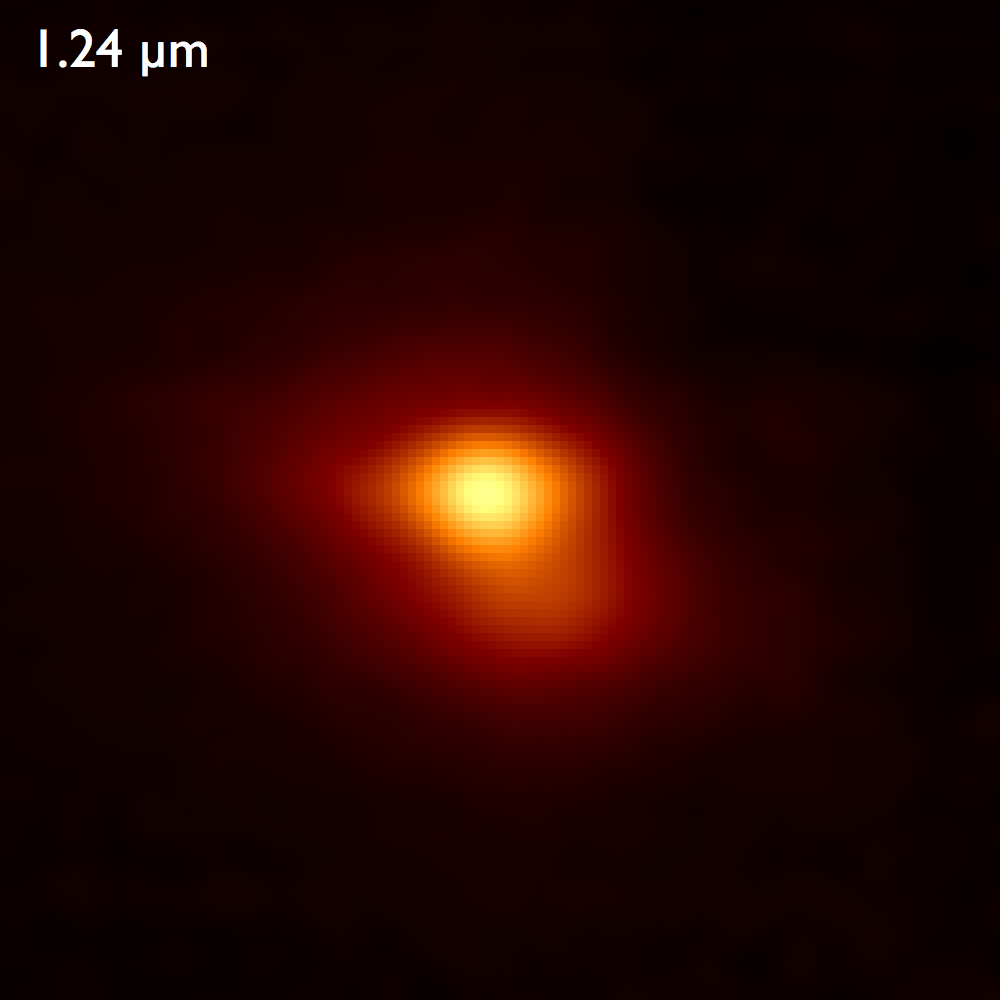} \includegraphics[width=4.4cm]{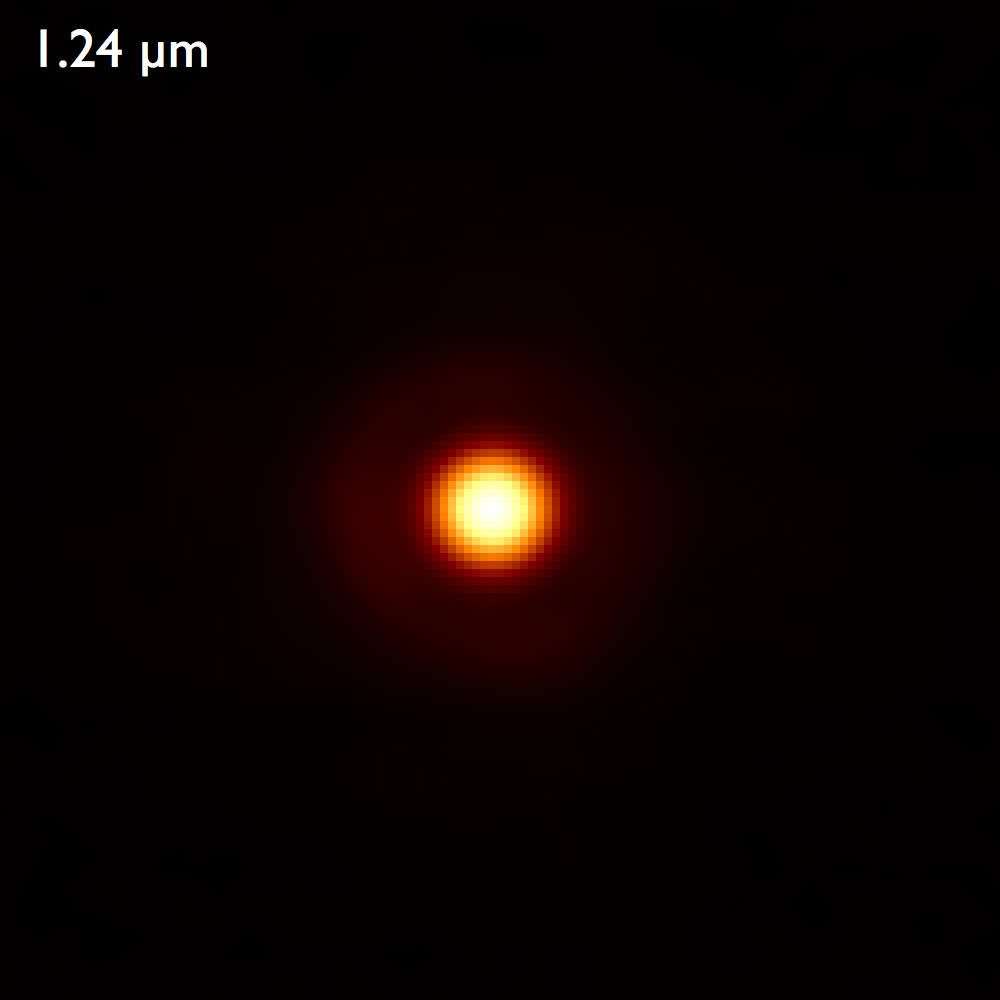} \vspace{1mm}
        \hspace{1mm}
        \includegraphics[width=4.4cm]{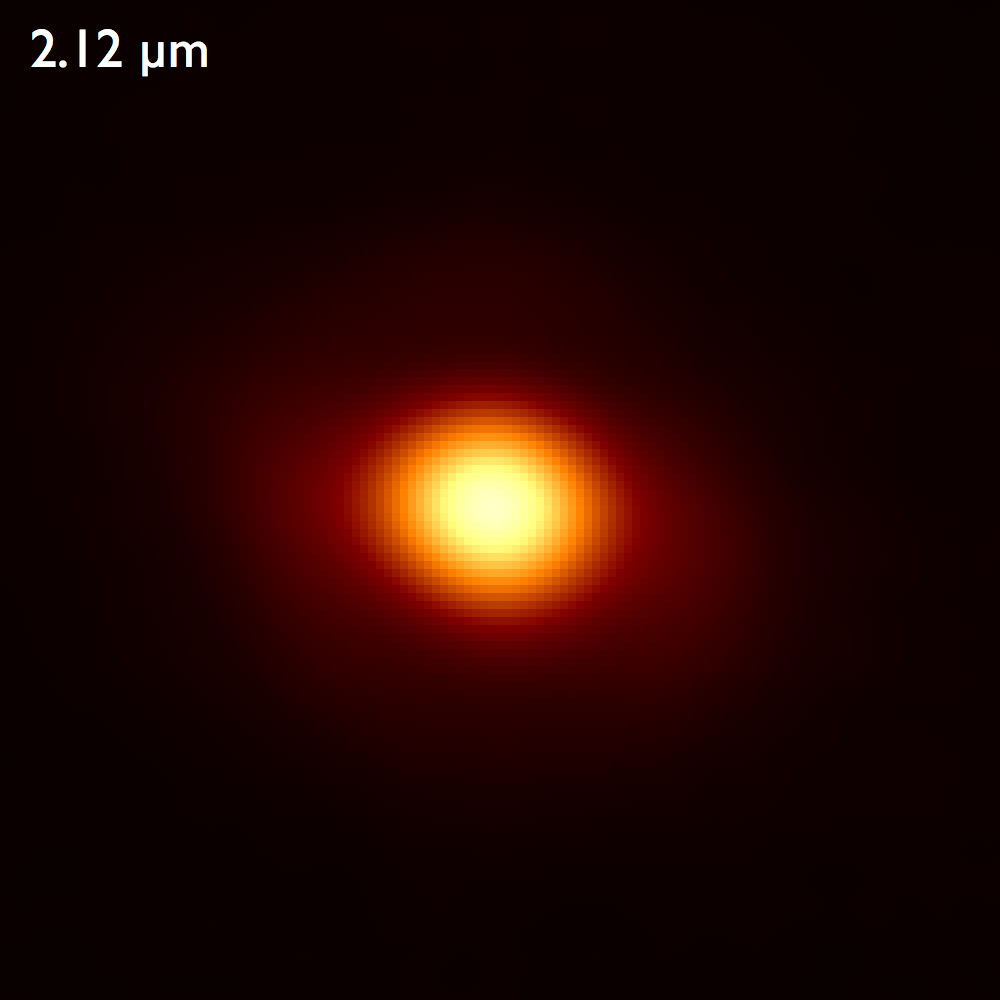} \includegraphics[width=4.4cm]{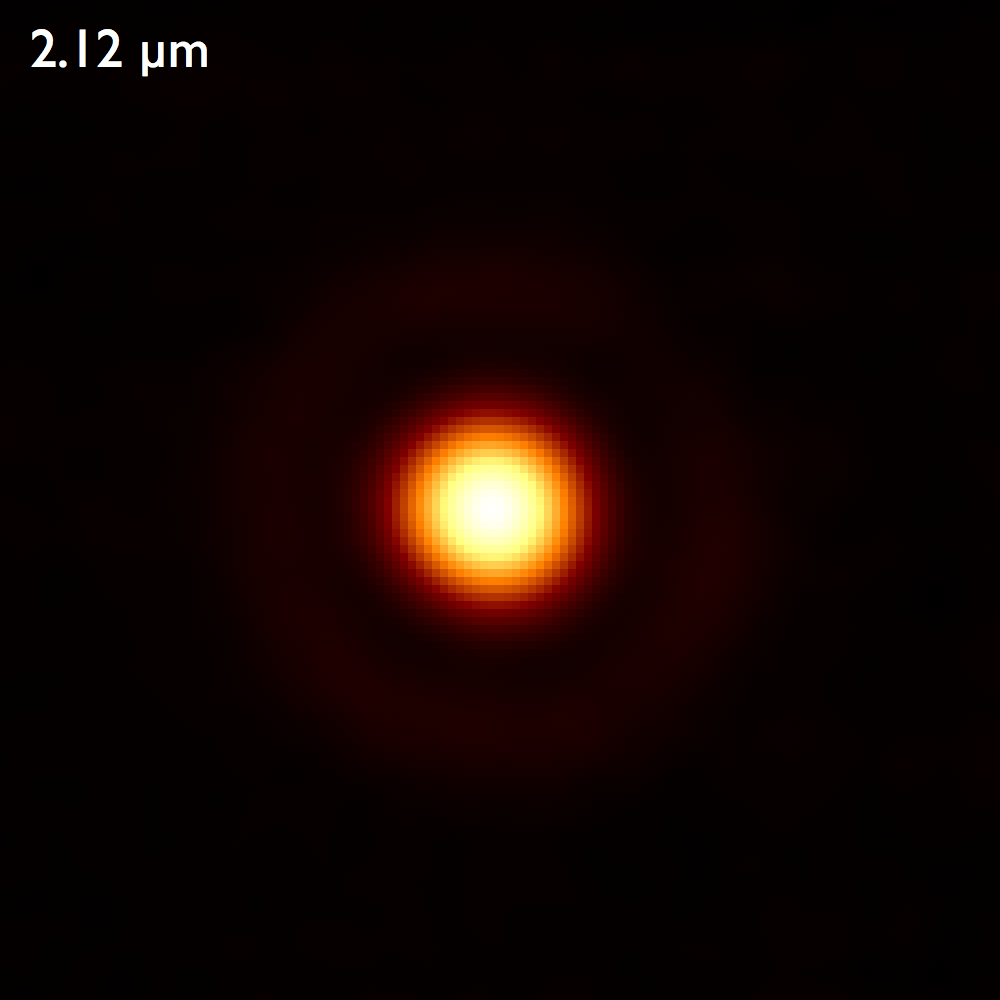}
        \includegraphics[width=4.4cm]{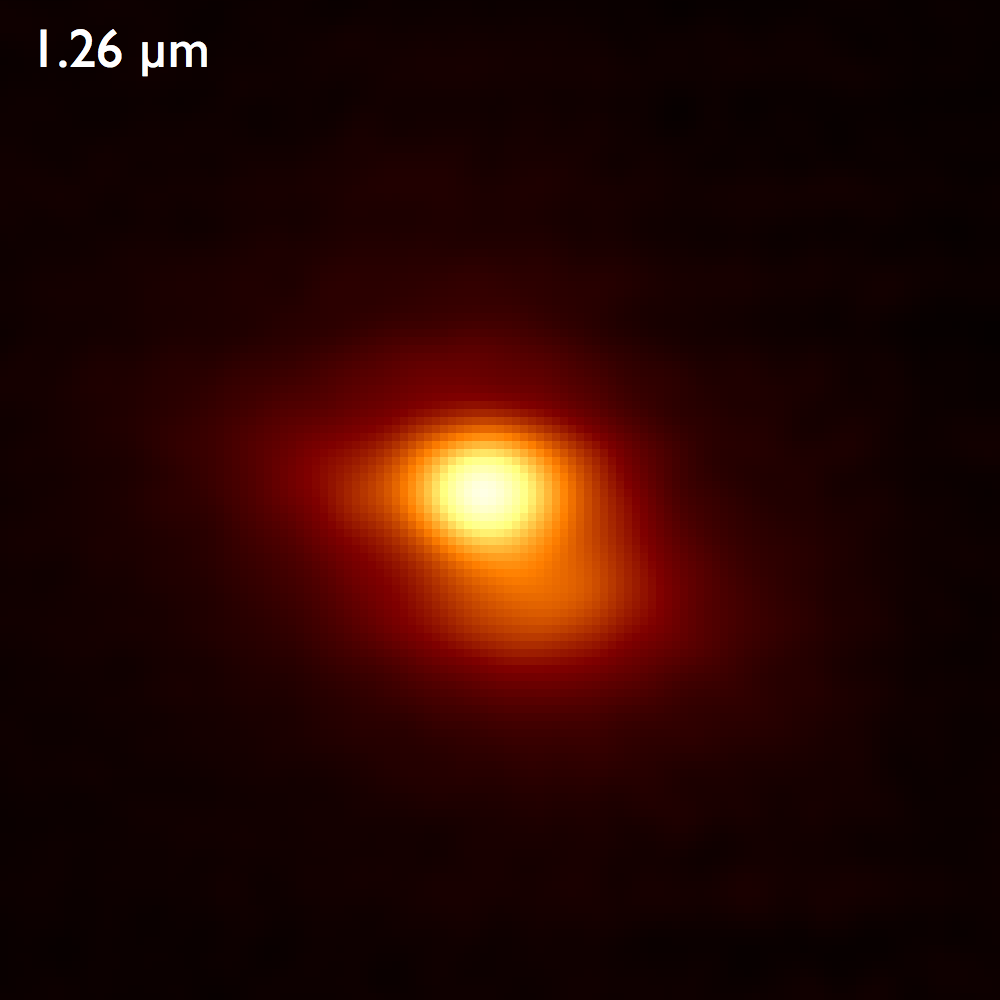} \includegraphics[width=4.4cm]{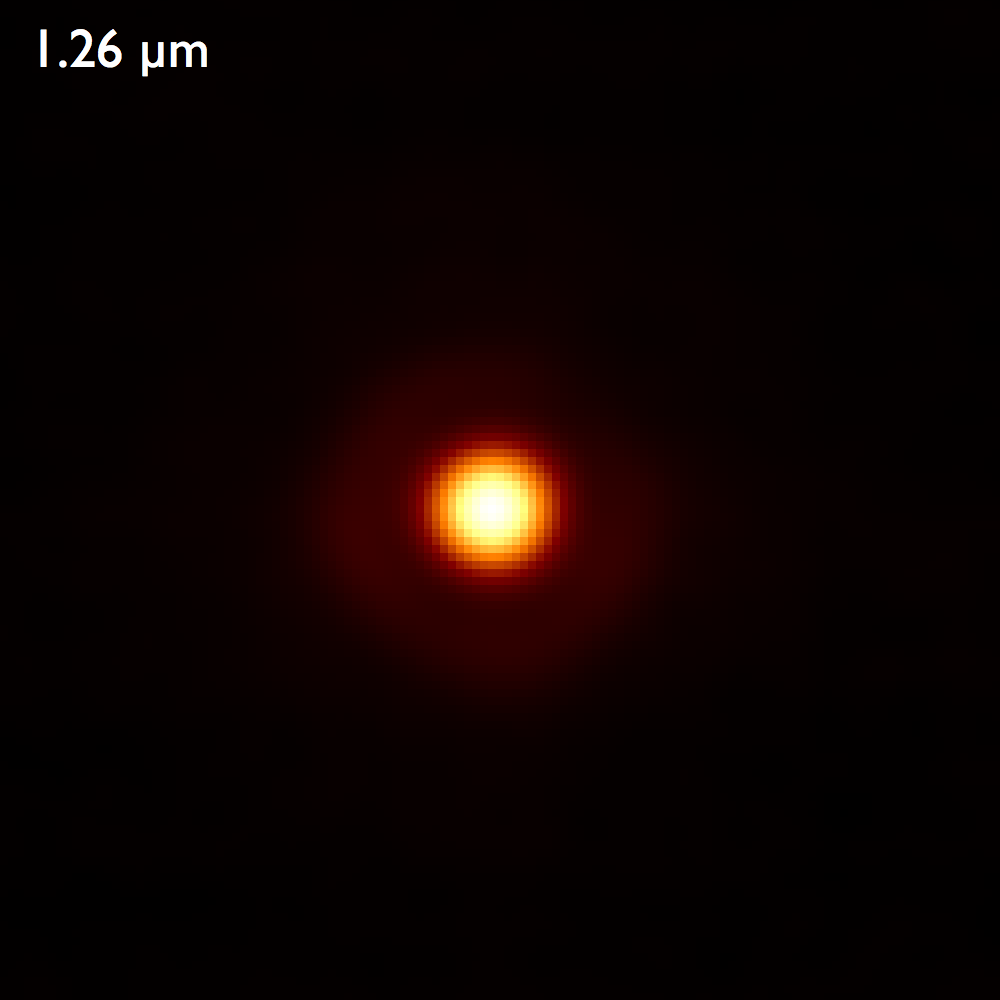} \hspace{1mm}
        \includegraphics[width=4.4cm]{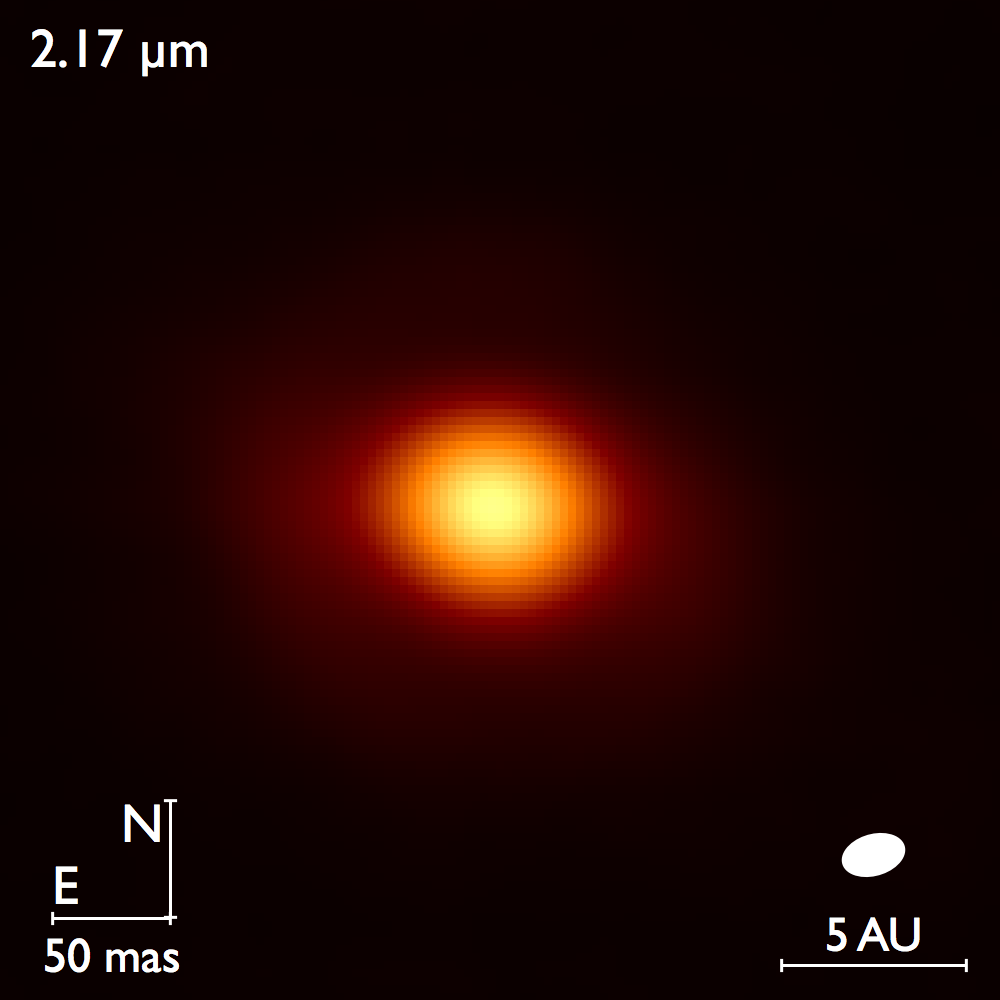} \includegraphics[width=4.4cm]{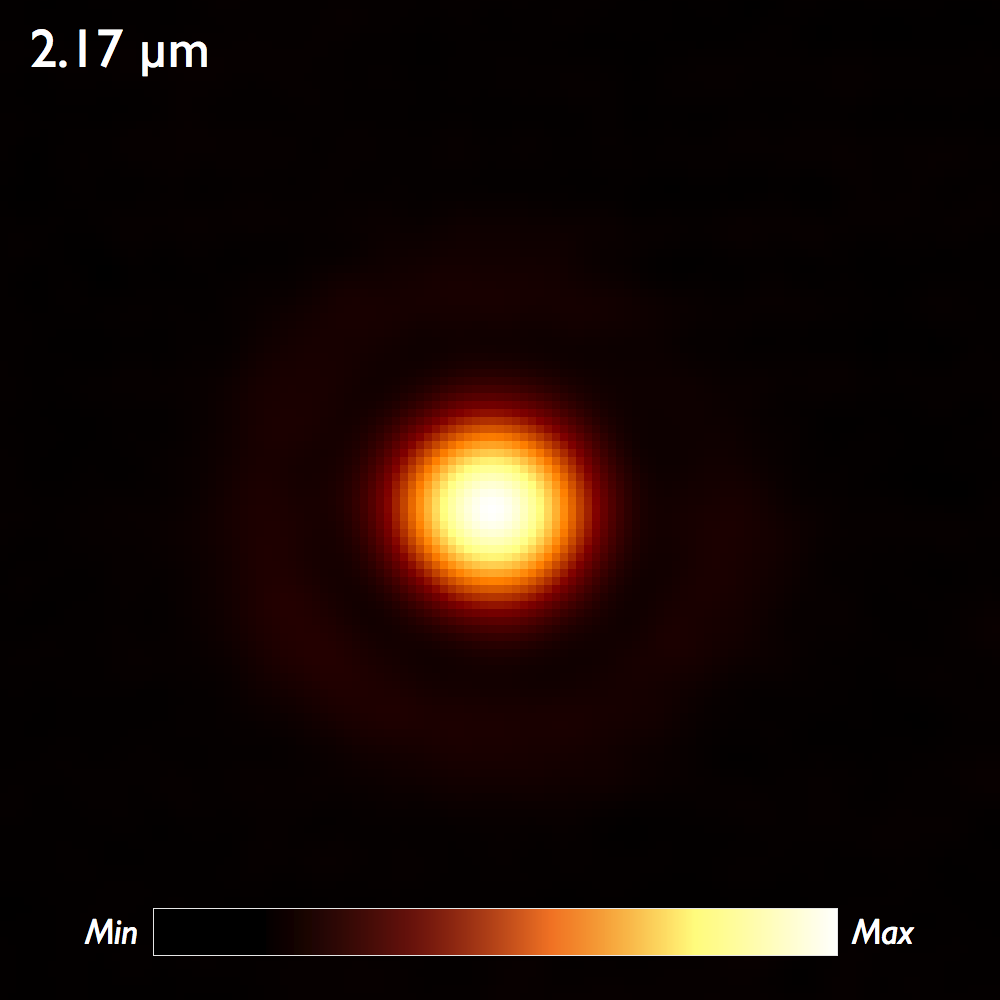}
        \caption{Non-deconvolved images of L$_2$\,Pup and its PSF calibrator ($\beta$\,Col) in the 10 narrow-band filters of CONICA covering the $JHK$ bands. The field of view is $0.414\arcsec$, with North up and East to the left. The color scale (bottom right panel) is a function of the square root of the irradiance, from the minimum to the maximum in each image. The white ellipse represented in the 2.17\,$\mu$m panel corresponds to the model derived from the VINCI data (Sect.~\ref{uniformellipse}). \label{nondeconvJHK}}
\end{figure*}

\begin{figure}[b]
        \centering
        \includegraphics[width=4.4cm]{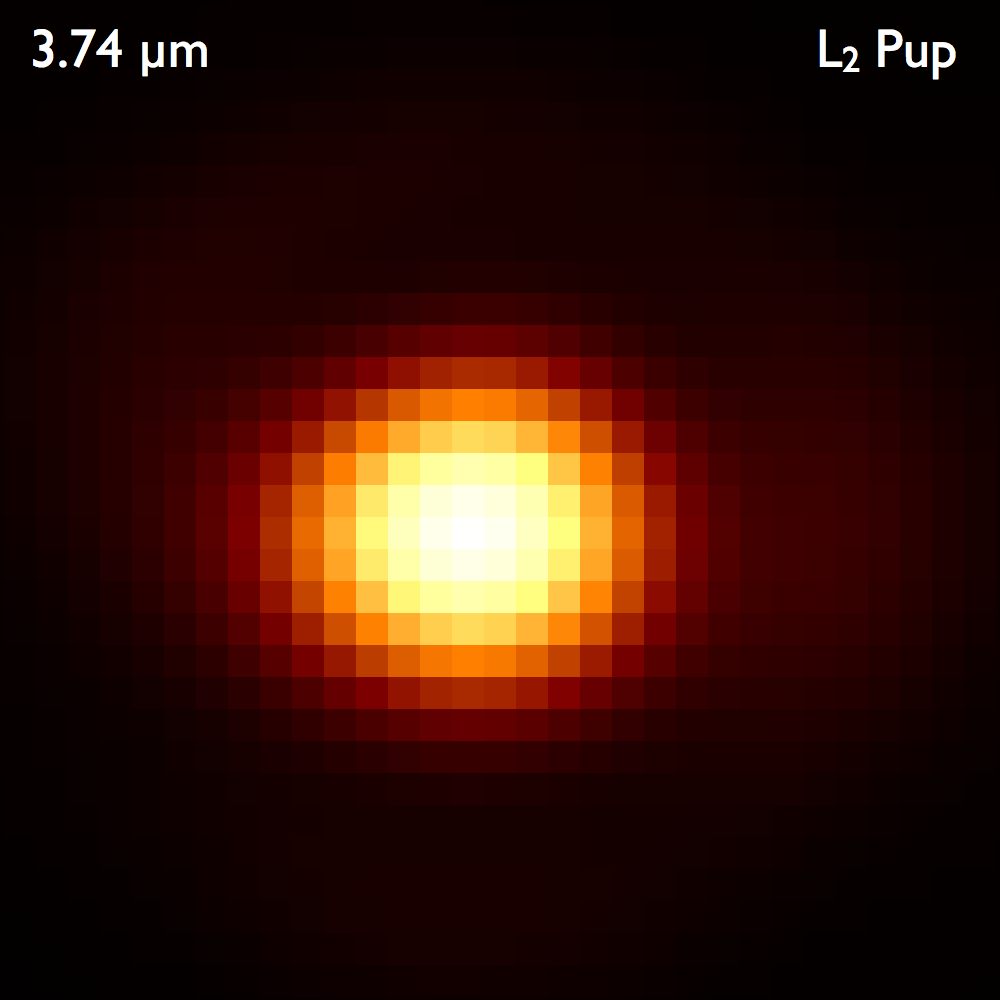} \includegraphics[width=4.4cm]{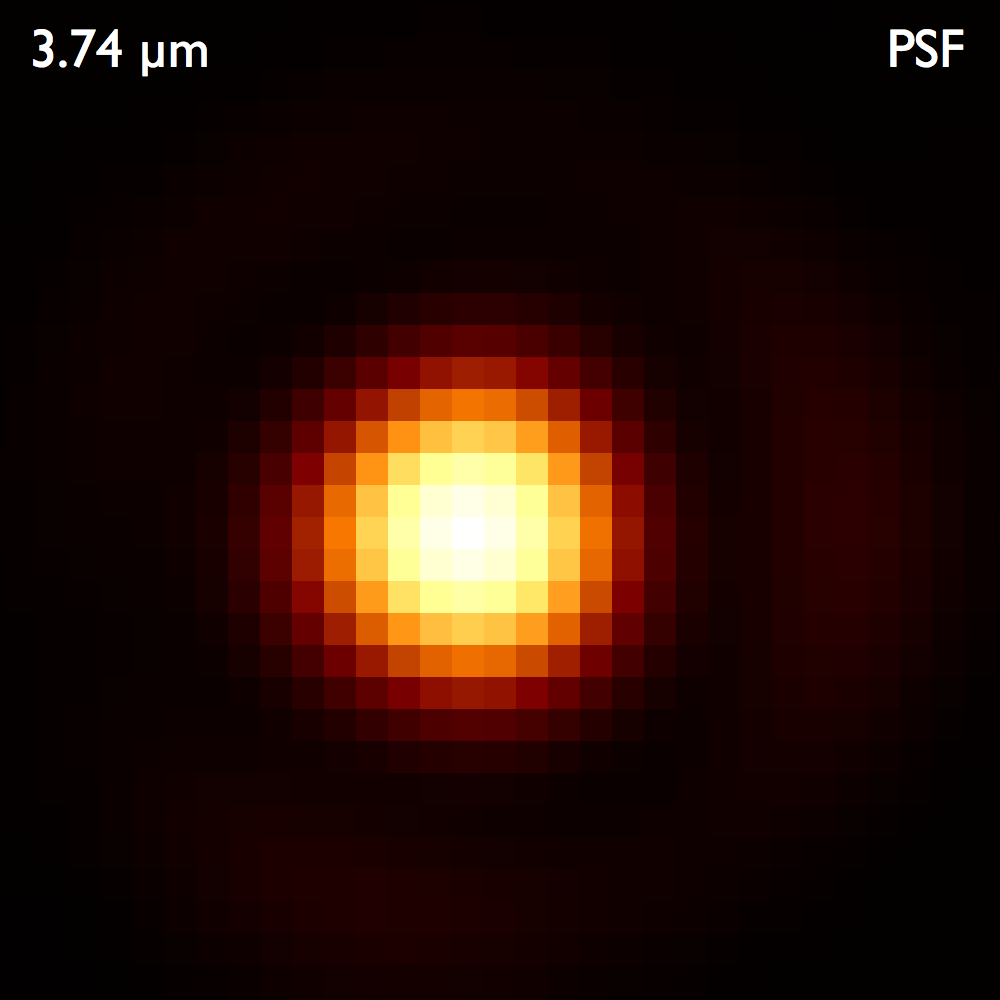}
        \includegraphics[width=4.4cm]{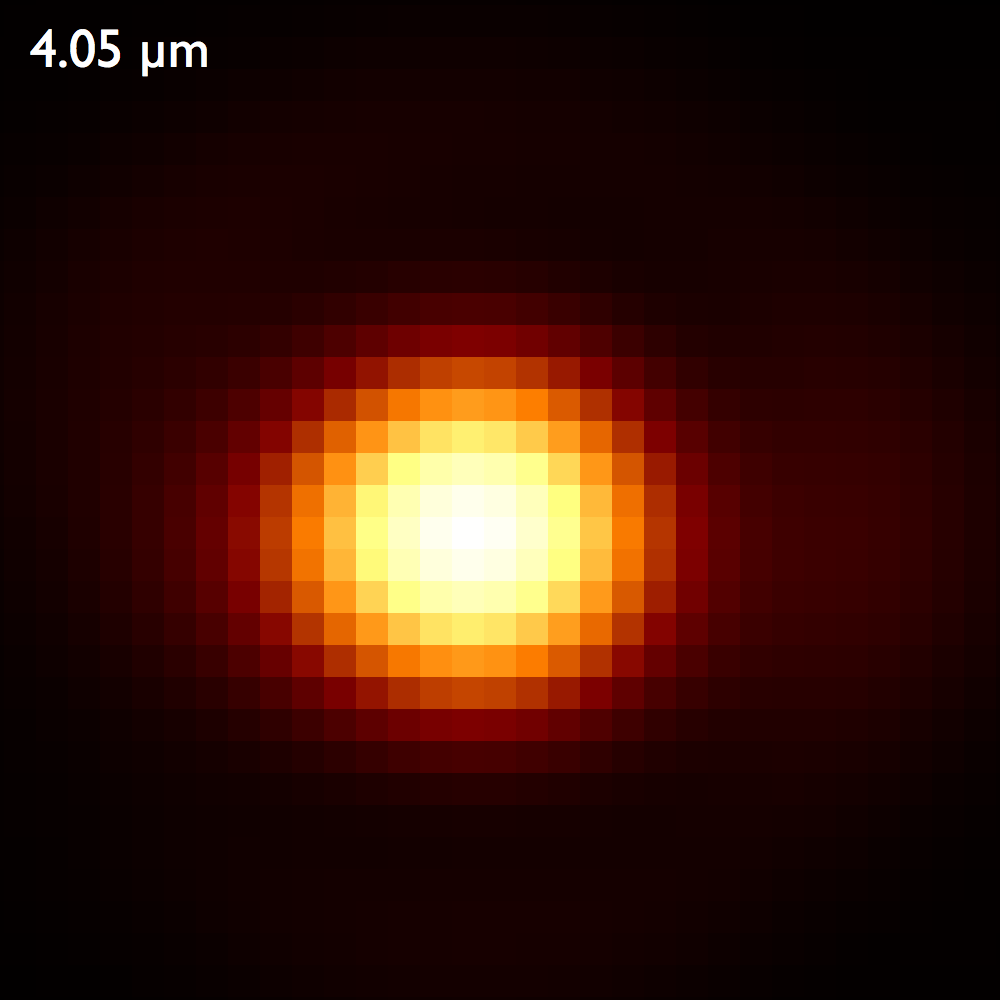} \includegraphics[width=4.4cm]{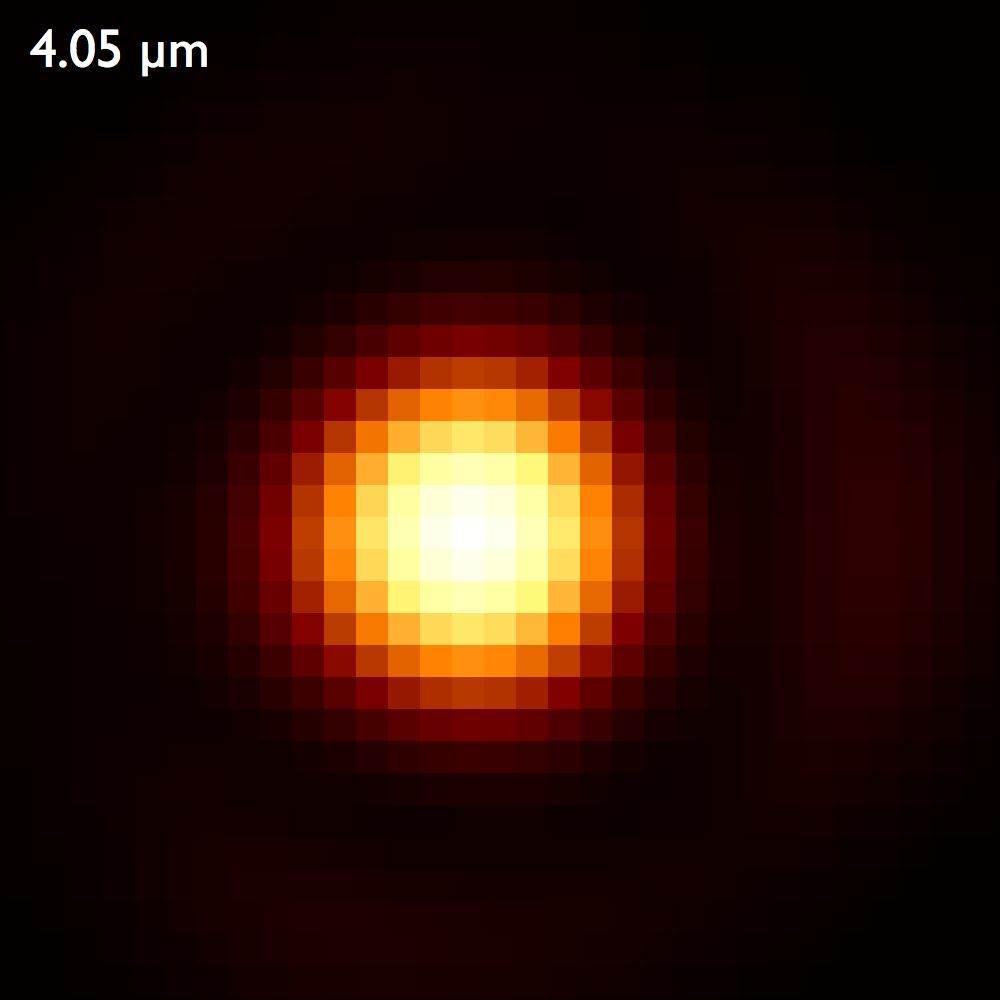}
        \caption{Non-deconvolved images of L$_2$\,Pup and its associated PSF ($\alpha$\,Lyn) in the two narrow-band filters of CONICA in the $L$ band. Orientation, field of view and color scale are the same as in Fig.~\ref{nondeconvJHK}.
        \label{nondeconvL}}
\end{figure}

We photometrically normalized the images using our PSF reference. We used the calibrators $\beta$\,Col ($JHK$ bands) and $\alpha$\,Lyn ($L$ band), whose images are shown in Fig.~\ref{nondeconvJHK} and \ref{nondeconvL}. 
We retrieved the spectral energy distribution (SED) from the Castelli-Kurucz catalogue \citepads{2004astro.ph..5087C, 2005MSAIS...8...14K} provided by the STScI\footnote{\url{http://www.stsci.edu/hst/observatory/cdbs/castelli\_kurucz\_atlas.html}}. The physical parameters used to characterize the calibrators are listed in Table \ref{Table_PSFparams}. We tested the reliability of the SED by comparing it with the existing broadband photometry measurements for both stars. Then we integrated the SED in each observed narrow-band using the NACO filter profiles presented in Fig.~\ref{nb_filters}. To calibrate photometrically the images of L$_2$\,Pup, we then multiplied their pixel values by the integrated SED of the PSF calibrator in each filter band and normalized then by the integrated flux measured on a disk of 33\,mas in radius, thus avoiding a contamination by the adaptive optics residual halo. The resulting photometry of L$_2$\,Pup in the 12 narrow band filters of NACO is listed in Table~\ref{table_naco_photometry}.

\begin{table}[h!]
        \caption{Physical parameters used to characterize the photometric/PSF calibrators}             
        \label{Table_PSFparams}      
        \centering          
        \begin{tabular}{l l l} 
        \hline\hline
        \noalign{\smallskip}
        Parameter & $\beta$\,Col & $\alpha$\,Lyn \\
        \hline
        \noalign{\smallskip}
        $\theta_{\rm LD}$ (mas) & $3.99 \pm 0.05$ & $7.30 \pm 0.08$ \\
        $T_{\rm eff}$ (K) & 4500 & 3750 \\
        $\log(g)$ & 2.0 & 0.0 \\
        $\log$(Fe/H) & $+0.5$ & $-0.5$ \\               
        \hline
        \end{tabular}
        \tablefoot{$\theta_{\rm LD}$ is the limb-darkened angular diameter.}
\end{table}

\begin{table}
        \caption{Photometry of L$_2$\,Pup in the NACO narrow-band filters for the observing epoch 2013.22.}
        \label{table_naco_photometry}
        \centering          
        \begin{tabular}{lccl}
                \hline \hline
                \noalign{\smallskip}
                Filter & $\lambda_0$ & $\Delta \lambda$ & Flux density \\
                 & $[\mu$m$]$ & $[\mu$m$]$ & [10$^{-10}$\,W/m$^2$/$\mu$m] \\
                \hline
                \noalign{\smallskip}
                NB1.04 & 1.040 & 0.015 & $7.15 \pm 0.93$ \\
                NB1.08 & 1.083 & 0.015 & $7.75 \pm 0.10$ \\
                NB1.09 & 1.094 & 0.015 & $6.46 \pm 0.84$ \\
                NB1.24 & 1.237 & 0.015 & $7.52 \pm 0.98$ \\     
                NB1.26 & 1.257 & 0.014 & $8.74 \pm 0.11$ \\     
                NB1.28 & 1.282 & 0.014 & $8.19 \pm 0.11$ \\     
                NB1.64 & 1.644 & 0.018 & $7.42 \pm 0.97$ \\
                NB1.75 & 1.748 & 0.026 & $7.69 \pm 0.10$ \\
                NB2.12 & 2.122 & 0.022 & $7.20 \pm 0.94$ \\
                NB2.17 & 2.166 & 0.023 & $6.42 \pm 0.84$ \\
                NB3.74 & 3.74 & 0.02 & $4.73 \pm 0.62$ \\
                NB4.05 & 4.05 & 0.02 & $5.13 \pm 0.67$ \\
                \hline
        \end{tabular}
\end{table}

\subsubsection{Image deconvolution}

%______________ Figure
\begin{figure*}[]
        \centering
        \includegraphics[width=4.5cm]{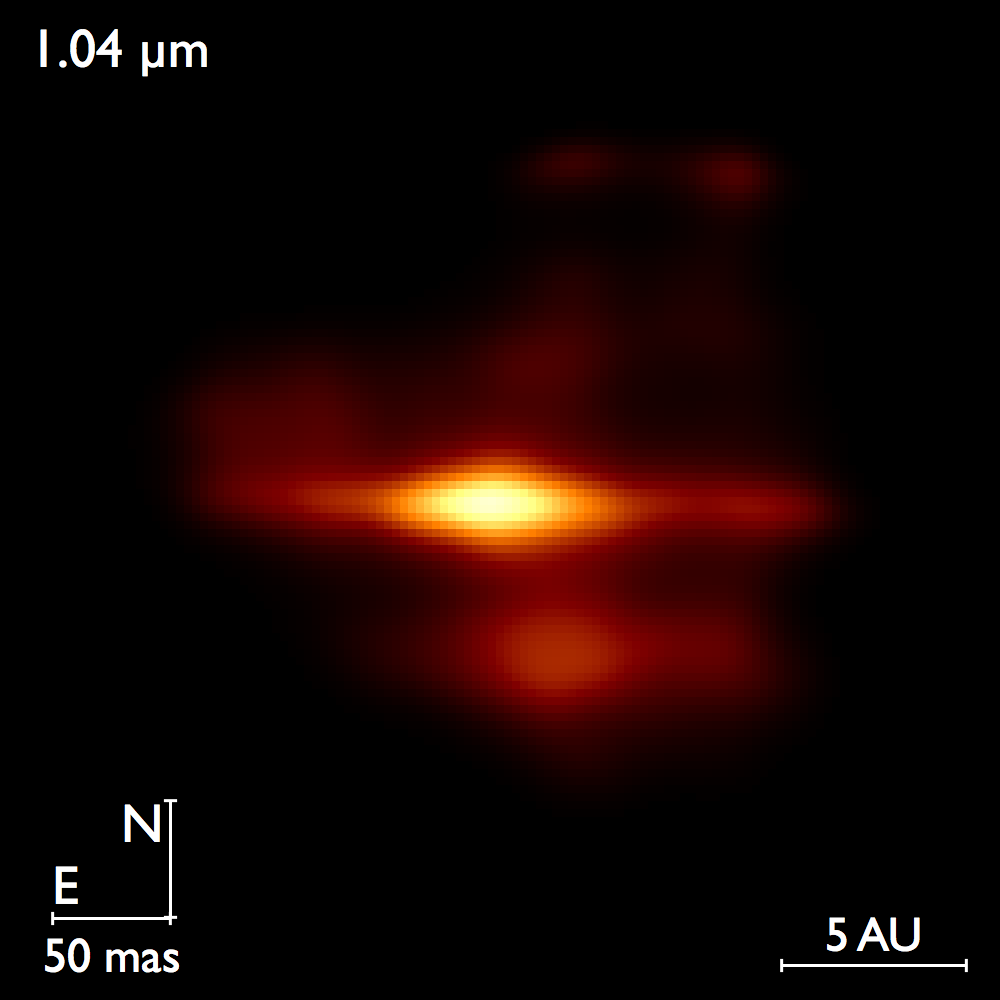}
        \includegraphics[width=4.5cm]{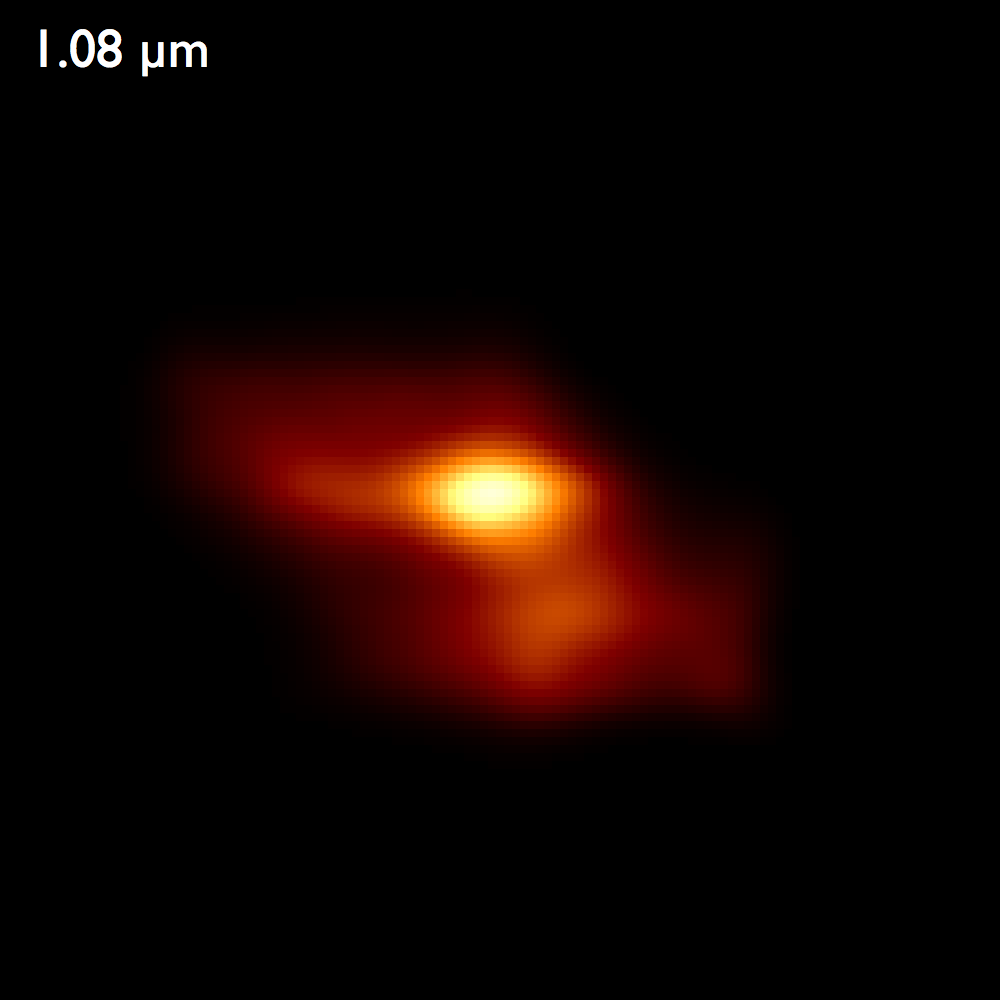}
        \vspace{1mm}
        \includegraphics[width=4.5cm]{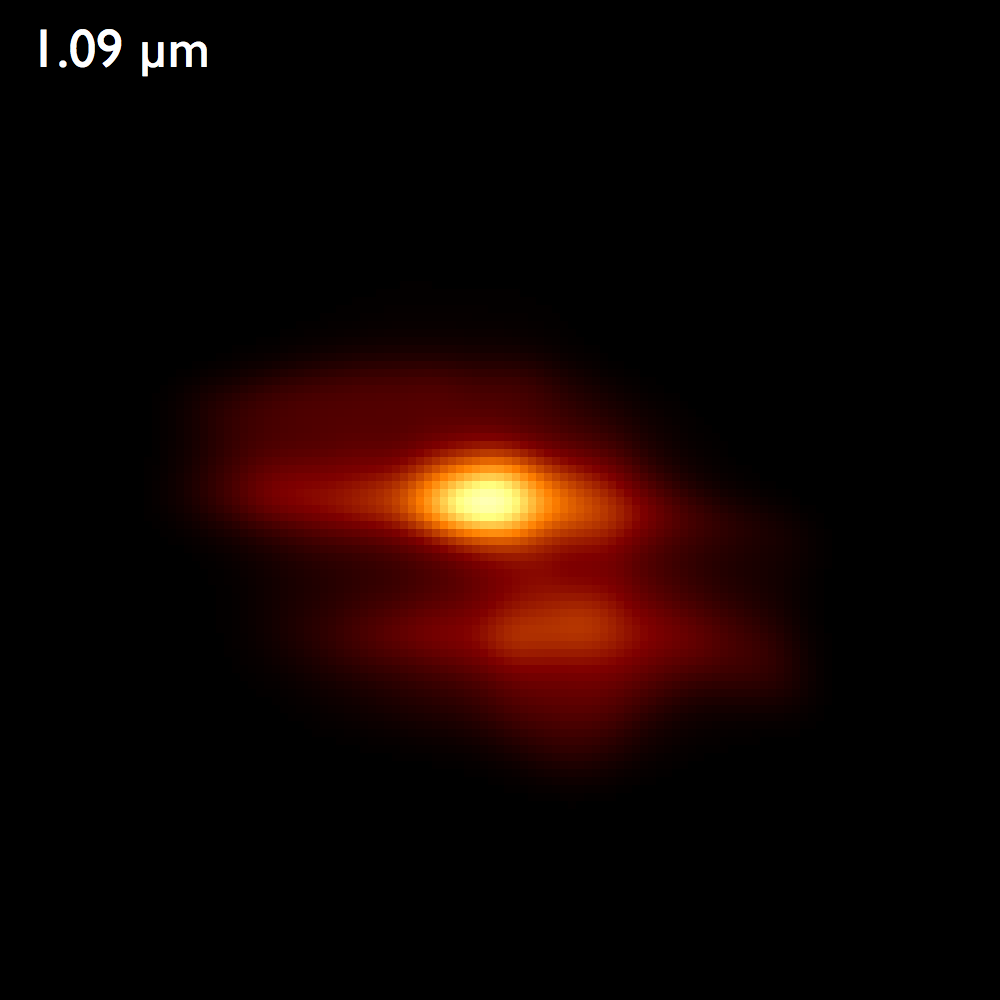}
        \includegraphics[width=4.5cm]{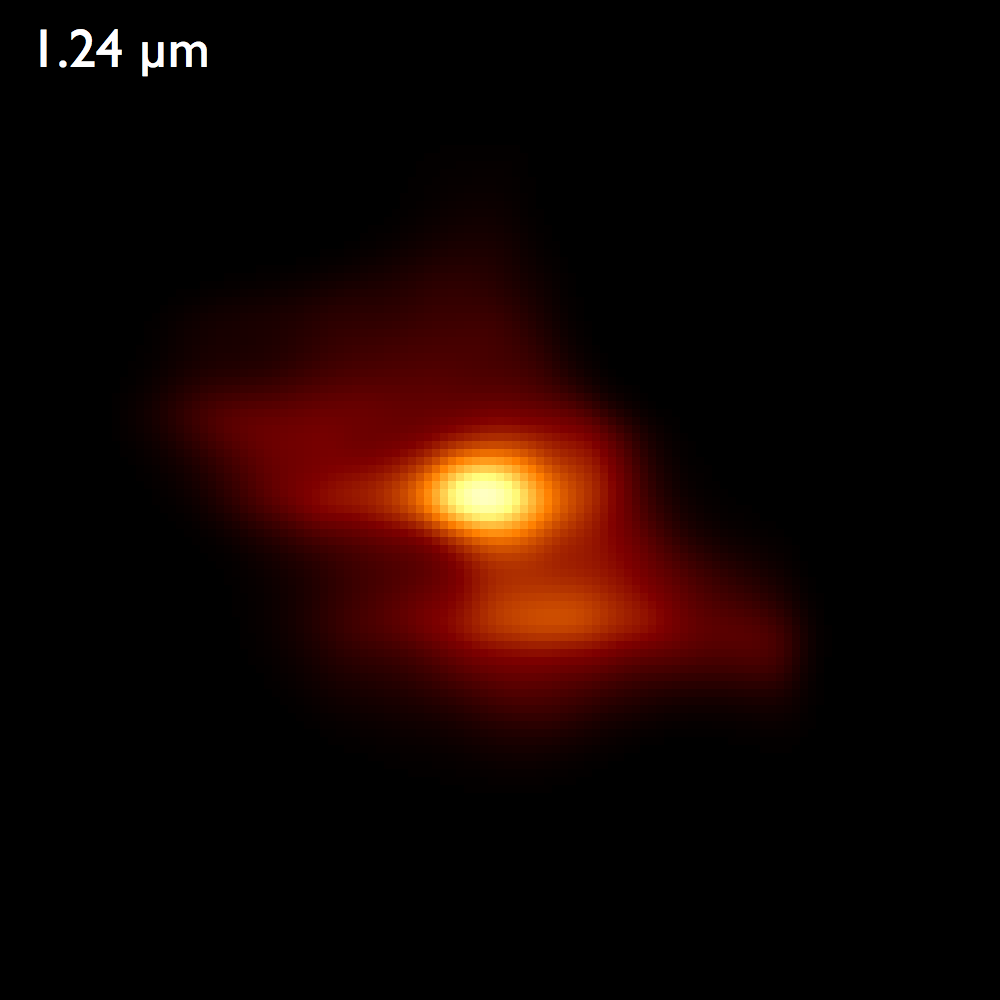}
        \includegraphics[width=4.5cm]{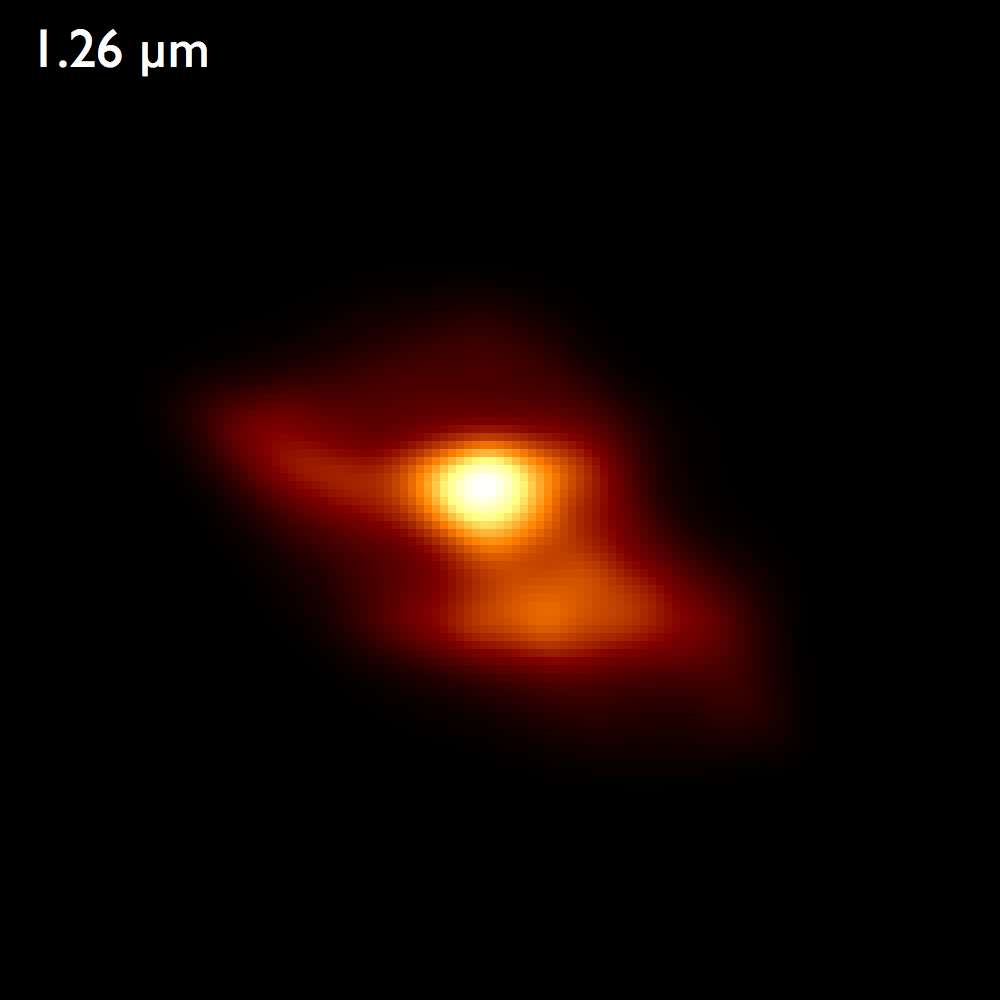}
        \includegraphics[width=4.5cm]{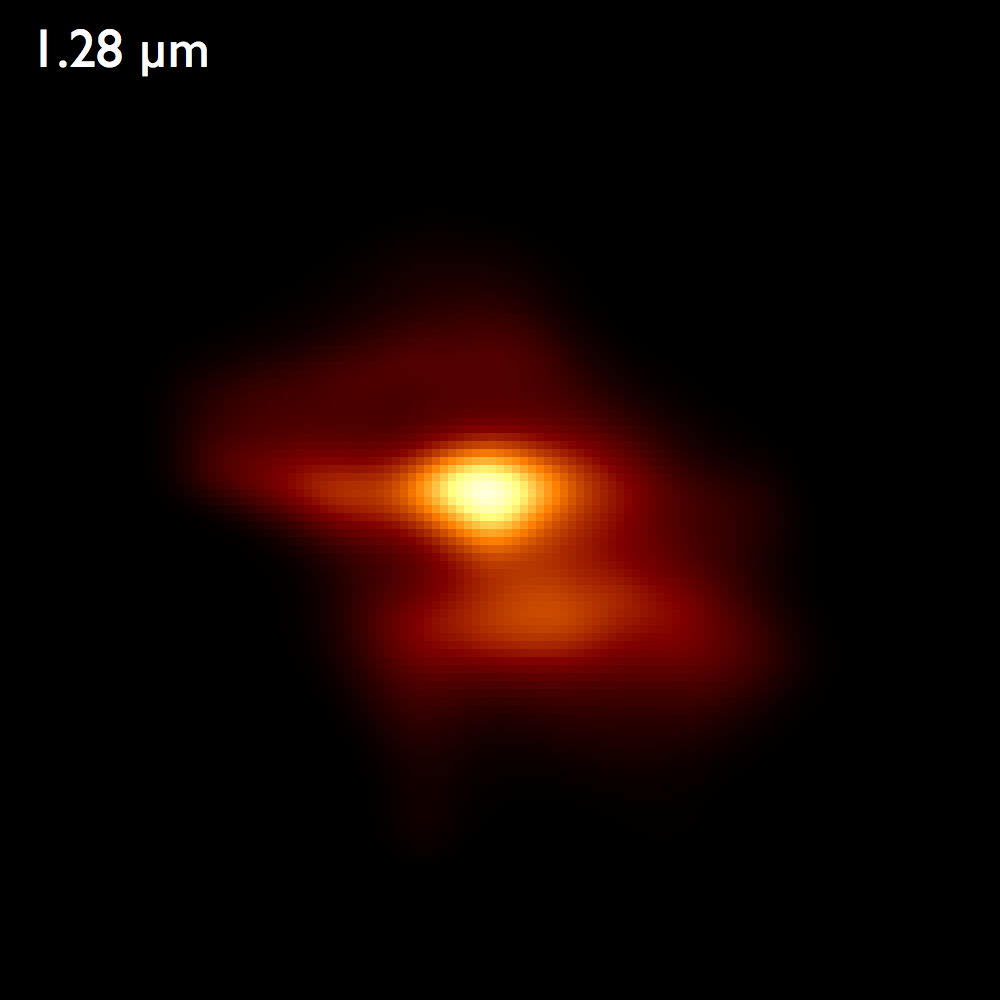}
        \vspace{1mm}
        \includegraphics[width=4.5cm]{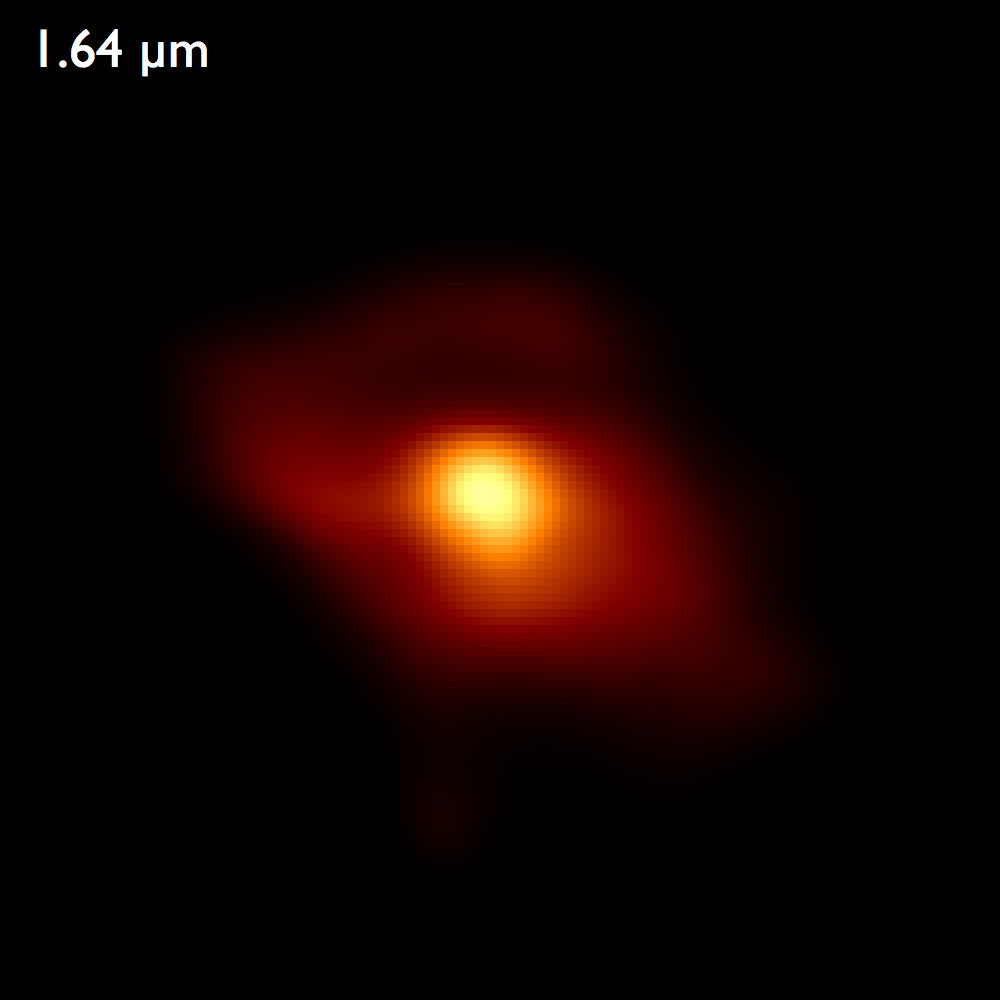}
        \includegraphics[width=4.5cm]{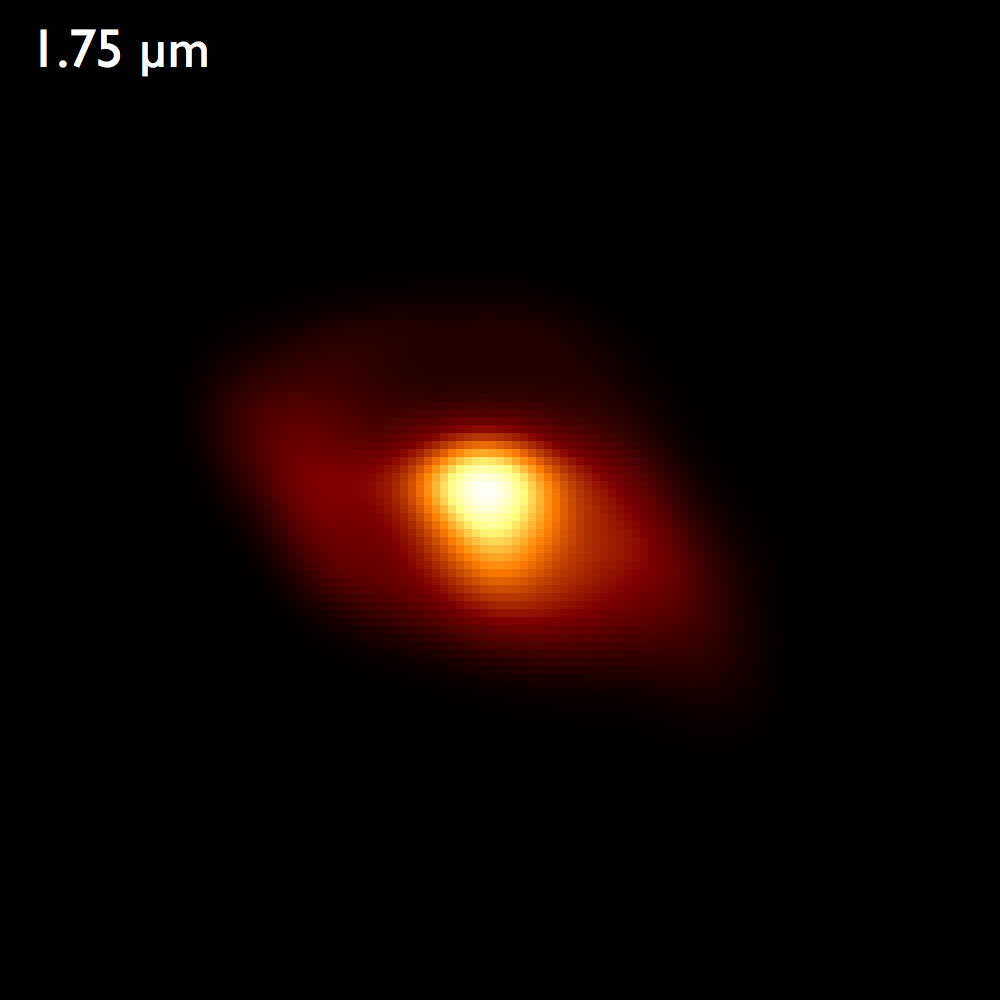}
        \includegraphics[width=4.5cm]{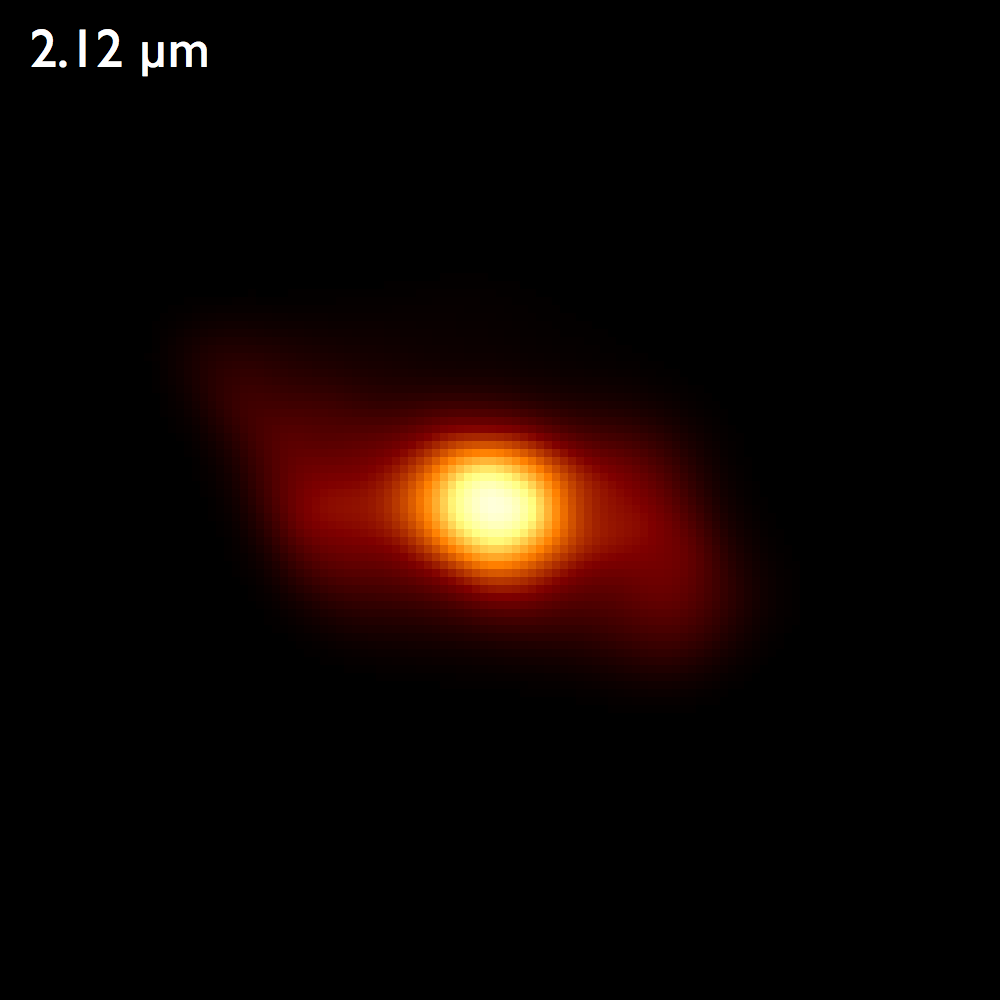}
        \includegraphics[width=4.5cm]{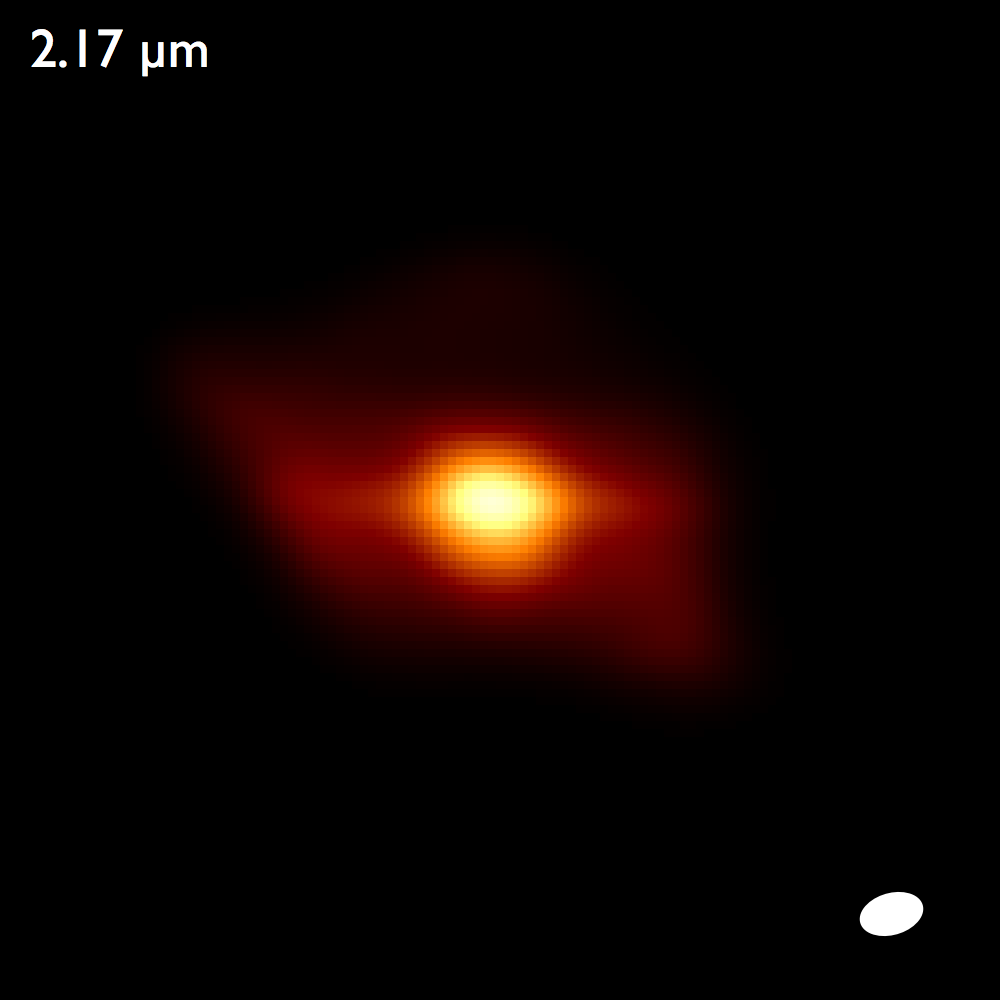}
        \includegraphics[width=4.5cm]{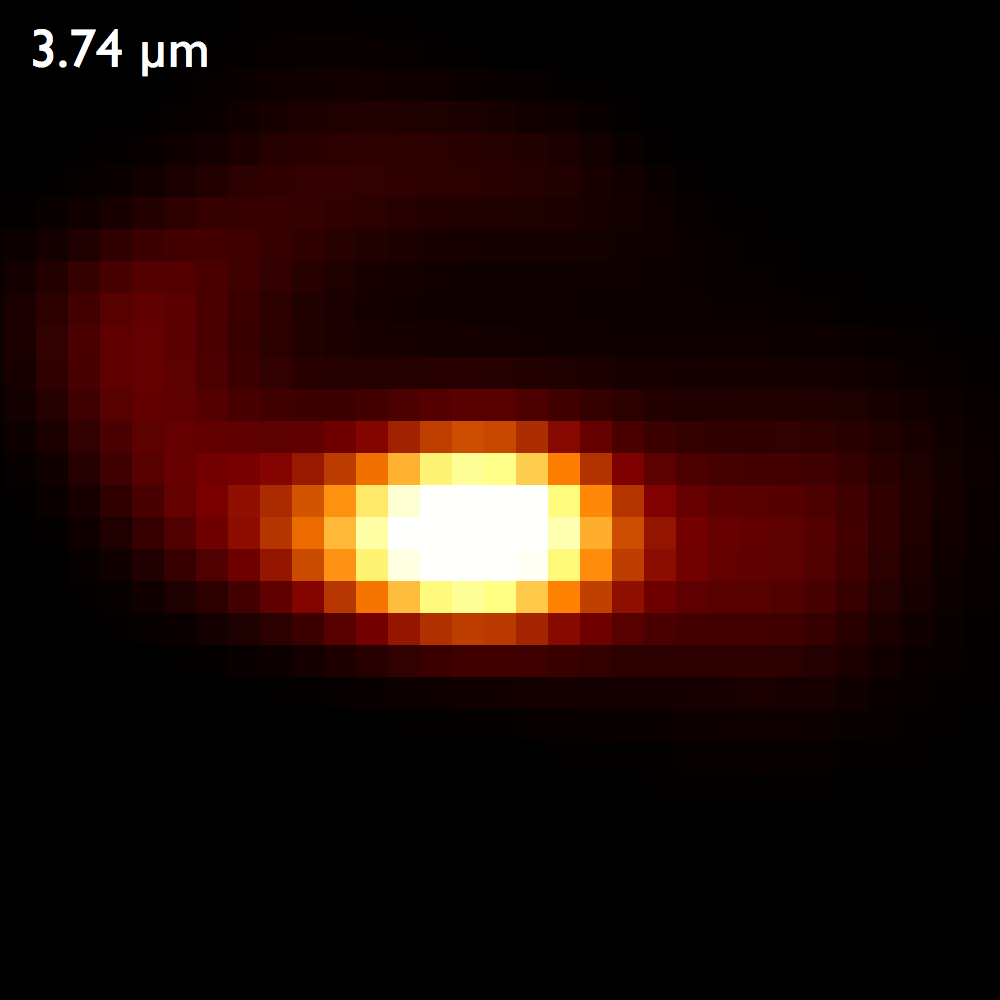}
        \includegraphics[width=4.5cm]{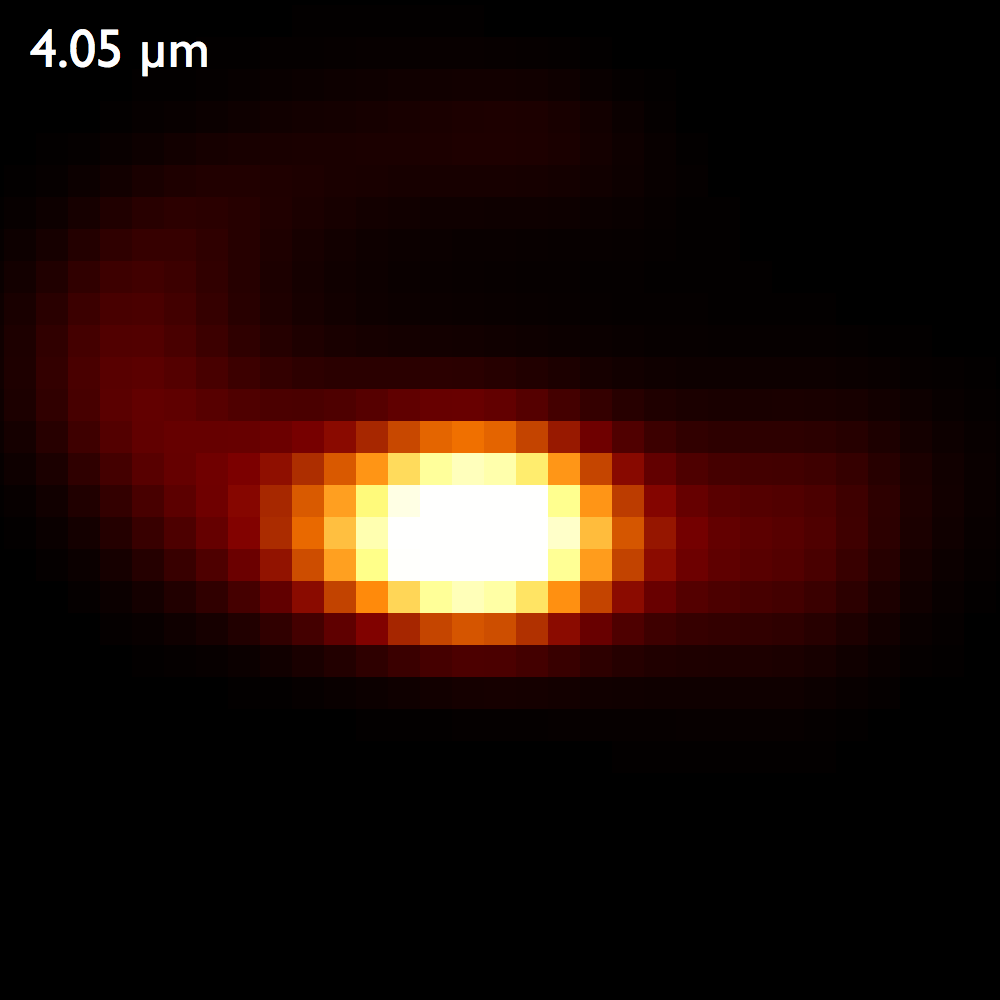}
        \caption{Deconvolved images of L$_2$\,Pup using a uniform, 80-step Lucy deconvolution. Orientation, field of view and color scale are the same as in Fig.~\ref{nondeconvJHK}. The white ellipse represented in the 2.17\,$\mu$m panel corresponds to the star model derived from the VINCI data (Sect.~\ref{uniformellipse}). \label{deconv80}}
\end{figure*}

As discussed by \citetads{2009A&A...504..115K}, the improvement in Strehl ratio from the serendipitous imaging approach compared with classical long exposures is particularly significant in the $J$ -band filters. This technique allows us to retrieve details at the diffraction limit of the telescope, and makes the deconvolution of the images much more efficient.

\begin{figure}[]
	\centering
	\includegraphics[width=\hsize]{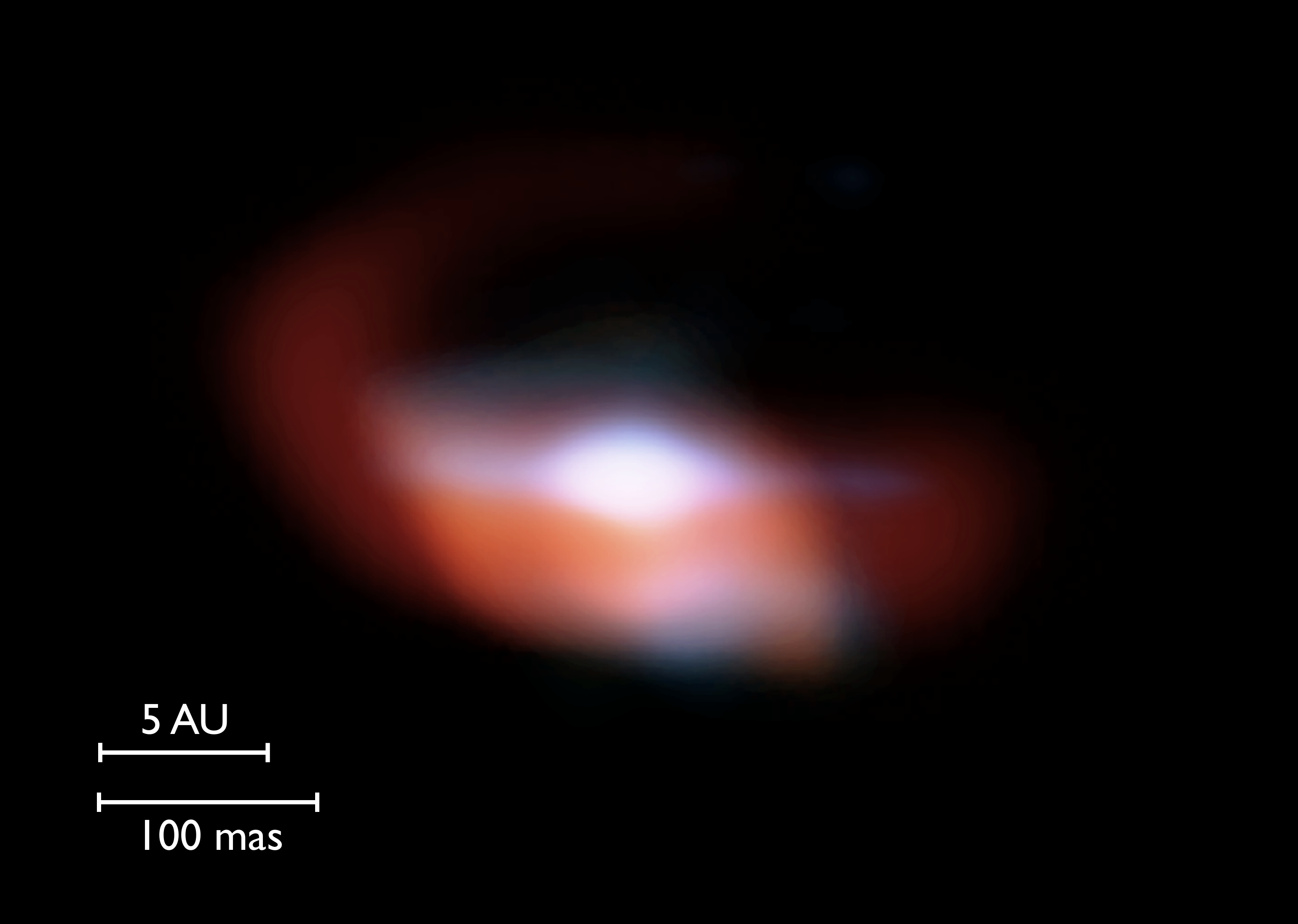}
	\caption{Color composite view of L$_2$\,Pup assembled from the 12 NACO narrow-band images in the JHKL bands. \label{L2PupColor}}
\end{figure}

We deconvolved the images of L$_2$\,Pup using the PSF images (Sect.~\ref{rawreduction}) as the dirty beams and the Lucy-Richardson (L-R) algorithm implemented in the IRAF software package. The resulting average deconvolved images in the 12 narrow-band filters are presented in Fig.~\ref{deconv80}, and a color composite image is presented in Fig.~\ref{L2PupColor}. We stopped the L-R deconvolution after 80 iterations (uniformly for all wavelengths), as the images do not show a significant evolution for additional processing steps.

%______________ Figure

\subsection{VLTI/VINCI interferometry \label{vinciobs}}

L$_2$\,Pup was observed on several occasions during the early commissioning of the VLTI in late 2001, using the VLT INterferometer Commissioning Instrument \citepads[VINCI,][]{2004A&A...425.1161K} and the two VLTI 0.35\,m test siderostats installed on the E0-G0 baseline (16\,m ground length). We used a regular $K$ -band filter ($\lambda = 2.0-2.4\,\mu$m) and processed the data using the standard VINCI data reduction software version 3.0. As VINCI is a single-mode instrument (i.e. it spatially filters the light beams using single-mode fibers), its field of view (FOV) is defined by a combination of the seeing spot and the acceptance function of the input fibers. The latter is matched to the Airy pattern of the telescope, whose typical size is $\approx 1.4\arcsec$. For extended objects larger than the Airy pattern of the telescopes, variable seeing conditions can result in a variable bias on the measured visibility \citepads{2002A&A...387..366G}. However, as shown in Sect.~\ref{nacoobs}, most of L$_2$\,Pup's flux is coming from an area on the sky that is well contained in the $1.4\arcsec$ Airy pattern of the 0.35\,m VLTI siderostats. The VINCI measurements are therefore not affected by this bias. L$_2$\,Pup's instrumental visibilities were calibrated using \object{Sirius} \citepads[$\theta_{\rm UD\,K} = 5.936 \pm 0.016$\,mas;][]{2003A&A...408..681K} and $\epsilon$\,Lep \citepads[\object{HD 32887}, $\theta_{\rm UD\,K} = 5.91 \pm 0.064$\,mas;][]{2002A&A...393..183B}.

%_________Table of VINCI observations
\begin{table}
        \caption{Table of the VINCI observations of L$_2$ Pup in the infrared $K$ band (effective wavelength $\lambda_{\rm eff} = 2.19\,\mu$m).}
        \label{vinci_log}
        \centering          
        \begin{tabular}{lccc}
        \hline \hline
        \noalign{\smallskip}
        MJD & $B$ [m] & PA [deg] & $V^2$ \\
        \hline
        \noalign{\smallskip}
        52195.318 & 15.18 & 39.89 & $ 0.1439 \pm 0.0064 $ \\
        52197.277 & 14.67 & 27.74 & $ 0.1799 \pm 0.0093 $ \\
        52197.319 & 15.28 & 41.92 & $ 0.1421 \pm 0.0080 $ \\
        52200.292 & 15.00 & 35.87 & $ 0.1250 \pm 0.0024 $ \\
        52200.338 & 15.63 & 49.93 & $ 0.0875 \pm 0.0019 $ \\
        52200.376 & 15.96 & 60.52 & $ 0.0678 \pm 0.0013 $ \\
        52201.281 & 14.88 & 33.11 & $ 0.1327 \pm 0.0027 $ \\
        52201.316 & 15.39 & 44.40 & $ 0.0938 \pm 0.0015 $ \\
        52201.356 & 15.84 & 55.81 & $ 0.0708 \pm 0.0012 $ \\
        52202.280 & 14.90 & 33.46 & $ 0.1278 \pm 0.0067 $ \\
        52202.326 & 15.56 & 48.17 & $ 0.0804 \pm 0.0042 $ \\
        52202.377 & 15.99 & 62.23 & $ 0.0664 \pm 0.0048 $ \\
        52203.278 & 14.91 & 33.75 & $ 0.1149 \pm 0.0057 $ \\
        52203.343 & 15.78 & 53.81 & $ 0.0687 \pm 0.0034 $ \\
        52203.375 & 15.99 & 62.33 & $ 0.0593 \pm 0.0030 $ \\
        52210.343 & 15.93 & 58.98 & $ 0.0449 \pm 0.0023 $ \\
        52218.360 & 15.98 & 69.02 & $ 0.0368 \pm 0.0009 $ \\
        52218.363 & 15.96 & 69.76 & $ 0.0369 \pm 0.0011 $ \\
        52218.367 & 15.95 & 70.62 & $ 0.0374 \pm 0.0011 $ \\
        \hline
        \end{tabular}
        \tablefoot{MJD is the average modified julian date of the exposures, $B$ is the projected baseline length in meters, PA is the position angle relative to north (E=$90^\circ$) and $V^2$ is the squared visibility of the fringes.}
\end{table}

\subsubsection{Resolved flux debiasing\label{resolvedflux}}

From the morphology of L$_2$\,Pup in our NACO images, it is clear that a significant fraction of the flux collected by the VINCI instrument in the $K$ band within its $\approx 1.4\arcsec$ field of view comes from spatially extended emission. Any flux contribution from structures larger than the resolving power of the interferometer will contribute essentially uncorrelated flux to the measurements. In other words, VINCI's $V^2$ values are displaced towards lower values by the contribution from the extended disk. To retrieve the properties of the central object, we therefore have to correct for this effect.

\begin{figure}[t]
        \centering
        \includegraphics[width=4.4cm]{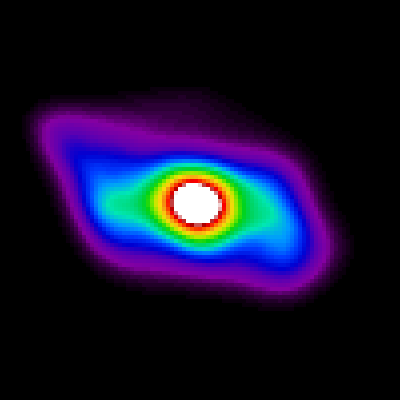}
        \includegraphics[width=4.4cm]{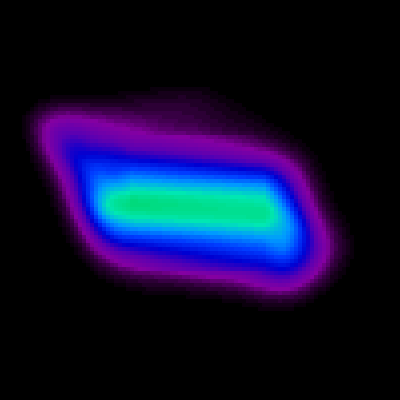}
        \caption{Deconvolved image of L$_2$\,Pup at $2.17\,\mu$m (left) and contribution from the disk alone (right). The field of view is $1.33\arcsec$. \label{UnresolvedFlux}}
\end{figure}

Because we do not have interferometric measurements on short baselines (0-15\,m), we used the NACO images at 2.12 and 2.17\,$\mu$m (Fig.~\ref{deconv80}, within the VINCI wavelength range) to measure the relative flux contribution that is unresolved by the Unit Telescope.
The NACO images were obtained in 2013, while the VINCI observations were recorded in late 2001. According to Fig.~2 in \citetads{2002MNRAS.337...79B}, L$_2$\,Pup was at this time in its current low photometric flux state.
We interpolated the L-R deconvolved $2.12$ and $2.17\,\mu$m images of L$_2$\,Pup over the central object to remove its flux contribution. We used the deconvolved images because the L-R deconvolution algorithm presents the advantage of preserving the photometric flux. The result of this operation is shown in Fig.~\ref{UnresolvedFlux}. We then estimated the unresolved flux fraction $\omega = 1 - f_\mathrm{resolved} / f_\mathrm{total}$ from the flux $f_\mathrm{resolved}$ in the image of the disk alone and the total flux in the original image $f_\mathrm{total}$. We obtained $\omega = 59.2\%$ for the $2.17\,\mu$m image and $\omega = 62.9\%$ for the $2.12\,\mu$m band.
We therefore considered a value of $\omega = 60 \pm 5\%$ to debias the VINCI visibilities using
\begin{equation}
V^2_{\rm debiased} = \frac{V^2_{\rm measured}}{\omega^2}.
\end{equation}

%______________ Figure

\begin{figure}[]
        \centering
        \includegraphics[width=\hsize]{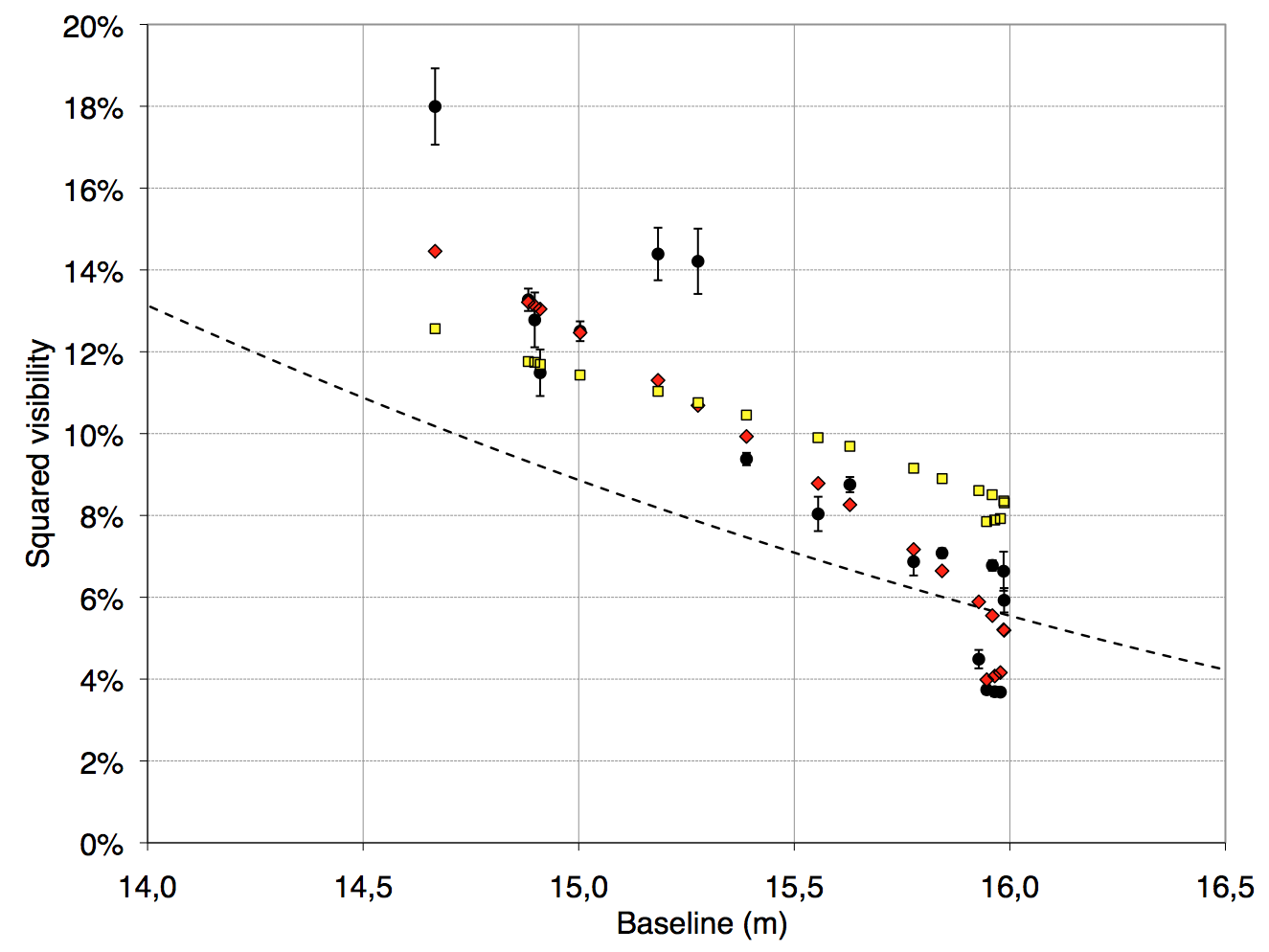} 
        \includegraphics[width=\hsize]{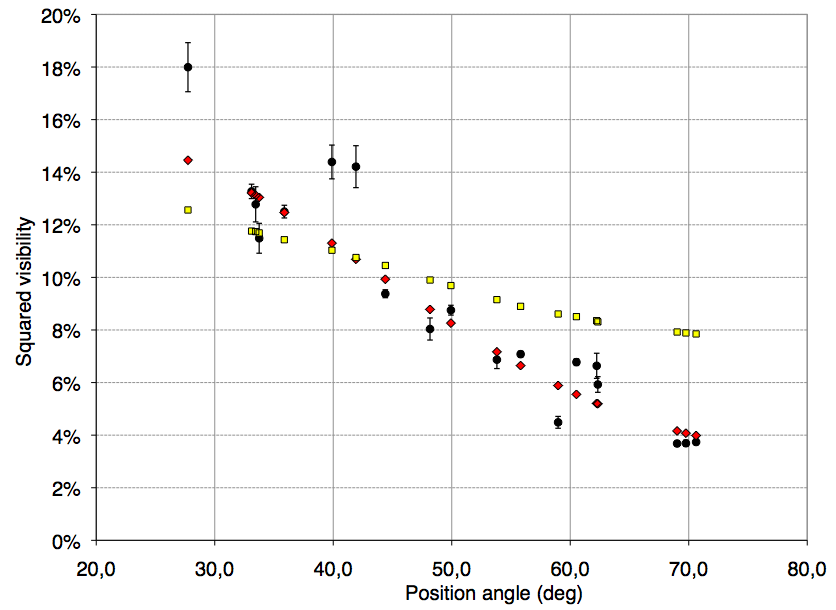} 
        \caption{VINCI squared visibilities measured on L$_2$\,Pup, as a function of the baseline length (top panel) and of the baseline position angle (bottom panel) at an effective wavelength of 2.19\,$\mu$m. The solid black dots are the VINCI visibility measurements, the red diamonds are the best fit uniform ellipse model visibilities (see Sect.~\ref{uniformellipse} for details), and the yellow squares show the visibility derived from the RADMC-3D model presented in Sect.~\ref{radmc}. The dashed curve in the top panel is a 26.6\,mas uniform disk squared visibility function (assuming $\omega=100\%$ unresolved flux), intended to guide the eye.\label{l2pupvinci}}
\end{figure}

The VINCI squared visibility measurements are presented in Fig.~\ref{l2pupvinci}. As shown in the top panel, the distribution of the $V^2(B)$ measurement points does not follow the general trend expected for a circularly symmetric object. Using a simple uniform disk model with 100\%  unresolved flux fraction, we obtained a best-fit uniform disk (UD) angular diameter $\theta_{\rm UD} = 26.6 \pm 1.5$\,mas with a poor reduced $\chi^2_{\rm red} \approx 150$. Considering $\omega=60\%$, we obtained $\theta_{\rm UD} = 22.4 \pm 2.5$\,mas with an even higher reduced $\chi^2_{\rm red} \approx 210$. This behavior is due to the fact that adding a resolved flux contribution to the model makes the slope of the visibility curve $V(B)$ shallower, while the observations show that it is actually steeper. We therefore test hereafter two simple geometrical models to better reproduce the VINCI data: 1) a uniform ellipse model and 2) a binary star model. 

\subsubsection{Uniform ellipse model \label{uniformellipse}}

In addition to the fixed resolved flux contribution $\omega = 60 \pm 5\%$, the adjusted model has three parameters: the uniform ellipse major-axis angular size $\theta_{\rm a}$, the minor-axis angular size $\theta_{\rm b}$ , and the position angle of the major axis $\alpha_{\rm a}$ (North=0$^\circ$, East=90$^\circ$). The formalism used to fit the data is presented in detail in \citetads{2006A&A...453.1059K}.

The best-fit parameters are $\theta_{\rm a} = 26.9 \pm 1.6\,\mathrm{mas}$, $\theta_{\rm b} = 17.5 \pm 1.6\,\mathrm{mas}$, and $\alpha_{\rm a} = 106 \pm 4^\circ$.
The minimum reduced $\chi^2_{\rm red}$ of the fit is 18, indicating that the dispersion of the measurements is larger than our simple ellipse model predicts. We therefore chose a conservative approach to estimate the error bars by solving the equation $\chi_{\rm red}^2(a,b,\alpha) = 2\, \chi_{\rm red\,min}^2$. The impact of the 5\% error bar on $\omega$ dominates the derived parameter uncertainties. The uniform ellipse model is certainly not physically realistic, but reasonable considering the limited available VINCI measurements.

As shown in the NACO images in the $K$ band (Fig.~\ref{deconv80}) and in Fig.~\ref{UnresolvedFlux}, the contribution of the circumstellar material of L$_2$\,Pup is important in the $K$ band. The distribution of the emitting material appears elongated approximately along an east-west axis, consistent with the position angle $\alpha_a$ of the major axis of the adjusted ellipse.
We took into account the resolved flux measured with NACO, including the part of the circumstellar material that is well inside the interferometric FOV of VINCI. But this contribution is extended spatially along an east-west axis, and it will decrease the measured visibilities (compared with a star without envelope) when the projected baseline azimuth is close to this direction. The measured east-west elongation is therefore probably caused by the circumstellar material and not by a physical elongation of the star itself.

We interpret the minor axis $\theta_{\rm b}  = 17.5 \pm 1.6$\,mas as an upper limit on the uniform disk angular diameter of the central star in the $K$ band. Considering that we do not detect a significant extension of L$_2$\,Pup's envelope in the $K$ band NACO images along the orthogonal direction (i.e. approximately North-South, see Fig.~\ref{deconv80}), this upper limit is most probably close to the actual angular size of the star within the stated uncertainty domain.
The linear limb-darkening coefficient computed by \citetads{2011A&A...529A..75C} using the flux conservation method is $u=0.3139$ in the $K$ band for $T_{\rm eff} = 3500$\,K, $\log g=1.5$ and solar metallicity. This value of $u$ gives a correction \citepads{1974MNRAS.167..475H}
\begin{equation}
\theta_{\rm LD} / \theta_{\rm UD} = \sqrt{\frac{1-u/3}{1-7u/15}} = 1.024.
\end{equation}
Applied to $\theta_{\rm UD} = 17.5 \pm 1.6$\,mas, this gives a maximum limb-darkened angular diameter $\theta_{\rm LD}= 17.9 \pm 1.6\,\mathrm{mas}$ for the central star. The angular diameter that we obtain agrees well with the size of the compact component observed by \citetads{2004MNRAS.350..365I} using aperture masking in the near-infrared. It is also consistent with the prediction by \citetads[][$\theta_{\rm LD} = 19.3$\,mas]{1998NewA....3..137D}, which was based on surface brightness considerations.

Combining the angular diameter with the parallax of L$_2$\,Pup \citepads[$\pi = 15.61 \pm 0.99$\,mas,][]{2007A&A...474..653V}, this translates into a maximum photospheric radius of R$(\mathrm{L}_2\,\mathrm{Pup}) = 123 \pm 14$\,R$_\odot$. This radius is consistent with the typical size of an M5III giant star \citepads{1998NewA....3..137D}. According to the AAVSO\footnote{\url{http://www.aavso.org}} light curve for L$_2$\,Pup, the VINCI observations were obtained at a phase close to the minimum apparent luminosity in its 141\,d cycle. For a pulsating star, this corresponds to a diameter slightly below its average value. 

%______________ Figure

\subsubsection{Binary star model}

It has been suggested by \citetads{2007ApJS..173..137G} that L$_2$\,Pup is a binary star, with a very close companion on a 141\,day orbit. Because the VINCI data have been obtained over a relatively short time (three weeks) compared to this orbital period, we considered all of them simultaneously to search for a binary star solution. In our adjusted model, we adopted an unresolved flux fraction $\omega=60\%$ (Sect.~\ref{resolvedflux}) and assumed that the central star $A$ is a uniform disk with an angular diameter $\theta_{\rm UD}$ (adjusted). We added an unresolved secondary source $B$ located at a position $(\Delta x, \Delta y)$ relative to the primary, with a flux ratio $\rho = f(B)/f(A)$. We mapped the parameter space over a broad range of parameter values ($\theta_{\rm UD}=10$ to 30\,mas, $\Delta x$ and $\Delta y= 0$ to 100\,mas, $\rho=0$ to 0.5), but found no satisfactory solution. The best-fit reduced $\chi^2$ values remained significantly higher (around 300) than the uniform ellipse model. Qualitatively, the reason for this behavior is that our simple binary star model does not reproduce properly the observed slope and azimuth dependence of the measured visibility curve. Moreover, in absence of closure phase data (VINCI is a two-telescope instrument providing only $V^2$ values), it is not possible to fit a more sophisticated model that for instance includes a better description of the extended dust disk.
From this analysis, we do not exclude the presence of a companion, however, because our VINCI data are too fragmentary to conclude.

\subsection{VLTI/MIDI spectroscopy and interferometry\label{midiobs}}

To complete the photometric and spectrophotometric data present in the literature, we retrieved VLTI/MIDI observations from the ESO Data Archive and processed them using the MIA+EWS\footnote{\url{http://home.strw.leidenuniv.nl/~nevec/MIDI/}} data processing pipeline \citepads{2007NewAR..51..666C}. MIDI is an interferometric beam combiner for two telescopes \citepads{2003Ap&SS.286...73L}, which provides fringe visibility and differential phase measurements in addition to classical spectroscopic capabilities.

The MIDI observation of L$_2$\,Pup was recorded on 3 January 2005 (MJD = 53373.24), together with one observation of the calibrator star \object{HD 67582}, which is part of the \citetads{1999AJ....117.1864C} reference catalog. The two stars were observed between UT 05:06 and UT 06:40 using the 8-meter Unit Telescopes UT2 and UT4. This configuration provided a projected baseline length of 88.9\,m along a position angle of PA=5.2$^\circ$ on L$_2$\,Pup.

%______________ Figure
\begin{figure}[]
        \centering
        \includegraphics[width=\hsize]{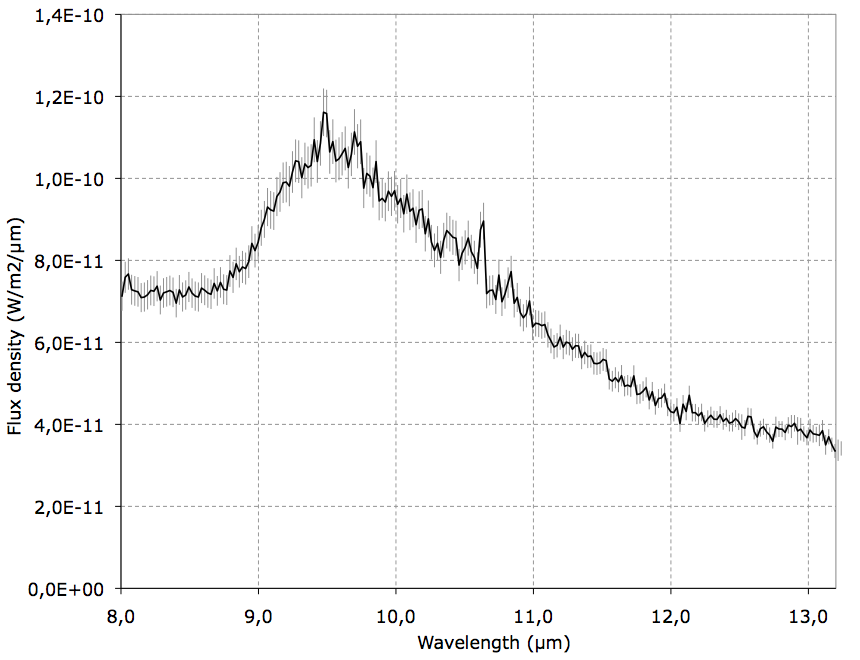} 
        \caption{MIDI photometric spectrum of L$_2$\,Pup.\label{l2pupmidispec}}
\end{figure}

The photometric spectrum is presented in Fig.~\ref{l2pupmidispec}. The classical silicate emission feature is clearly visible longward of $\lambda=9\,\mu$m. This feature was already present in the Japanese Infrared Telescope in Space (IRTS) spectrum presented by \citetads{2002MNRAS.337...79B}.

\begin{figure}[]
        \centering
        \includegraphics[width=\hsize]{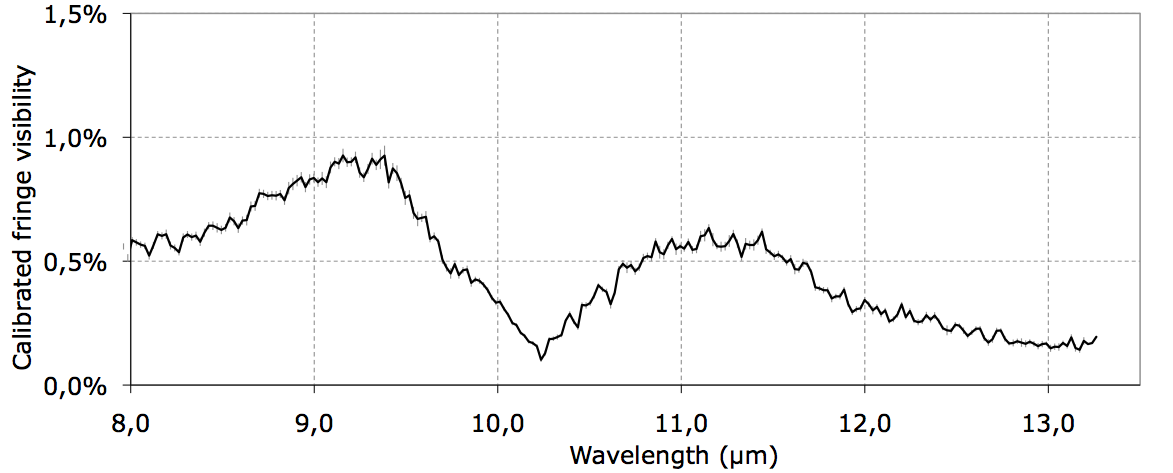} 
        \includegraphics[width=\hsize]{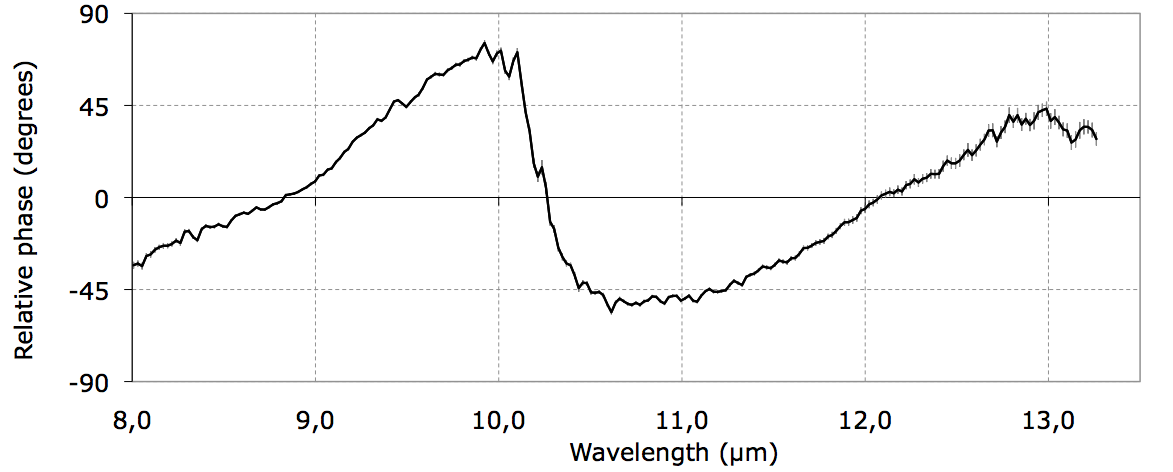} 
        \caption{MIDI fringe visibility spectrum (top panel) and relative phases (bottom panel) of L$_2$\,Pup.\label{l2pupmidiinterf}}
\end{figure}

\begin{figure}[]
        \centering
        \includegraphics[width=\hsize]{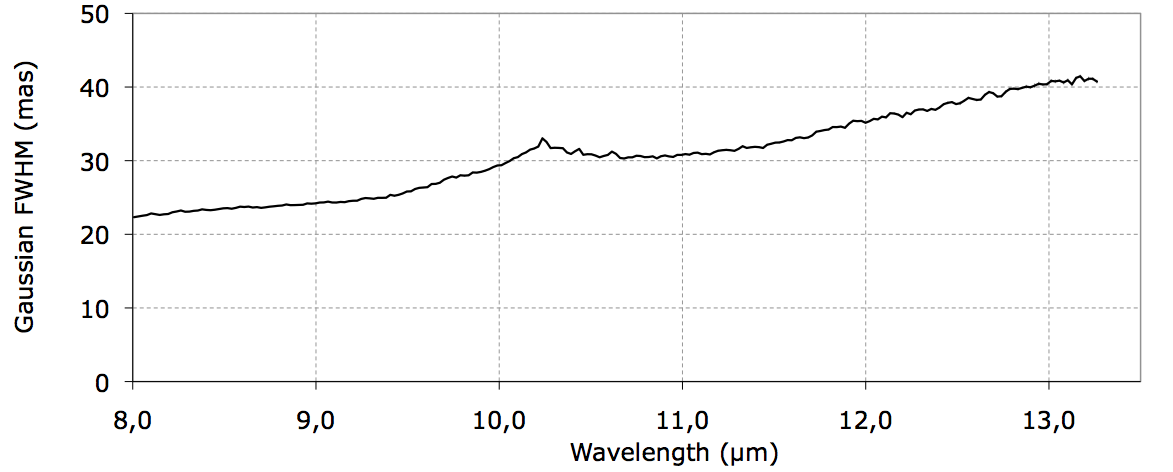} 
        \caption{Equivalent Gaussian FWHM of L$_2$\,Pup derived from the MIDI visibilities.\label{l2pupmidifwhm}}
\end{figure}

The interferometric visibility spectrum and differential phases are presented in Fig.~\ref{l2pupmidiinterf}. The visibility level is extremely low, indicating that the thermal emission from the dust is fully resolved on the relatively long UT1-UT4 observation baseline, and therefore more extended than the resolution of the interferometer ($\approx 25$\,mas). A simple fit of a Gaussian following the formalism presented by \citetads{2004A&A...423..537L} results in the FWHM angular size as a function of wavelength presented in Fig.~\ref{l2pupmidifwhm}.
The $\approx 18$\,mas angular size derived from the VINCI measurements (Sect.~\ref{vinciobs}) corresponds reasonably well to the FWHM derived at $\lambda = 8\,\mu$m ($\approx 22$\,mas). At this wavelength, the flux contribution of the photosphere is still higher
than that of the dust. Around $10\,\mu$m and longward, the measured FWHM increases significantly up to 40\,mas because the thermal emission becomes dominant.

From the spectral energy distribution discussed in Sect.~\ref{photom}, the relative flux contribution of the central star is $\approx 10\%$ at $10\,\mu$m, and based on the VINCI angular diameter, its expected photospheric visibility is $\approx 40\%$.
These two figures imply that the photosphere probably contributes a $\approx 4\%$ visibility to the MIDI measurements. Because we observe a much lower visibility value (below 1\%), the morphology of the dust envelope in this wavelength range is probably more complex than a simple Gaussian. This conclusion is strengthened by the presence of a dust band in front of the star and of a large loop in the $L$ band images (Sect.~\ref{morphology}).

Interferometric measurements with MIDI at shorter baselines and covering a broader range of position angles are necessary to better constrain the true extension and morphology of the dust envelope at $10\,\mu$m. The position angle of the presented MIDI measurement is almost aligned with the minor axis of the disk-like structure observed with NACO. As a consequence, the derived FWHM extension should be considered a lower limit for the actual extension of the thermal infrared emission.

The differential phase signal shows a significant variation with wavelength that is commonly associated with a drift of the photocenter of the object as the wavelength changes.
This phase behavior is commony interpreted as caused by a disk \citepads[see e.g.][]{2007A&A...474L..45D, 2008A&A...490..173O}, and few alternative explanations are possible.
The photospheric stellar flux decreases as wavelength increases, while the silicate emission feature at 10\,$\mu$m and longward increases the relative contribution from the dust
considerably. This change in flux ratio between the star and the disk as a function of wavelength directly affects the interferometric phase, as observed on the MIDI phase signal (Fig.~\ref{l2pupmidiinterf}).

However, considering the complexity of the geometry of the observed dust distribution and the scarcity of the MIDI sampling of the spatial frequency $(u,v)$ plane, a different interpretation of the visibility and phase signals appears to be very difficult. We therefore did not consider the MIDI interferometric observables (visibility and phase) any longer in the present study. We used the photometric spectrum to constrain the photometry of our radiative transfer model, however.

%______________ Figure

%______________ Figure

\subsection{Spectral energy distribution of L$_2$\,Pup \label{photom}}

We searched the literature for photometry of L$_2$\,Pup and the retrieved measurements are listed in Table~\ref{photometryliterature}. Because it is a variable star, there is some scattering in the measured fluxes, particularly in the visible. The WISE bands 1, 2, and 3 \citepads{2012yCat.2311....0C} are discrepant compared to the other literature measurements, possibly because of detector saturation on this extremely bright source, and were excluded from this analysis. The Planck data points were retrieved from the recently released \citetads{2013yCat.8091....0P} catalog, considering the PSF fitting method. Compared with the other listed measurements, the angular resolution of the Planck instruments is low, with a beam size of $\approx 5\arcmin$ at the considered frequencies (217-857\,GHz). The millimeter wavelength emission from L$_2$\,Pup and its envelope (millimeter excess) are clearly detected, however.

%_________Table of photometry from the literature
\begin{table}
        \caption{Photometric measurements of L$_2$ Pup from the literature.}
        \label{photometryliterature}
        \centering          
        \begin{tabular}{llll}
        \hline \hline
        \noalign{\smallskip}
        System & $\lambda_0 \pm \Delta \lambda$ & Flux density & Ref. \\
         & $[\mu$m$]$ & [W/m$^2$/$\mu$m] & \\
        \hline
        \noalign{\smallskip}
        ANS 330\,nm & $ 0.33 \pm 0.01 $ & $ 5.52 \pm 0.13 \times 10^{-12} $ & W82 \\
        Johnson U & $ 0.36 \pm 0.02 $ & $ 3.01 \pm 0.14 \times 10^{-11} $ & D02 \\
        Johnson U & $ 0.36 \pm 0.02 $ & $ 4.26 \pm 0.20 \times 10^{-11} $ & M94 \\
        Johnson B & $ 0.44 \pm 0.05 $ & $ 1.56 \pm 0.07 \times 10^{-10} $ & D02 \\
        Johnson B & $ 0.44 \pm 0.05 $ & $ 1.65 \pm 0.08 \times 10^{-10} $ & M78 \\
        Johnson B & $ 0.44 \pm 0.05 $ & $ 2.22 \pm 0.11 \times 10^{-10} $ & M94 \\
        Johnson V & $ 0.55 \pm 0.04 $ & $ 3.58 \pm 0.07 \times 10^{-10} $ & D02 \\
        Johnson V & $ 0.55 \pm 0.04 $ & $ 3.58 \pm 0.07 \times 10^{-10} $ & M78 \\
        Johnson V & $ 0.55 \pm 0.04 $ & $ 5.02 \pm 0.09 \times 10^{-10} $ & M94 \\
        Johnson R & $ 0.70 \pm 0.11 $ & $ 1.78 \pm 0.08 \times 10^{-9} $ & D02 \\
        Johnson R & $ 0.70 \pm 0.11 $ & $ 1.97 \pm 0.09 \times 10^{-9} $ & M78 \\
        Johnson I & $ 0.90 \pm 0.11 $ & $ 7.78 \pm 0.37 \times 10^{-9} $ & D02 \\
        Johnson I & $ 0.90 \pm 0.11 $ & $ 8.22 \pm 0.39 \times 10^{-9} $ & M78 \\
        Johnson J        & $ 1.25 \pm 0.15 $ & $ 6.52 \pm 0.43 \times 10^{-9} $ & B02 \\
        Johnson J & $ 1.25 \pm 0.15 $ & $ 5.75 \pm 0.50 \times 10^{-9} $ & D02 \\
        Johnson J & $ 1.25 \pm 0.15 $ & $ 4.91 \pm 0.43 \times 10^{-9} $ & M78 \\
        2MASS J & $ 1.25 \pm 0.15 $ & $ 8.33 \pm 0.89 \times 10^{-9} $ & C03 \\
        COBE 1.25\,$\mu$m & $ 1.27 \pm 0.16 $ & $ 7.51 \pm 0.50 \times 10^{-9} $ & S04 \\
        Johnson H & $ 1.62 \pm 0.10 $ & $        5.82 \pm 0.37 \times 10^{-9} $ & B02  \\
        Johnson H & $ 1.62 \pm 0.10 $ & $ 5.76 \pm 0.50 \times 10^{-9} $ & D02 \\
        2MASS H & $ 1.62 \pm 0.10 $ & $ 6.36 \pm 0.94 \times 10^{-9} $ & C03 \\
        Johnson K & $ 2.20 \pm 0.30 $ & $       2.76 \pm 0.14 \times 10^{-9} $ & B02 \\
        2MASS K & $ 2.16 \pm 0.30 $ & $ 3.59 \pm 0.61 \times 10^{-9} $ & C03 \\
        Johnson K & $ 2.20 \pm 0.30 $ & $ 2.39 \pm 0.21 \times 10^{-9} $ & D02 \\
        Johnson K & $ 2.20 \pm 0.30 $ & $ 2.01 \pm 0.17 \times 10^{-9} $ & M78 \\
        COBE 2.20\,$\mu$m & $ 2.22 \pm 0.19 $ & $ 3.52 \pm 0.22 \times 10^{-9} $ & S04 \\
        Johnson L & $ 3.40 \pm 0.90 $ & $ 1.05 \pm 0.04 \times 10^{-9} $ & B02 \\
        Johnson L & $ 3.40 \pm 0.90 $ & $ 1.12 \pm 0.10 \times 10^{-9} $ & D02 \\
        COBE 3.50\,$\mu$m & $ 3.53 \pm 0.49 $ & $ 9.34 \pm 0.46 \times 10^{-10} $ & S04 \\
        COBE 4.90\,$\mu$m & $ 4.88 \pm 0.36 $ & $ 3.33 \pm 0.08 \times 10^{-10} $ & S04 \\
        IRAS 12\,$\mu$m & $ 11.5 \pm 3.5 $ & $ 5.49 \pm 0.17 \times 10^{-11} $ & B88 \\
        COBE 12\,$\mu$m & $ 12.3 \pm 4.0 $ & $ 4.24 \pm 0.10 \times 10^{-11} $ & S04 \\
        COBE 25\,$\mu$m & $ 20.8 \pm 4.5 $ & $ 9.62 \pm 0.29 \times 10^{-12} $ & S04 \\
        WISE 22.1\,$\mu$m & $ 22.1 \pm 5.0 $ & $ 6.68 \pm 0.32 \times 10^{-12} $ & C12 \\
        IRAS 25\,$\mu$m & $ 23.5 \pm 2.7 $ & $ 4.59 \pm 0.18 \times 10^{-12} $ & B88 \\
        COBE 60\,$\mu$m & $ 56 \pm 14 $ & $ 1.56 \pm 0.65 \times 10^{-13} $ & S04 \\
        IRAS 60\,$\mu$m & $ 62 \pm 17 $ & $ 7.34 \pm 0.59 \times 10^{-14} $ & B88 \\
        AKARI 65\,$\mu$m & $ 65 \pm 15 $ & $ 6.06 \pm 0.20 \times 10^{-14} $ & I10 \\
        AKARI 90\,$\mu$m & $ 90 \pm 30 $ & $ 1.57 \pm 0.15 \times 10^{-14} $ & I10 \\
        IRAS 100\,$\mu$m & $ 103 \pm 18 $ & $ 7.86 \pm 0.94 \times 10^{-15} $ & B88 \\
        AKARI 140\,$\mu$m & $ 140 \pm 30 $ & $ 2.02 \pm 0.12 \times 10^{-15} $ & I10 \\
        AKARI 160\,$\mu$m & $ 160 \pm 20 $ & $ 1.09 \pm 0.25 \times 10^{-15} $ & I10 \\
        Planck 857 GHz & $ 350 \pm 59 $ & $ 3.37 \pm 0.71 \times 10^{-17} $ & P13 \\
        Planck 353 GHz & $ 850 \pm 144 $ & $ 9.39 \pm 1.58 \times 10^{-19} $ & P13 \\
        Planck 217 GHz & $ 1382 \pm 234 $ & $ 1.13 \pm 0.50 \times 10^{-19} $ & P13 \\
        \hline
        \end{tabular}
        \tablefoot{The references are B02 = \citetads{2002MNRAS.337...79B}, D02 = \citetads{2002yCat.2237....0D},
        M78 = \citetads{1978A&AS...34..477M}, M94 = \citetads{1994cmud.book.....M}, W82 = \citetads{1982A&AS...49..427W},
        B88 = \citetads{1988iras....1.....B}, C03 = \citetads{2003tmc..book.....C}, I10 = \citetads{2010A&A...514A...1I},
        C12 = \citetads{2012yCat.2311....0C}, P13 = \citetads{2013yCat.8091....0P}, S04 = \citetads{2004ApJS..154..673S}.}
\end{table}

\begin{figure*}[]
        \centering
        \includegraphics[width=\hsize]{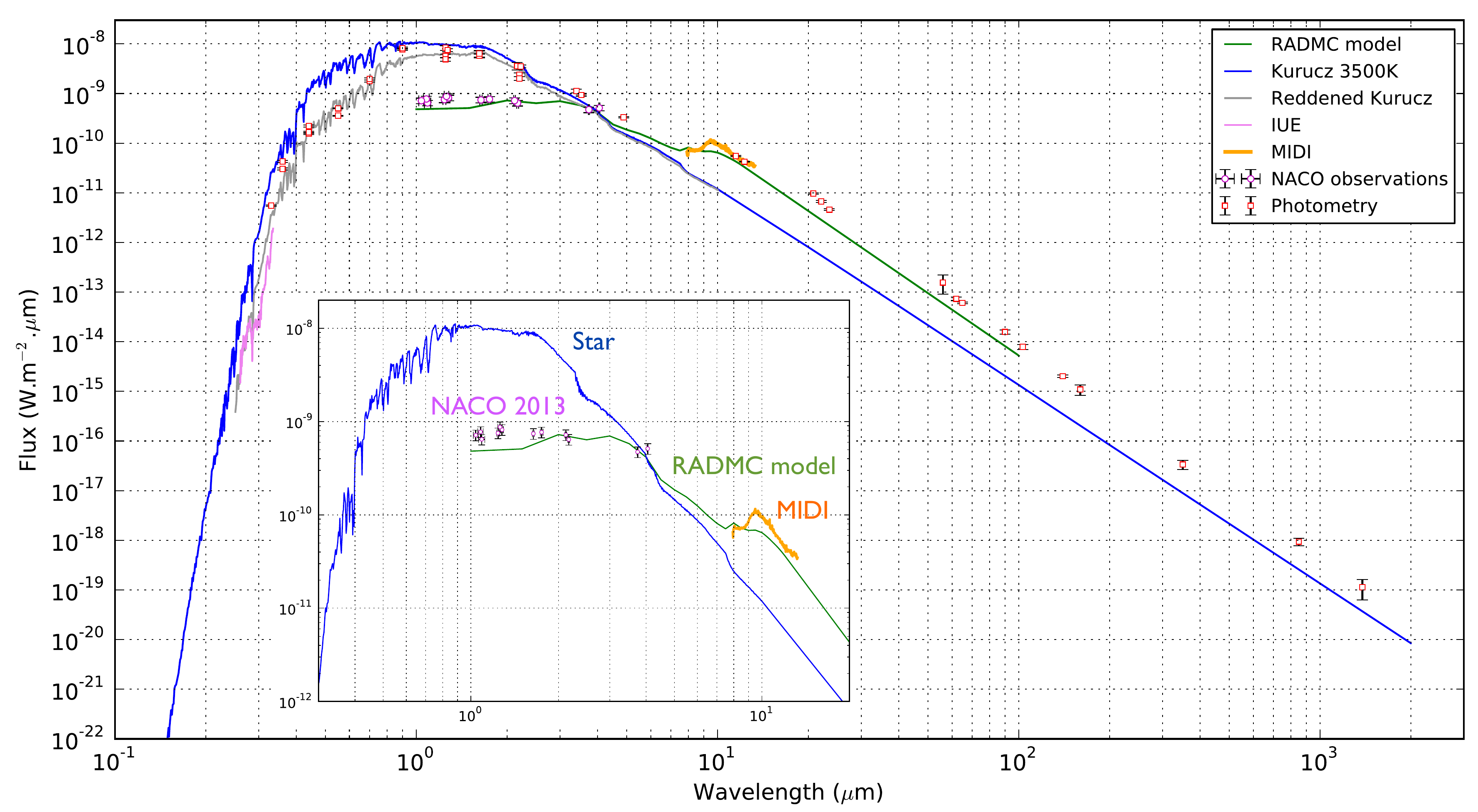} 
        \caption{Spectral energy distribution of L$_2$\,Pup. The blue curve represents the ATLAS model of the central star, and the gray curve is a reddened version of this spectrum assuming $E(B-V)=0.6$. The available photometry from IUE, MIDI, NACO and the literature are also represented for reference. The green curve represents the photometry extracted from our RADMC-3D model of a circumstellar dust disk around L$_2$\,Pup. The inset shows an enlargement of the visible to thermal infrared section of the spectrum.
        \label{l2pupsed}}
\end{figure*}

Figure~\ref{l2pupsed} presents the available photometric measurements from the literature as red squares, the archival IUE spectrum with reference LWP22250\footnote{Retrieved from the IUE database at MAST
\url{http://archive.stsci.edu/iue/search.php}} as a green curve, and the MIDI spectrum (Sect.~\ref{midiobs}) as an orange curve.

Many different values of the effective temperature of L$_2$\,Pup are found in the literature, from 2800\,K or below up to 3500\,K or more. Adding to the difficulty of the measurement, this temperature is also variable over the 14-\,day cycle of the star \citepads{2009Ap.....52...88R}. \citetads{2002MNRAS.337...79B} discussed the choice of effective temperature for this star. The main difficulty in estimating this parameter is that the photometric flux of L$_2$\,Pup itself is affected by two bias sources, depending on the wavelength range. At visible and very near-infrared wavelengths (shortward of the $H$ band), there is strong circumstellar absorption (as shown in Fig.~\ref{deconv80}) and scattering that results in a negative bias. Longward of the $K$ band, there is a significant contribution from the thermal emission of the dust, which creates a positive bias on the measured flux. The near-infrared $H$ and $K$ bands provide the best window to avoid these two bias sources, as the scattering is much less efficient in these bands than in the visible, while the thermal emission from the dust is still minor compared with the photospheric flux. We here took advantage of the $JHKL$ photometry published by \citetads{2002MNRAS.337...79B} to derive the mean magnitudes of L$_2$\,Pup: $m_J = -0.71 \pm 0.07$, $m_H = -1.66 \pm 0.07$, $m_K = -2.13 \pm 0.05$, $m_L = -2.78 \pm 0.04$. We now combine these $H$ and $K$ mean magnitudes with the angular diameter $\theta_\mathrm{LD} = 17.9 \pm 1.6$\,mas we derived in Sect.~\ref{uniformellipse} to apply the surface brightness-temperature relations derived by \citetads{2004A&A...426..297K}. We obtain effective temperatures in $H$ and $K$  of 3250 and 3500\,K, with an uncertainty of 500\,K. We therefore selected for L$_2$\,Pup a model atmosphere SED with an effective temperature of 3500\,K. We note that this is close to the choice of \citetads{2002ApJ...569..964J} (3400\,K).

In Fig.~\ref{l2pupsed}, the thin gray curve represents an ATLAS model for $T_{\rm eff} = 3500$\,K, $\log g = 1.5$, $[M/H]=0.0$, which was retrieved from the \citetads{2004astro.ph..5087C} grid\footnote{\url{http://www.stsci.edu/hst/observatory/cdbs/castelli\_kurucz\_atlas.html}} 
and normalized to a limb-darkened angular diameter of  $\theta_{\rm LD}=17.9$\,mas. The thin gray curve is a reddened version of this ATLAS model spectral energy distribution for a selective absorption $E(B-V)=0.6$, using the prescription by \citetads{1999PASP..111...63F} for a standard $R_V=A_V/E(B-V)=3.1$ interstellar dust model. This value of $E(B-V)$ was adjusted to reproduce the observed SED of L$_2$\,Pup before its recent dimming (i.e., considering the photometric measurements from the literature).

The integrated flux from the Kurucz model gives a bolometric luminosity $L = 2000 \pm 700 $\,L$_\odot$, assuming the \emph{Hipparcos} parallax \citepads[$\pi = 15.61 \pm 0.99$\,mas,][]{2007A&A...474..653V} and a solar bolometric luminosity of $L_\odot = 3.846 \times 10^{26}$\,W \citepads{2010AJ....140.1158T}.

%______________ Figure

%__________________________________Modeling
\section{Modeling of the nebula \label{radmc}}

\subsection{Overall morphology of the nebula and chosen model\label{morphology}}

\begin{figure}[]
\centering
\includegraphics[width=\hsize]{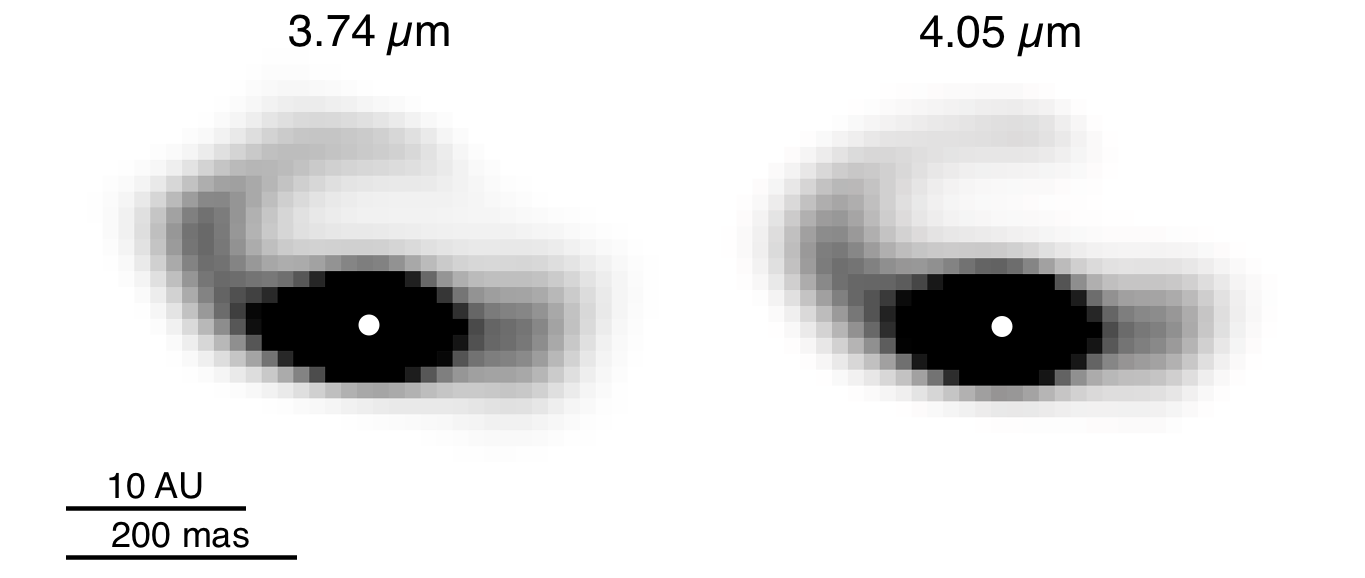} 
\caption{Deconvolved NACO images of L$_2$\,Pup at 3.74 and 4.05\,$\mu$m. The position and angular size of the star is represented with a white disk.
% The square root intensity scale is expressed in W\,m$^{-2}$\,$\mu$m$^{-1}$\,arcsec$^{-2}$. The field of view is $0.60 \times 0.60 \arcsec$.
\label{dec80loop}}
\end{figure}

The NACO images presented in Fig.~\ref{deconv80} show that the morphology of the nebula around L$_2$\,Pup is dominated by a dark band in the $J$ band and elongated or loop-like structures at longer wavelengths. As the scattering of the stellar light by the dust is more efficient at shorter wavelengths, the dust band masks the star very efficiently in the $J$ band, and we observe essentially the light scattered above and below the dust disk. This results in an apparent ``double source'' with one component to the north and one to the south of the star itself, which essentially remains hidden behind the dust. As wavelength increases, the scattering becomes less efficient, and the dust becomes more translucent.
In the $K$ band, the star is visible as a single source, together with an east-west segment approximately 2-3\,AU in radius. In the $L$ band (around 4\,$\mu$m), the thermal emission of the close-in warm dust becomes important. We observe an elongated central source and a large loop extending up to more than 10\,AU in radius from the star in the northeast quadrant (Fig.~\ref{dec80loop}). The innermost section of this remarkable feature is also detected at shorter wavelengths ($H$ and $K$ bands). The symmetry of the nebula is essentially along the NS and EW directions, although a noticeable level of warping is visible in the NE-SW direction. This asymmetry may be related to the formation of the loop.

%______________ Figure

The general structure of L$_2$\,Pup's envelope appears consistent with a dust disk seen almost edge-on. Under this hypothesis, we computed a model of the disk that is presented in the next paragraphs. However, this is certainly not the only possible interpretation of the NACO images. The dust band might for instance be transiting temporarily in front of the star without pertaining to a complete circumstellar disk. We also assumed in our model that the dust distribution in the disk follows an axial symmetry. The long-term photometric variability of L$_2$\,Pup presented by \citetads{2002MNRAS.337...79B} could be interpreted as the consequence of variable obscuration of the star by circumstellar dust. In this scenario, the relatively large amplitude of the photometric variation in the visible ($\Delta m_V \approx +2.5$\,mag since 1995) would imply a very inhomogeneous dust distribution across the disk surface, which is currently not reflected in our model.

The dust disk model we chose should therefore be understood as a simplified, first-order attempt to reproduce the observed morphology of the envelope of L$_2$\,Pup and its present photometry.
%The asymmetries (loop, warp) and inhomogeneities are not part of this model.

\subsection{RADMC-3D radiative transfer model}

We used the RADMC-3D code\footnote{Available at \url{http://www.ita.uni-heidelberg.de/\~dullemond/software/radmc-3d/}} \citepads{2012ascl.soft02015D} to model the dusty nebula around L$_2$ Pup. RADMC-3D is designed for astrophysical radiative transfers calculations. It is working in two steps: first it computes the dust temperature by running a thermal Monte Carlo simulation \citepads{2001ApJ...554..615B}, then it produces an image of the dust continuum. The model was first created using the protoplanetary disk GUI provided in the RADMC-3D package and developed by \citetads{2010ApJ...721..431J},
which  we adapted to suit our needs. We used a regular grid without refinement, whose parameters are listed in Table~\ref{Table_Grid_Parameters}. 

        \begin{table}
                \caption{Grid parameters of the RADMC-3D simulation}
                \label{Table_Grid_Parameters}      
                \centering          
                \begin{tabular}{l l} 
                \hline\hline
                Parameter & Value \\
                \hline
                \noalign{\smallskip}
                Inner radius & 3~AU \\
                Outer radius & 150~AU \\
                Number of points in radius direction & 250 \\
                Maximum meridional angle & 0.8~rad \\
                Number of points in meridional angle & 120 \\
                Number of points in azimuthal angle & 360 \\    
                Wavelength interval limits & $\left[0.1,1,10,100 \right]~\mu$m \\
                Number of points in each interval & $\left[20,50,20 \right]$ \\
                \hline
                \end{tabular}
        \end{table}

Consistent with the VLTI/VINCI observations (see Sect.~\ref{vinciobs}), we considered a radius of 123\,R$_\odot$ for the central star. The spectrum of the photosphere was taken from an ATLAS model with $T_\mathrm{eff}$ = 3500~K, $\log g$ = 0.0 and $\left[M/H\right]$ = 0.0 (Fig.~\ref{l2pupsed}). 
We considered an isotropic scattering model for the dust.
The disk was considered to be a flared disk, its height being described by
\begin{equation}
H(r) = H_\mathrm{out}\, \left( \frac{r}{R_\mathrm{out}} \right)^\beta
,\end{equation}
where $\beta$ is the flaring index and $H_\mathrm{out}$ the external height. The dust surface density for the species $i$ at a radius $r$ from the star is defined by
\begin{equation}
\sigma^i(r) = \sigma^i_\mathrm{out} \, \left( \frac{r}{R_\mathrm{out}} \right)^\alpha
,\end{equation}
where $\sigma^i_\mathrm{out}$ is its surface density at $R_\mathrm{out}$ and $\alpha$ the dust density power-law exponent.

To compare our RADMC-3D model with the VINCI observations, we computed interferometric visibilities from our model image at 2.17\,$\mu$m.
The results are presented in Fig.~\ref{l2pupvinci} (yellow squares). A multiplicative factor of $75\%$ was applied to match the model visibilities to the mean value of the VINCI visibilities.
This indicates that a significant resolved flux contribution of $\approx 25\%$ of the total flux in the $K$ band is not included in our RADMC-3D model.
The overall slope of the RADMC-3D visibility curve also appears too shallow compared with the VINCI data, which are  better reproduced by the uniform ellipse model presented in Sect.~\ref{uniformellipse}.

These differences may be caused by the simplicity of our dust disk model. It assumes that the scattering material is confined close to the plane of the disk and that the dust distribution follows a simple flared-disk geometry.
Discrepancies with the observations may be caused for example by extended emission above or below the plane of the disk, inhomogeneous dust clumps, or simply by a dust distribution in the disk different from the law assumed in our model. We note that the long-term variability of L$_2$\,Pup reported by \citetads{2002MNRAS.337...79B} indicates an inhomogeneous dust distribution. Our NACO images also show that emission is indeed present outside of the disk plane, in particular through its distinct warp and the extended $L$ -band loop (Sect.~\ref{morphology}). Our current limited interferometric data set does not allow us to directly constrain the dust distribution law of our model. 
The difference in slope of our model visibility and the VINCI measurements may be caused by dust emission within a few tens of milliarcseconds from the star that is not taken into account in our RADMC-3D model. Such material could mimic an elliptical elongation of the central object in the interferometric visibilities.

%______________ Figure
\begin{figure*}[]
\centering
\includegraphics[width=4.5cm]{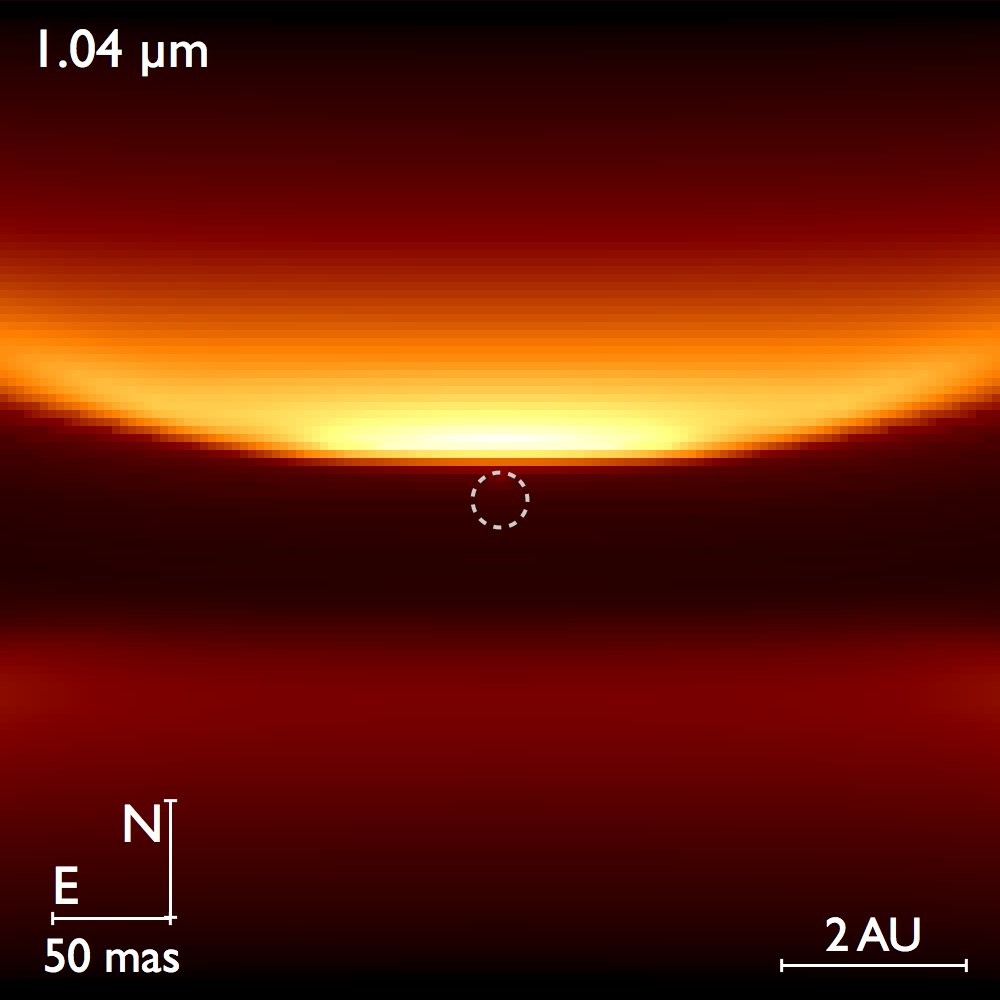}
\includegraphics[width=4.5cm]{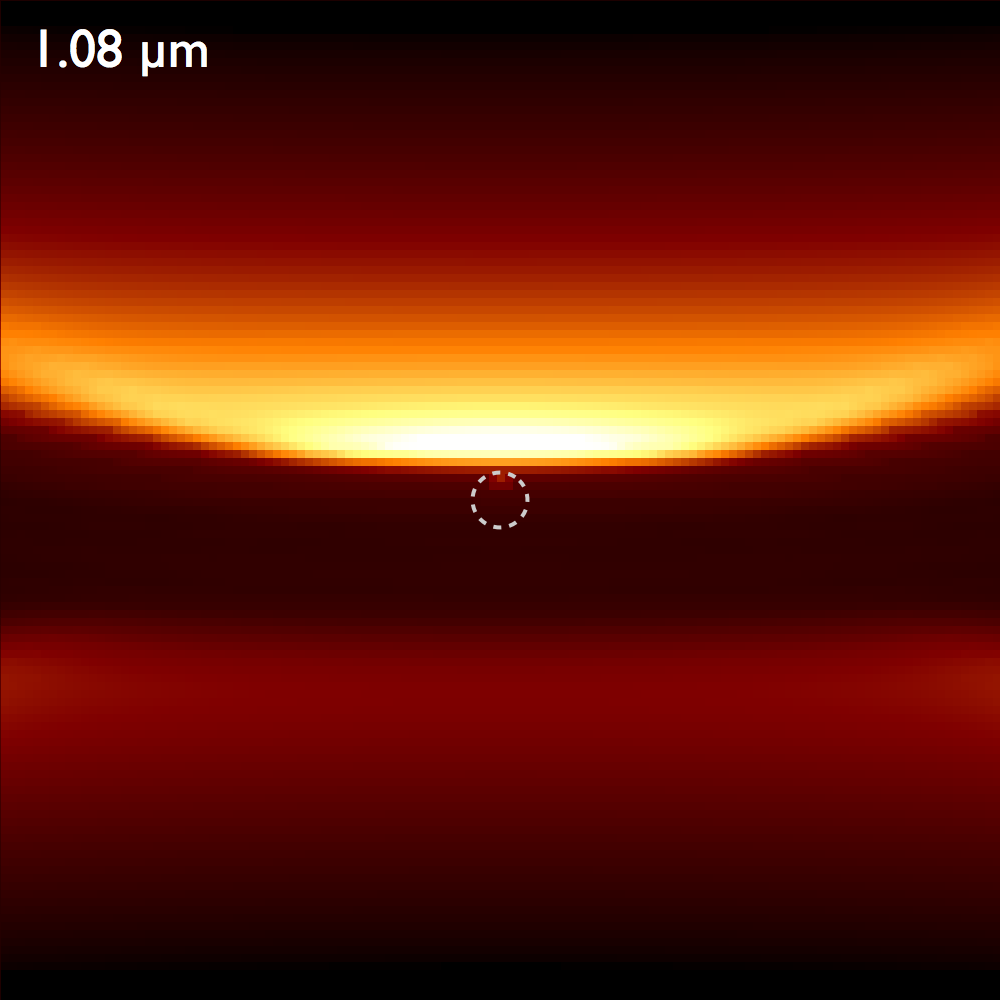}
\vspace{1mm}
\includegraphics[width=4.5cm]{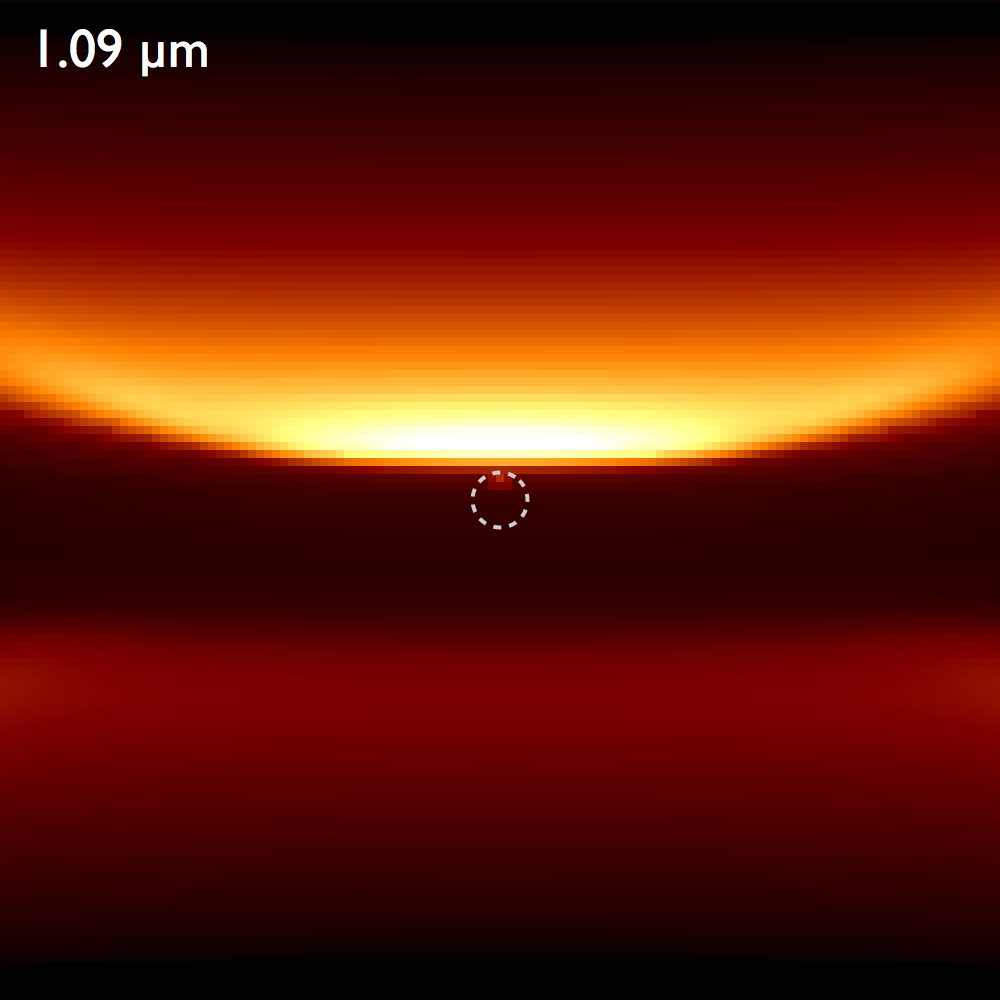}
\includegraphics[width=4.5cm]{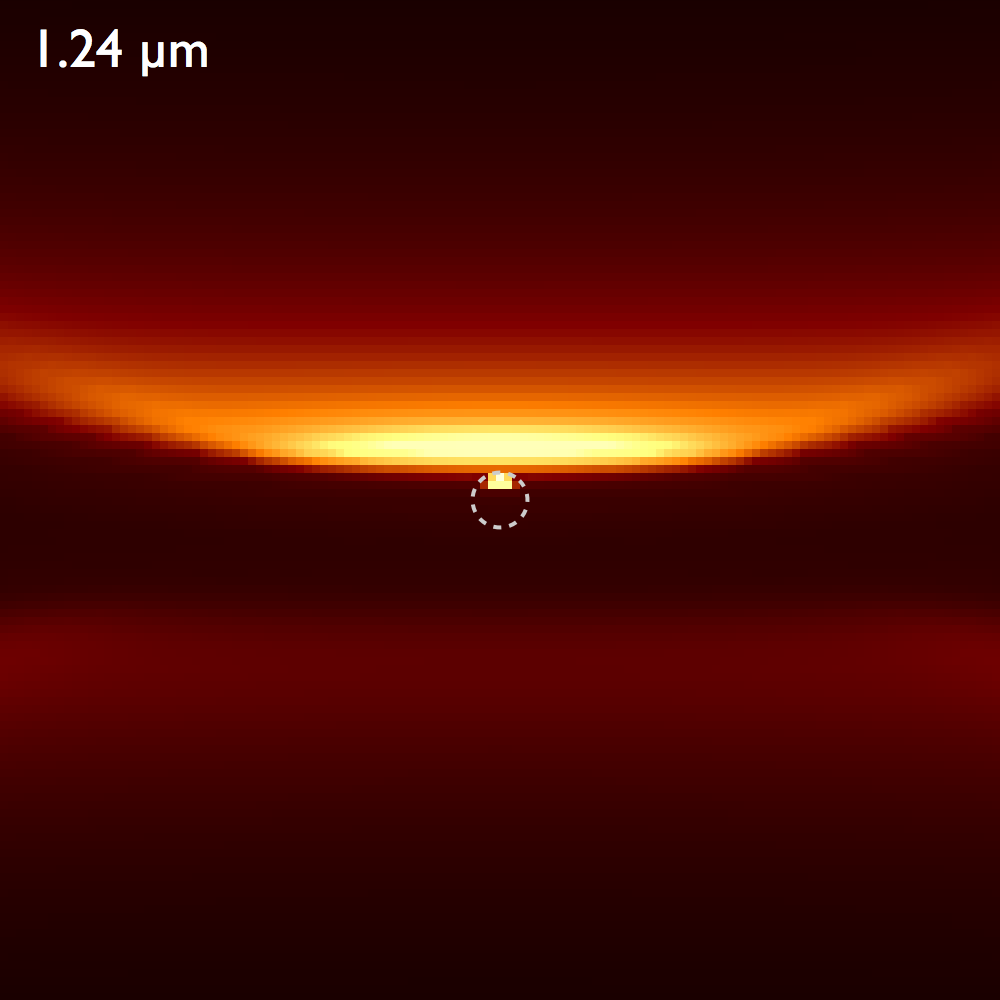}
\includegraphics[width=4.5cm]{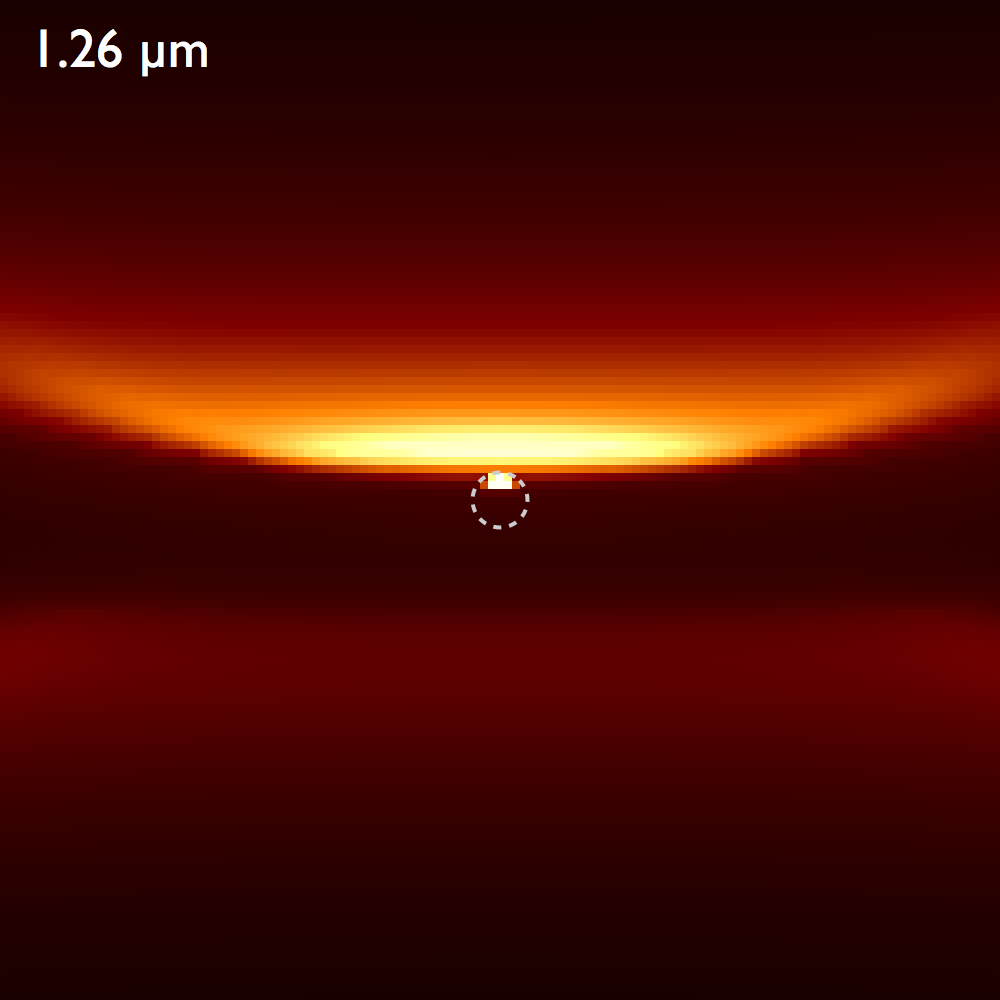}
\includegraphics[width=4.5cm]{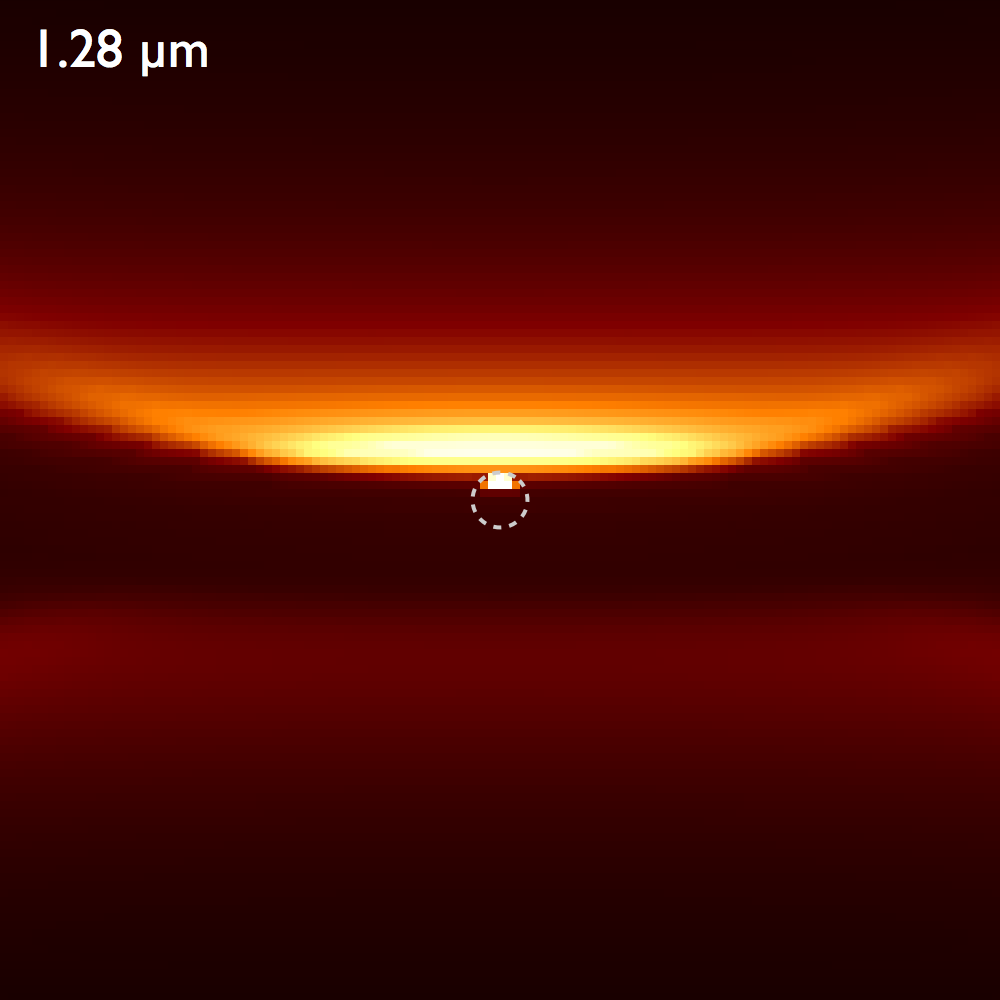}
\vspace{1mm}
\includegraphics[width=4.5cm]{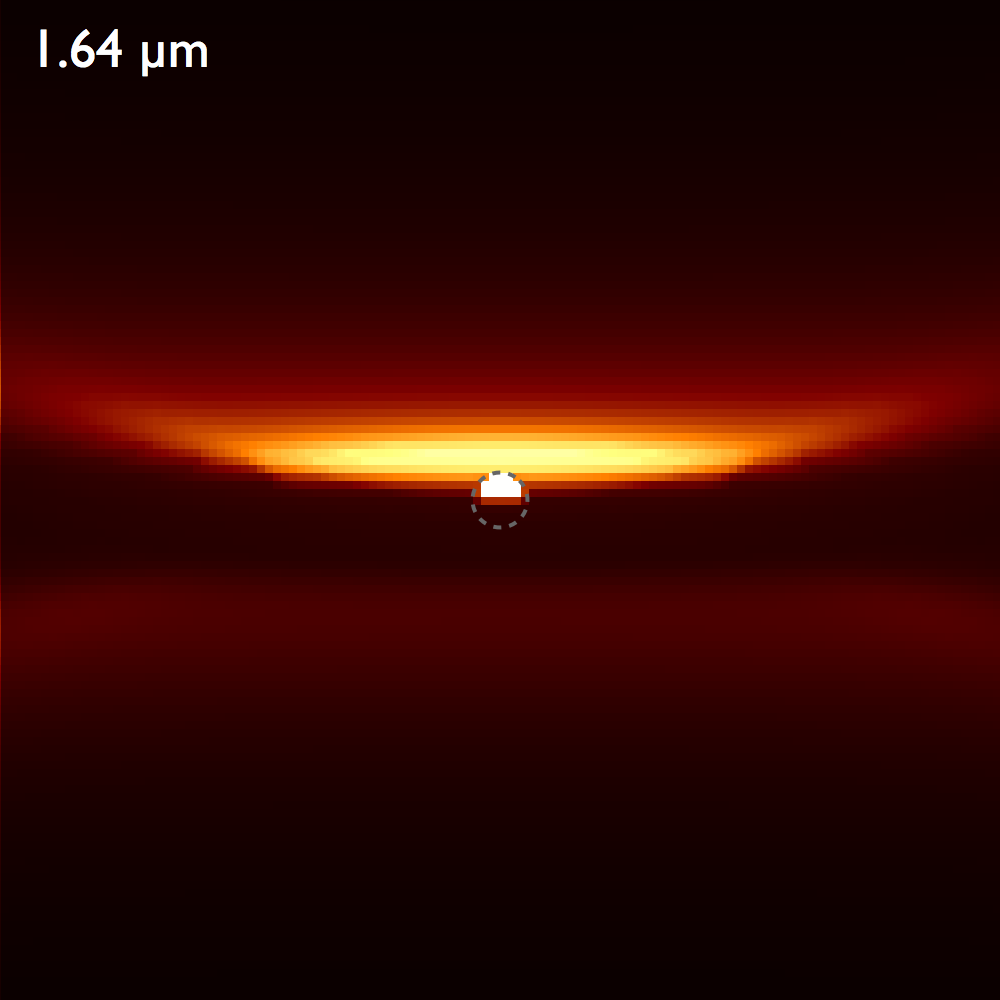}
\includegraphics[width=4.5cm]{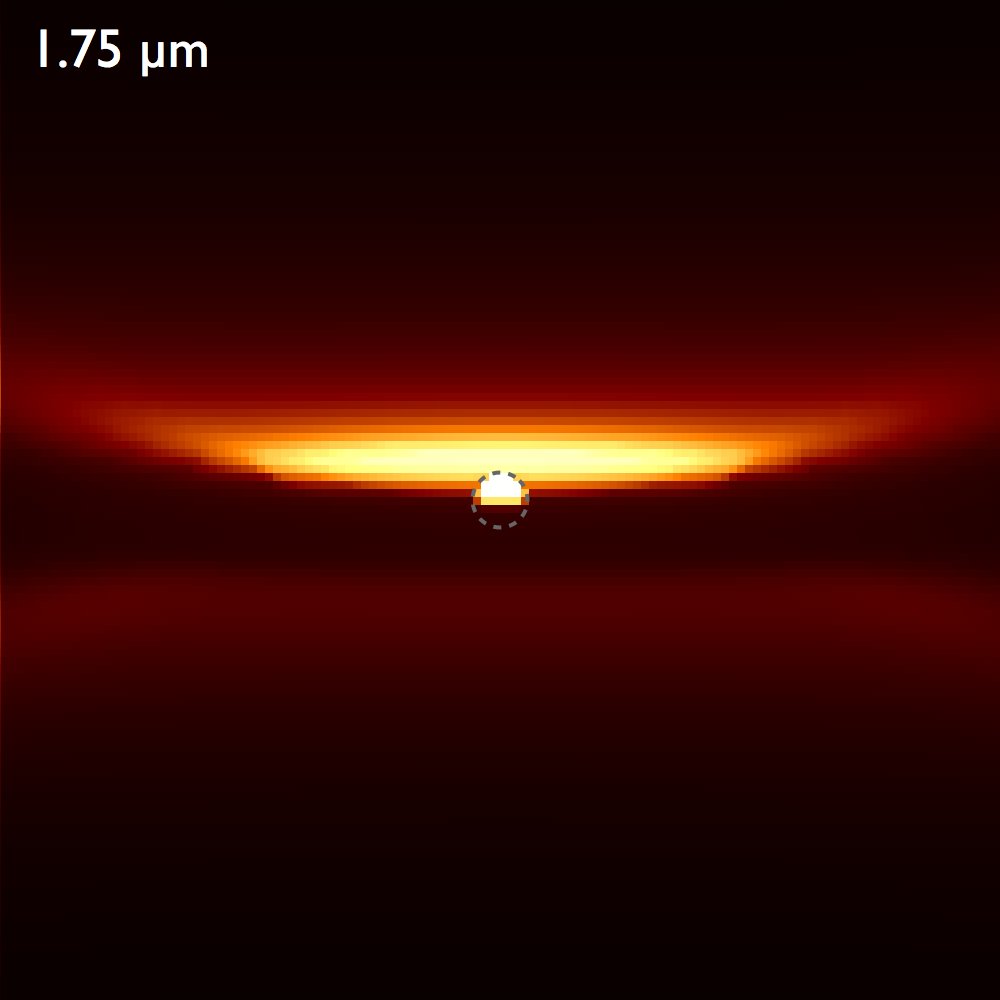}
\includegraphics[width=4.5cm]{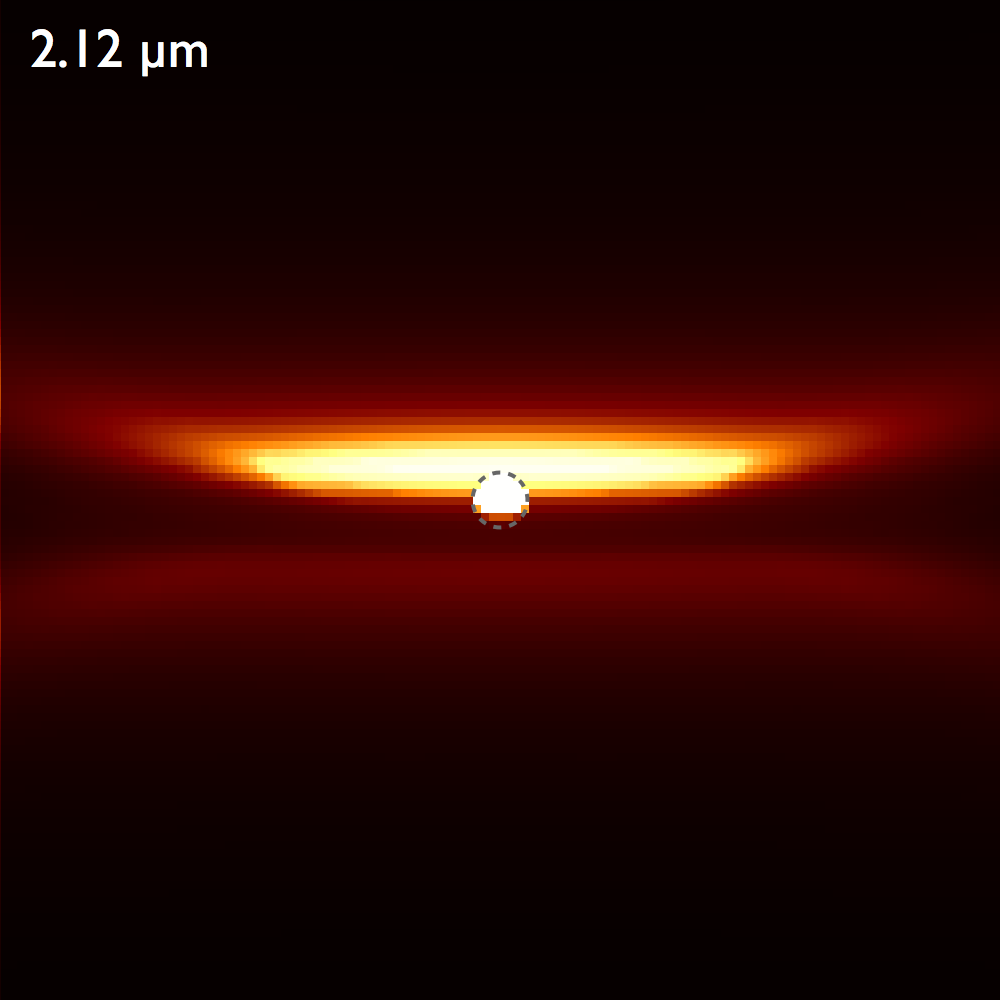}
\includegraphics[width=4.5cm]{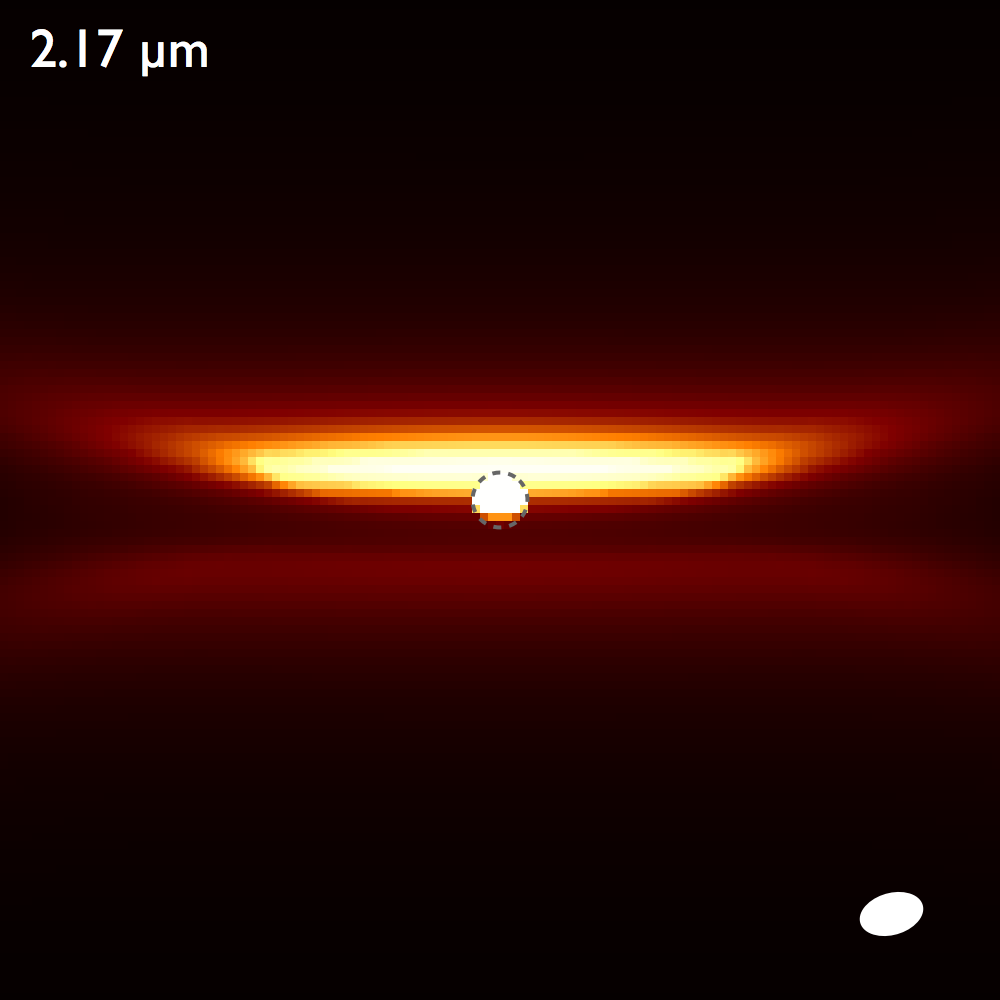}
\includegraphics[width=4.5cm]{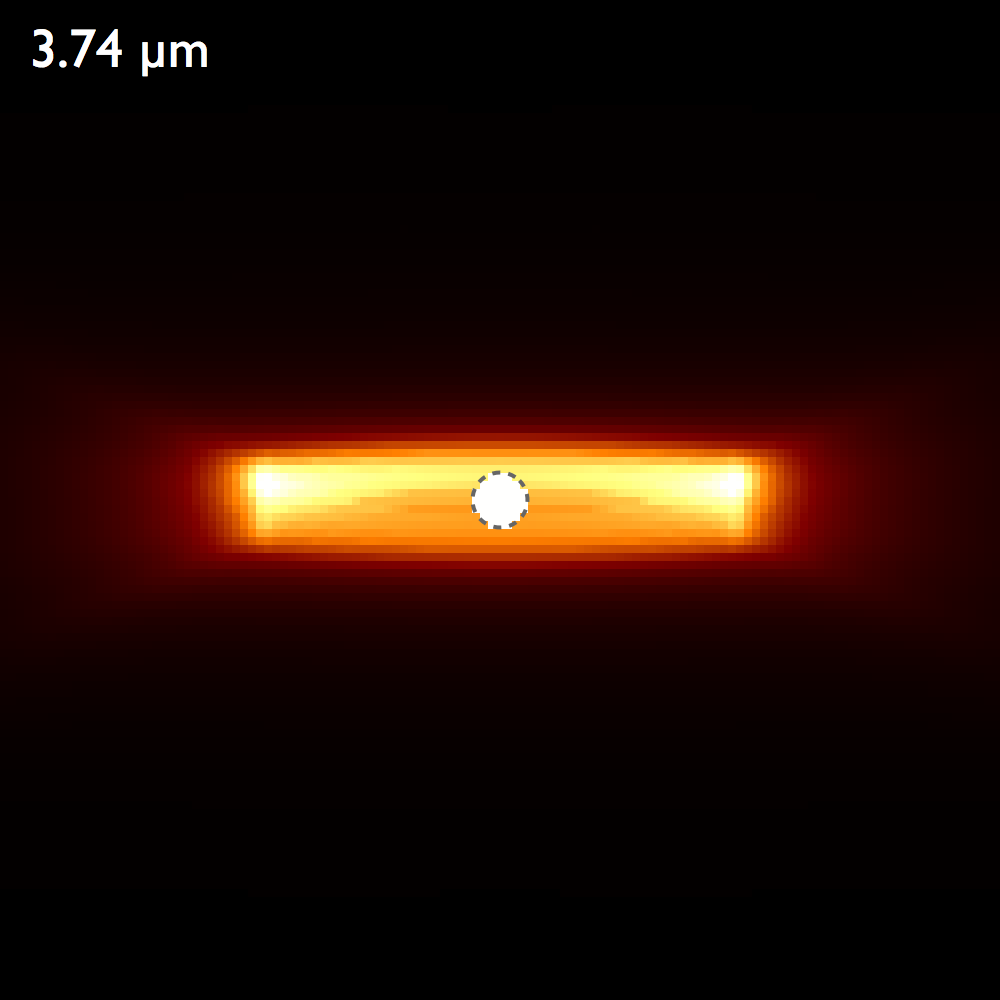}
\includegraphics[width=4.5cm]{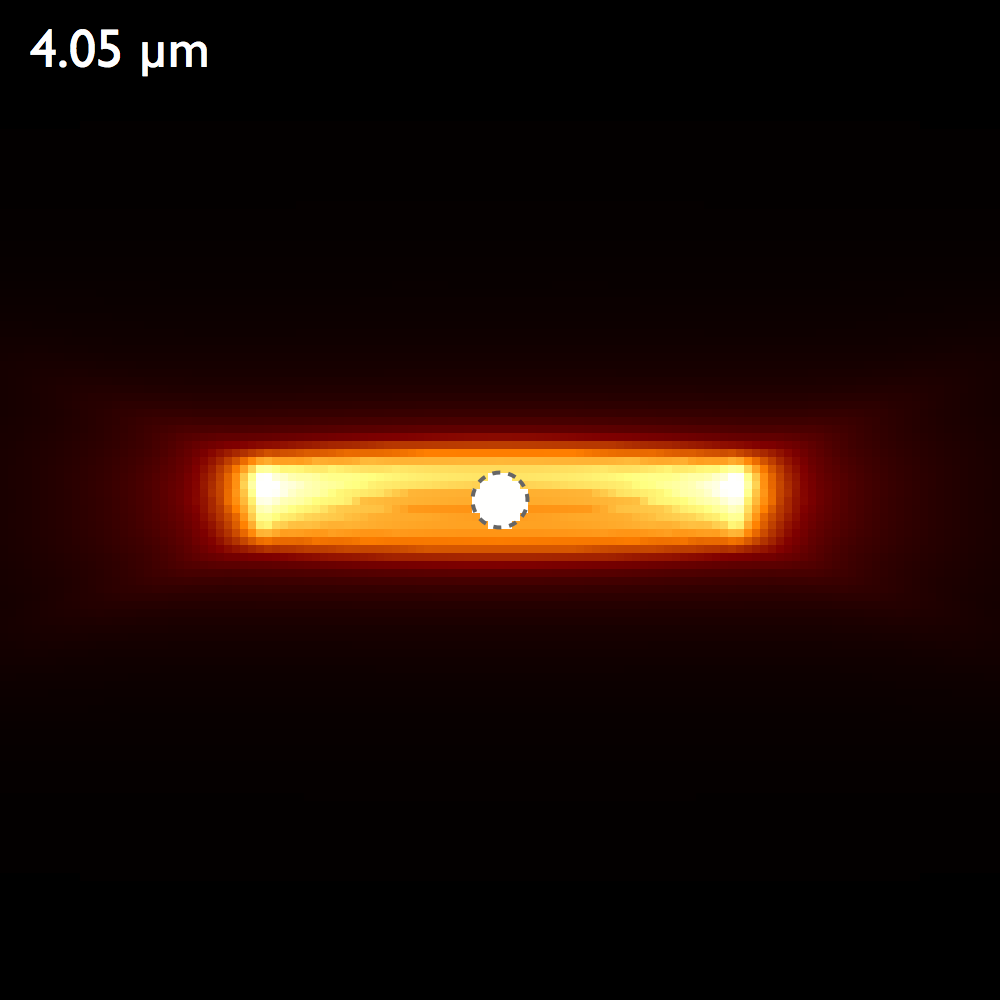}
\caption{RADMC-3D model images of L$_2$\,Pup. The field of view of each image is $0.414\arcsec$ (identical to Figs.~\ref{nondeconvJHK}, \ref{nondeconvL} and \ref{deconv80}), with north up and east to the left. The position and adopted angular size of the star is represented with a dashed circle. The color scale is a function of the square root of the irradiance, from 100 to 8\,000\,Jy\,arcsec$^{-2}$ in the $1.04-1.09\,\mu$m bands, 0 to 25\,000\,Jy\,arcsec$^{-2}$ in the $1.24-1.28\,\mu$m bands and 0 to 100\,000\,Jy\,arcsec$^{-2}$ in the $HKL$ bands. The white ellipse represented in the 2.17\,$\mu$m panel corresponds to the central object model derived from the VINCI data (Sect.~\ref{uniformellipse}). \label{modell2pup}}
\end{figure*}

\subsection{Derived parameters}

Because the morphology of the circumstellar disk depends on many parameters, we did not directly fit the model to the data. Considering the observed structure of the envelope in our VLT/NACO observations (Sect.~\ref{morphology}), we approximated the initial model parameters to obtain comparable images. To refine these parameters, we computed the photometry from the model at the wavelengths of the NACO filters and MIDI photometry and compared it with the observations. We adjusted the inclination of the system, the dust density, and the dust composition to match the measured fluxes. We considered that the disk was made of amorphous silicate, namely olivine MgFeSiO$_4$ and pyroxene MgFeSi$_2$O$_3$. Their physical constants were obtained from \citetads{1994A&A...292..641J} and \citetads{1995A&A...300..503D} at the Astrophysical Institute and University Observatory Jena\footnote{\url{http://www.astro.uni-jena.de/Laboratory/OCDB/amsilicates.html}}. The specific weight was 3.7 and 3.2~g.cm$^{-3}$ for the olivine and for the pyroxene, respectively. The derived parameters are listed in Table \ref{Table_Derived_Parameters} and the photometry given by the model is compared with the observed values in Fig.~\ref{l2pupsed} (green curve).

\begin{table}
	\caption{Derived parameters of the RADMC-3D dust disk model.}
	\label{Table_Derived_Parameters}      
	\centering          
	\begin{tabular}{l l} 
		\hline\hline
		Derived parameter & Value \\
		\hline
		\noalign{\smallskip}
		R$_\mathrm{in}$ & 6~AU \\
		R$_\mathrm{out}$ & 120~AU \\
		$\alpha$ & -3.5 \\
		MgFeSiO$_4$ density at R$_\mathrm{out}$ & $7.10^{-8}$~cm$^{-2}$ \\
		MgFeSiO$_4$ grain size & 0.1~$\mu$m \\
		MgFeSi$_2$O$_3$ density at R$_\mathrm{out}$ & $1.10^{-7}$~cm$^{-2}$ \\
		MgFeSi$_2$O$_3$ grain size & 0.3~$\mu$m \\
		H/R at R$_\mathrm{out}$ & 0.8~rad \\
		$\beta$ & 0.8\\
		Inclination & $84^\circ$\\
		\hline
	\end{tabular}
\end{table}
 
\subsection{Polarimetric signature of the disk}

\citetads{1986A&A...154....1M} observed that L$_2$\,Pup shows a significant overall polarization that slowly varies with time. They found a typical average polarization vector orientation of 160-170$^\circ$ in the visible, with a degree of linear polarization around $\gamma_V = 5\%$. This is consistent with the results in the $V$ and $I$ bands reported by \citetads{2001AJ....122.2017S}, but a clear variability is observed in the $UBV$ bands. In the infrared $JHK$ bands, \citetads{1971AJ.....76..901D} found degrees of linear polarization of $\gamma_H=0.8-1.5\%$, with orientations of 160-004$^\circ$. \citetads{2011Msngr.146...18L} recently presented promising observations of L$_2$\,Pup using the polarimetric sparse aperture masking mode of NACO (SAMPol) that show a clear polarization signal. However, few details are given in this article, and we do not discuss these observations here in more detail.

The RADMC-3D code enabled us to compute the polarimetric observables of our dust disk model. We obtain an integrated degree of linear polarization in the $V$ band of $\gamma_V = 6.7 \pm 2.1$\%. This model value  agrees well with the existing measurements. The associated position angle of the polarization vector is north-south
because we assumed an east-west direction for our disk model. While we did not fit this parameter, this direction (modulo 180$^\circ$)  also agrees reasonably well with the observed values.

A similar polarization behavior was observed by \citetads{2014IAUS..301..463N} for the Mira star V\,CVn in the visible, leading these authors to conclude that this star is surrounded by a circumstellar disk. L$_2$\,Pup is mentioned in this article as the only other example of a Mira surrounded by circumstellar disk, indicating that this may be a relatively rare configuration for this class of stars.

%__________________________________Discussion
\section{Discussion \label{discussion}}

\subsection{Evolutionary status of L$_2$\,Pup \label{evolutionarystatus}}

\begin{figure}[]
%\centering
\includegraphics[width=\hsize]{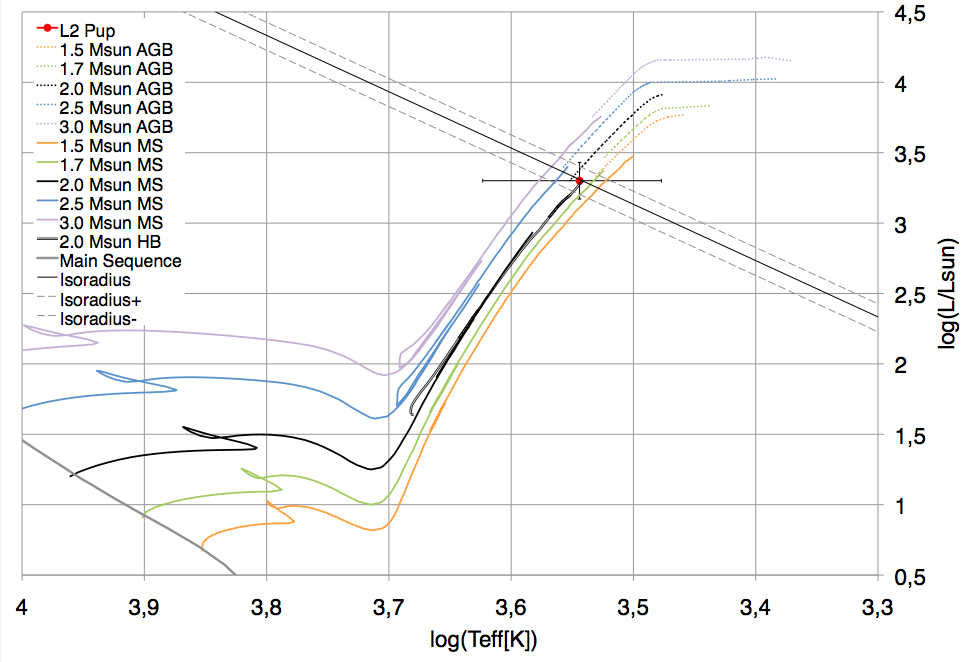} 
\caption{Position of L$_2$\,Pup in the Hertzsprung-Russell diagram (red dot) and evolutionary tracks from \citetads{2008A&A...484..815B}.\label{evolution}}
\end{figure}

Based on our interferometric estimate of the linear radius of L$_2$\,Pup presented in Sect.~\ref{uniformellipse} ($R= 123 \pm 14\,R_\odot$)
and the luminosity derived in Sect.~\ref{photom} ($L = 2000 \pm 700\,L_\odot$), it is possible to place L$_2$\,Pup in the Hertzsprung-Russell diagram. We plot in Fig.~\ref{evolution} the position of L$_2$\,Pup and the evolutionary models computed by \citetads{2008A&A...484..815B}. The constraint from the radius measurement combining the interferometric angular diameter and the parallax is shown as a diagonal line with its associated uncertainty domain.

%______________ Figure

L$_2$\,Pup appears to be at the beginning of the AGB phase, with a mass of approximately $2_{-0.5}^{+1.0}\,M_\odot$ and an age of $1.5_{-1.0}^{+1.5}$\,Gyr. The luminosity, effective temperature, and mass we derive for L$_2$\,Pup are consistent with those inferred by \citetads{2002A&A...388..609W}. \citetads{2005MNRAS.361.1375B} observed that L$_2$\,Pup's period shows small-scale random fluctuations of the phase and amplitude of its photometric variations.
Together with the fact that the brightness of L$_2$\,Pup in the $K$ band is very close to the predicted period-luminosity of Mira stars \citepads{2002MNRAS.337...79B}, this behavior is consistent with a star entering the Mira phase, but with a not yet stabilized pulsation period.
Another argument in this direction is the probable detection by \citetads{1987AJ.....94..981L} and \citetads{1999A&A...351..533L} of technetium in the atmosphere of L$_2$\,Pup. Thanks to its short lifetime ($\approx 2 \times 10^5$\,yr), the presence of the $s$-process $^{99}$Tc isotope in L$_2$\,Pup's spectrum is evidence for the recent occurrence of the third dredge-up.
The evolutionary status of L$_2$\,Pup therefore appears to be at an interesting, short-lived phase at the beginning of the Mira state.

\subsection{Mass-loss geometry}

\begin{figure}[]
\centering
\includegraphics[width=\hsize]{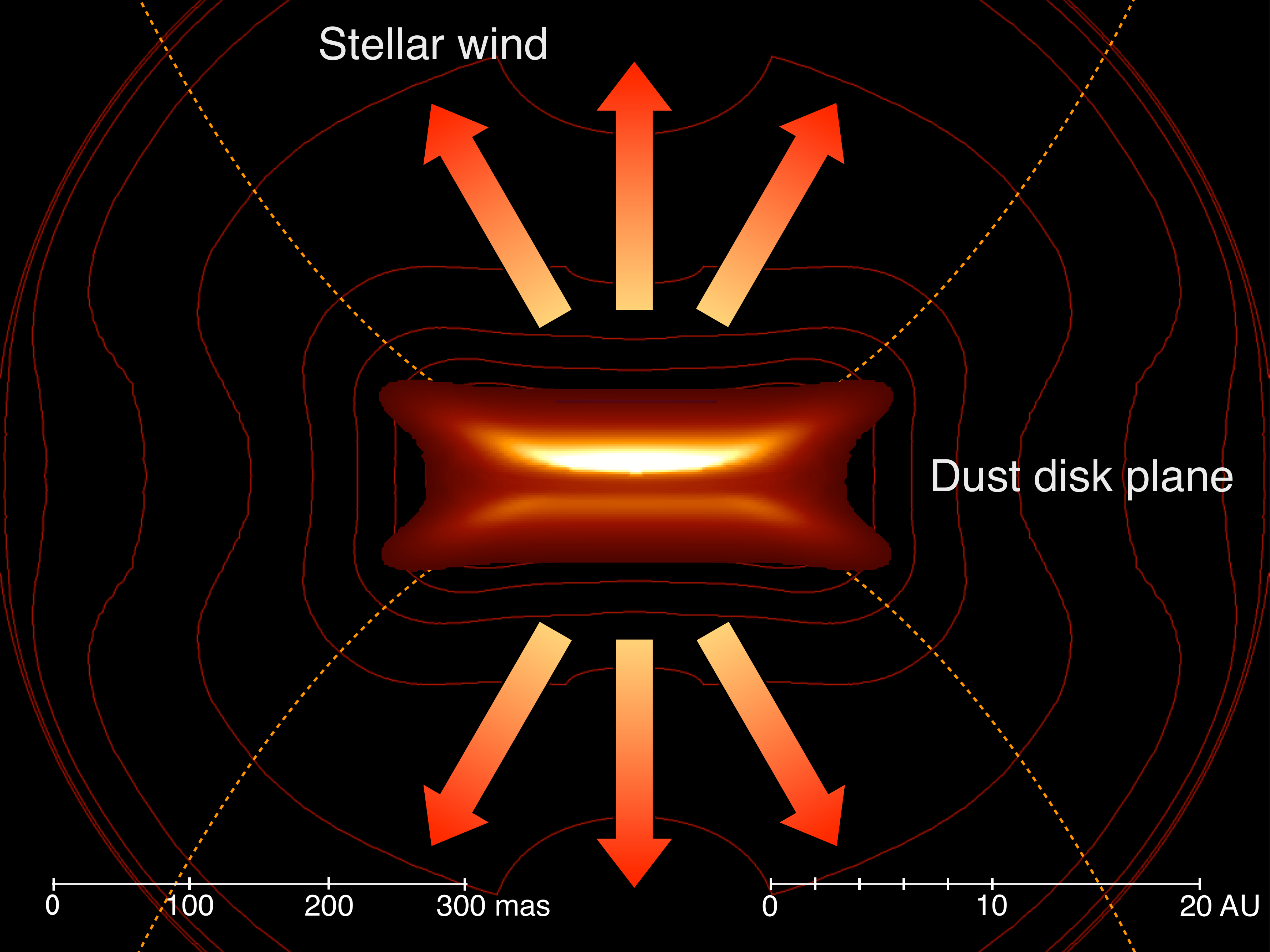} 
\caption{Diagram of a possible mass-loss geometry for L$_2$\,Pup. The image at the center was computed using our RADMC-3D disk model at $\lambda=10\,\mu$m. \label{l2pup10mic}}
\end{figure}

The observation of CO line emission in the radio domain by \citetads{2002A&A...391.1053O} and \citetads{2002A&A...388..609W} led these authors to propose that L$_2$\,Pup is losing mass at a rather low rate of $10^{-8}-10^{-7}\,M_\odot \mathrm{yr}^{-1}$, and with a slow expansion velocity $v_e < 5$\,km s$^{-1}$. This was also observed by \citetads{2004A&A...422..651S} and \citetads{1999A&AS..138..299K} in SiO and CO emission. The disk could provide an explanation for this peculiar behavior, as it could act as a collimator of mass loss in the plane of the sky. Under this hypothesis, the wind apparent velocity would appear to be lower than its true value because of the projection effect. A diagram of the configuration of our RADMC-3D disk model and the possible associated wind geometry is presented in Fig.~\ref{l2pup10mic}.

The best-fit RADMC-3D model implies that the inner rim of the circumstellar dust disk is located at a radius of $\approx 6$\,AU from the central star. This scale is clear in the NACO images (Figs.~\ref{deconv80} and \ref{UnresolvedFlux}). This relatively large distance of the dust from the star also results in a lack of warm dust emission in the spectral energy distribution (Fig.~\ref{l2pupsed}). This might mean that the distribution of the dust is a ring and not a flared disk. The presence of a companion appears as a natural explanation for this geometry \citepads{2009A&A...505.1221V}, but if we assume the absence of a binary companion (yet undetected), it is unclear which mechanism could create such a high equatorial and low polar density. A possibility that would allow the envelope to break the spherical symmetry is a relatively fast differential internal rotation that would create a temperature and wind velocity gradient between the equator of the star (cooler, slower wind) and its polar caps (hotter, faster wind). It could also affect the structure of the convection. But the asteroseismic observation of the spin-down of the core rotation in red giant stars at the last stages of the RGB \citepads{2012A&A...548A..10M} and the slow rotation rate in low-mass white dwarfs \citepads{1999ApJ...516..349K} tend to weaken this hypothesis. We note, however, that L$_2$\,Pup has a relatively high mass of $\approx 2\,M_\odot$, and \citetads{2012A&A...548A..10M} conclude that these stars keep a more rapidly rotating core after the tip of the RGB than lower mass stars. In addition, main-sequence stars in this mass range (i.e., A0-A1 spectral types and hotter) are more often faster rotators than lower mass objects \citepads{2007A&A...463..671R}. The presence of a magnetic field (possibly coupled to the star's pulsation) may also be a contributor, particularly because a weak surface magnetic field has recently been detected in Mira by \citetads{2014A&A...561A..85L}.

An important consequence of the wind collimation by the disk (or ring) is the possibility that L$_2$\,Pup will evolve into a bipolar nebula of the hourglass type. Such bipolar structures have already been observed in the visible \citepads[e.g. in \object{MyCn 18} by ][]{1999AJ....118..468S}, and in the millimeter domain \citepads[e.g. in the \object{Red Rectangle} by ]{2013A&A...557L..11B}. These objects present a remarkable axial symmetry, whose origin is often attributed to the presence of a central disk. However, only very few of these disks have been observed to date, for
example, in the \object{Red Rectangle} \citepads{2005A&A...441.1031B}, the \object{Ant} \citepads{2007A&A...473L..29C}, or M2-9 \citepads{2011A&A...527A.105L}. \citetads{2013A&A...557A.104B} recently reported CO emission from rotating disks around most sources of their sample of 24 post-AGB stars. These disks were probably already present during the AGB phase. For the \object{Boomerang} nebula, \citetads{2013ApJ...777...92S} also proposed that the central dusty structure may be a flared disk, although it has not yet been observed. From its position in the HR diagram, L$_2$\,Pup appears as a relatively ``young" AGB star, in the first stages of its evolution towards an hourglass planetary nebula, and therefore a possible progenitor of these spectacular structures.

%______________ Figure

\subsection{Binarity of L$_2$\,Pup}

\subsubsection{Binarity vs. pulsations \label{binarypulsation}}

 \citetads{2009A&A...498..489J} discussed and dismissed the possibility of a stellar companion close to L$_2$\,Pup with a 141-day orbital period.
 %Their reasoning is based on geometrical arguments (the companion would orbit very close to the giant star's surface) and the fact that the mass of the system would have to be very large to explain the short orbital period.
 We consider the classical relation from Kepler's third law for a binary star,
\begin{equation}
P^2 = \frac{4 \pi^2 r^3}{G\left( m_1 + m_2 \right)} 
,\end{equation}
where $m_1$ and $m_2$ are the two object masses, $P$ the orbital period, and $r$ the separation between the stars (the orbits are assumed to be circular). The application of this relation with $m_1 = 2\,M_\odot$ (Sect.~\ref{evolutionarystatus}) and $P = 141$\,d gives a solution with $r = 155\,R_\odot$ and $m_2 = 0.5\,M_\odot$. For edge-on circular orbits, the radial velocity amplitude $K_1$ of the primary star's displacement is given by
\begin{equation}
K_1 = \left( \frac{2 \pi G}{P} \right)^{1/3} \frac{m_2}{m_1} \left( m_1 + m_2 \right)^{1/3}
.\end{equation}
For the above solution, the radial velocity amplitude $K_1 \approx 14$\,km\,s$^{-1}$, which agrees with the radial velocity amplitude of 12\,km\,s$^{-1}$ measured by \citetads{2005A&A...431..623L}. Therefore, in principle, a low-mass stellar companion could explain the observed radial velocity amplitude and the photometric period, with an orbit close to but above the infrared photosphere of L$_2$\,Pup ($R = 120\,R_\odot$, Sect.~\ref{uniformellipse}).
%Such a very high mass is clearly inconsistent with the location of L$_2$\,Pup in the H-R diagram.  
We stress that this solution poses other difficulties. In particular, the amplitude ($\approx 2$\,mag), period, and random phase fluctuations of the photometric variation in the visible cannot be explained by a rotating ellipsoidal primary star. It thus appears unlikely that a stellar mass companion is present with a 141-day orbital period and can explain both the astrometric and photometric signals measured on L$_2$\,Pup. However, we cannot formally exclude the existence of a very close-in companion of L$_2$\,Pup from Kepler's laws alone.

As discussed by \citetads{2002MNRAS.337...79B}, the photometric variation period and luminosity of L$_2$\,Pup agree well with the period-luminosity relation of Mira stars. For instance, \citetads{2008MNRAS.386..313W} gave $M_K = -6.44 \pm 0.08$ for a period of 141\,d, which agrees excellently with the absolute magnitude $M_K = -6.36 \pm 0.05$ we derive from the average magnitude $m_K = -2.13 \pm 0.05$ assuming $E(B-V)=0.6$, (Sect.~\ref{photom}). L$_2$\,Pup therefore appears as a probable pulsating Mira star. Another way to look at this star is by using the surface brightness color relations by \citetads{2004A&A...426..297K} to estimate its angular diameter at its extremal phases. From the $(V-K)$ relations and considering the $\approx 2$\,mag visible photometric amplitude, we obtain a pulsation amplitude of $30-40$\% of the star's radius, or approximately $40\,R_\odot$. Assuming a sinusoidal radial velocity curve, the displacement of the star's surface over the 141\,d period translates into a radial velocity amplitude of $\approx 14$\,km\,s$^{-1}$, which is consistent with the 12\,km\,s$^{-1}$ value measured by \citetads{2005A&A...431..623L}.

From \citetads{2002A&A...393..563L}, the typical radial velocity amplitude of semiregular variable stars is $\approx 4$\,km\,s$^{-1}$ (short interval) to 8\,km\,s$^{-1}$ (long interval) with a typical shift of about 4\,km\,s$^{-1}$ blue from the center-of-mass velocity.  L$_2$\,Pup's measured radial velocity amplitude of 12\,km\,s$^{-1}$ is therefore too large to be consistent with semiregular pulsators. \citetads{1999A&A...341..224L} obtained radial velocity measurements of the Mira star R\,Vir in the infrared. This star has a similar period (145.5\,d) as L$_2$\,Pup, although with a larger photometric amplitude in the visible \citepads[6\,mag, ]{1985IBVS.2681....1K}, and it does not show technetium in its spectrum.
\citetads{2007ApJ...654L..77E} observed R\,Vir using the Keck Interferometer and obtained a radius of $130\,R_\odot$. \citetads{1994ApJ...422..102J} found a low mass-loss rate of $10^{-7}\,M_\odot\,\mathrm{yr}^{-1}$ for this star. Both quantities are very similar to L$_2$\,Pup. Considering the $130\,R_\odot$ radius and the relative pulsation amplitude listed in Table~3 of \citetads{1999A&A...341..224L}, we obtain a pulsation amplitude for R\,Vir of $\approx 40$\% of the star's radius, which is consistent with what we derive for L$_2$\,Pup from surface brightness considerations. Although its photometric amplitude in the visible is significantly larger, R\,Vir therefore appears as a reasonably good analog of L$_2$\,Pup, strengthening the association of L$_2$\,Pup to the Mira class.

\subsubsection{Interferometry and adaptive optics}

The presence of a long orbital period companion brighter than $\approx 5\%$ of the flux of L$_2$\,Pup in the near-infrared is unlikely in the field of view of our NACO images. Our VINCI interferometric observations also do not show the signature of a stellar companion very close to the central star. However, the dust disk and its complex morphology makes the interferometric detection of a secondary star particularly difficult. Due to the very high brightness of L$_2$\,Pup, the translation of a $\approx 5\%$ relative flux limit in terms of companion mass is in fact not constraining (assuming that it is less massive than L$_2$\,Pup itself and coeval).

Another possibility is that L$_2$\,Pup's companion is an evolved stellar remnant. Initially more massive than the giant star, it would have evolved faster and would therefore most likely be a white dwarf. Due to the extreme flux contrast with the giant star, such a companion would be undetectable through its flux contribution alone. Assuming it already went through the AGB phase, it might be at the origin of part of the dust surrounding L$_2$\,Pup, however.
The asymmetric and thin curved loop in the NACO images, particularly visible in the $L$ band (Fig.~\ref{dec80loop}) but also in the $H$ band, is an indication that an interaction between the wind from L$_2$\,Pup and a companion star may be occurring at present.

 \subsubsection{Apparent astrometric displacement}
 
The astrometric displacement of the center-of-light of L$_2$\,Pup observed in the \emph{Hipparcos} data by \citetads{2007ApJS..173..137G} has a period almost identical to the pulsation of the star and a semi-amplitude of 9.5\,mas.
This unlikely coincidence weakens the proposed explanation of its origin by the presence of a companion star. We develop in this paragraph the idea that the variable lighting of the circumstellar disk over the pulsation cycle of the star appears as a more plausible hypothesis.

Because the disk is seen almost edge-on, its aspect at the observing wavelength of \emph{Hipparcos} is probably similar to the $J$ band images presented in Sect.~\ref{nacoobs}, that is, it should exhibit two distinct scattering lobes above and below the plane of the disk. As the star changes in brightness and angular diameter, the flux ratio between the two lobes will also change and displace the photocenter. More distant dusty features (such as the loop), depending on their linear (de-projected) distance from the star and shadowing effects in the star's envelope, will also show
a changing relative brightnesses.  A combination of these different lighting effects in the close environment of L$_2$\,Pup could well shift the center-of-light position by the relatively large observed semi-amplitude of 9.5\,mas.

As an alternative explanation, during the expansion phase of the pulsation, the northern hemisphere of the stellar disk may ``emerge'' above the dust disk, as the star's radius inflates by approximately $40\,R_\odot$ (Sect.~\ref{binarypulsation}). Because the southern hemisphere of the star will remain darkened by the dust, this will induce a shift of the center-of-light toward the north. Depending on its amplitude, such a displacement could mimick the orbital reflex motion caused by a companion. This scenario alone, however, is insufficient to explain the 9.5\,mas amplitude of the \emph{Hipparcos} displacement observed by \citetads{2007ApJS..173..137G} because it is larger than the stellar radius. It could play a role in combination with lighting effects in the inhomogenous nebula, however.

It is interesting to note that a time-variable lighting effect was also invoked by \citetads{2013ApJ...774...21M} to explain their observation of variable SiO maser emission. This interpretation is consistent with our model of an edge-on disk around L$_2$\,Pup and the presence of dusty material in front of the star.

\subsubsection{X-ray emission}

The presence of a close companion orbiting in L$_2$\,Pup's wind will result in accretion of the AGB stellar wind, for example, through the wind Roche-lobe overflow process \citepads{2013A&A...552A..26A}. The low mass-loss rate and slow wind of L$_2$\,Pup \citepads{2002A&A...391.1053O} would likely result in a low accretion rate on a main-sequence or compact companion, and consequently little X-ray production. This X-ray flux would be additionally reduced by the strong absorption by the dust disk. An X-ray emission of L$_2$\,Pup was recently reported by \citetads{2012A&A...543A.147R} from archival XMM data, but it was attributed by these authors to a leak of optical (red) photons and not to true emission from the star. So the X-ray emission does not provide a clear diagnostic on the existence of a companion of L$_2$\,Pup.

%__________________________________Conclusion
\section{Conclusion}

The NACO observations we presented in Sect.~\ref{nacoobs} show that L$_2$\,Pup is veiled by a large dust band. Its morphology appears consistent with a circumstellar dust disk seen almost edge-on. Its aspect changes significantly from 1.0 to 4.0\,$\mu$m. At shorter wavelengths ($J$ band), the scattering by dust grains is very efficient, resulting in a high opacity and the presence of a dark band obscuring the stellar light. In the $L$ band, the dust scattering is much less efficient, and the thermal emission from the inner edge of the disk becomes dominant. In the intermediate near-infrared bands ($HK$), the disk is translucent and the central star becomes progressively more visible as the wavelength increases. Our simulations using the RADMC-3D radiative transfer code strengthened this interpretation because the model we developed reproduced both the aspect of the NACO images as a function of wavelength and the observed spectral energy distribution.
If we consider that the dust density is inhomogenous in the disk, the variable dust obscuration scenario proposed by \citetads{2002MNRAS.337...79B} to explain the long-term variability of L$_2$\,Pup is consistent with the proposed edge-on configuration. In this framework, the long minimum flux phase observed since $\approx 1995$ would result from the transit of a dense part of the dust disk in front of the star, creating the morphology observed in the NACO images.

A simple evolutionary analysis shows that L$_2$\,Pup is probably in an early phase of the AGB, with a mass of approximately $2\,M_\odot$, and an age of about 1.5\,Gyr. Its physical properties are very similar to those of the short-period Mira star R\,Vir, and its luminosity is consistent with the period-luminosity relation of this class of stars. We propose that L$_2$\,Pup should be classified as a short-period Mira star instead of a semiregular variable. 

We did not detect any stellar companion to L$_2$\,Pup in our NACO images or in our VINCI interferometric observations.  We propose that the astrometric wobble observed by \citetads{2007ApJS..173..137G} is caused by time-variable lighting effects on L$_2$\,Pup's circumstellar nebula and not by an orbiting companion. The measured radial velocity amplitude is also consistent with the pulsation of the central star, without the need to invoke a secondary object. We stress, however, that we do not exclude the possibility that a companion is present, particularly if its orbital period is significantly longer than the pulsation period of L$_2$\,Pup. In addition, the loop structure we detected in our $L$ -band images points at a possible interaction of a hidden companion with the dusty wind from the central star. As discussed by \citetads{2009A&A...505.1221V}, the mere existence of a dusty disk around L$_2$\,Pup is also in itself an indication that a companion may be present, as disks in post-AGB stars appear inherently connected to binarity.

L$_2$\,Pup presents interesting challenges for the modeling of the envelope of an evolved, moderately massive star. The geometric configuration that we propose, with a large disk seen almost edge-on, is particularly promising to test and develop our understanding of the formation of bipolar planetary nebulae in the post-AGB phase. Such a disk could also be a favorable environment to form large dust grains, and possibly planetesimals. The presence of circumstellar material around white dwarfs  \citepads{2005ApJS..161..394F} could be the final result of this secondary planet formation episode in AGB dust disks such as L$_2$\,Pup's.

%__________________________________Acknowledgements
\begin{acknowledgements}
% The authors thank Dr. Olivier Chesneau (Observatoire de la C\^ote d'Azur) for stimulating discussions on the nature of L$_2$\,Pup.
This research received the support of PHASE, the high angular resolution
partnership between ONERA, Observatoire de Paris, CNRS and University Denis Diderot Paris 7.
We acknowledge financial support from the ``Programme National de Physique Stellaire" (PNPS) of CNRS/INSU, France.
STR acknowledges partial support by NASA grant NNH09AK731.
AG acknowledges support from FONDECYT grant 3130361.
We acknowledge with thanks the variable star observations from the AAVSO International Database contributed by observers worldwide and used in this research.
This research made use of Astropy\footnote{Available at \url{http://www.astropy.org/}}, a community-developed core Python package for Astronomy \citepads{2013A&A...558A..33A}.
We used the SIMBAD and VIZIER databases at the CDS, Strasbourg (France), and NASA's Astrophysics Data System Bibliographic Services.
We used the IRAF package, distributed by the NOAO, which are operated by the Association of Universities for Research in Astronomy, Inc., under cooperative agreement with the National Science Foundation.
Some of the data presented in this paper were obtained from the Multimission Archive at the Space Telescope Science Institute (MAST). STScI is operated by the Association of Universities for Research in Astronomy, Inc., under NASA contract NAS5-26555. Support for MAST for non-HST data is provided by the NASA Office of Space Science via grant NAG5-7584 and by other grants and contracts.
\end{acknowledgements}

%__________________________________Bibliography
\bibliographystyle{aa} % style aa.bst
\bibliography{biblio}
\end{document}